\documentclass[11pt,twocolumn]{article}
\usepackage[
  top=2cm,
  bottom=2cm,
  left=2cm,
  right=2cm
]{geometry}
\usepackage[hidelinks]{hyperref}
\usepackage[table,xcdraw]{xcolor}
\usepackage[normalem]{ulem}  % for underlines

\definecolor{darkblue}{RGB}{0,0,139}

\renewcommand{\href}[2]{\textcolor{darkblue}{\uline{#2}}}
\usepackage{graphicx} 
\usepackage{makecell}
\usepackage{multirow}
\usepackage{array}
\usepackage{url}
\usepackage{amsmath}
\usepackage{booktabs}
\usepackage{longtable} 
\usepackage{tabularx} 
\usepackage{ltablex} 
\usepackage{threeparttable}
\usepackage{caption} 
\usepackage[T1]{fontenc}
\usepackage{float}
\usepackage{booktabs, array, ragged2e}

\usepackage{courier}  % or \usepackage{beramono}
\usepackage{xspace}
\newcommand{\mypara}[1]{\vspace{4pt}\noindent{\textbf{{#1}}\xspace}}

\usepackage{enumitem}
\setlist[itemize]{leftmargin=*, nosep} 
\usepackage{verbatim}

\title{Evidence of political bias in search engines and language models before major elections}
\author{Íris Damião$^{1,2}$ \and Paulo Almeida$^{1}$ \and João Franco$^{1}$\footnote{Current address: Vermont Complex Systems Institute, Burlington, Vermont, USA} \and Nuno Santos$^{2,3}$ \and Pedro C. Magalhães$^{4}$  \and Joana Gonçalves-Sá $^{1,5}$}
\date{}

\begin{document}

\maketitle
\small
\noindent $^{1}$ Social Physics and Complexity Lab - SPAC, LIP - Laboratório de Instrumentação e Física Experimental de Partículas, Lisbon, Portugal \\
$^{2}$ Instituto Superior Técnico - Universidade de Lisboa, Lisbon, Portugal \\
$^{3}$ INESC-ID, Lisbon, Portugal \\
$^{4}$ Instituto de Ciências Sociais da Universidade de Lisboa, Lisbon, Portugal \\
$^{5}$ NOVA LINCS - NOVA Laboratory for Computer Science and Informatics, NOVA School of Science and Technology, NOVA
University Lisbon, Lisbon, Portugal \\
\\
\noindent Corresponding author: Joana Gonçalves-Sá (e-mail: joanagsa at lip dot pt)\\

\begin{abstract}

%Alternative abstract, closer to 150 words:

Search engines (SEs) and large language models (LLMs) are central to political information access, yet their algorithmic decisions and potential underlying biases remain underexplored. We developed a standardized, privacy-preserving, bot-and-proxy methodology to audit four SEs and two LLMs before the 2024 European Parliament and US presidential elections. We collected answers to approximately 4,360 queries related to elections in five EU countries and 15 US counties, identified political entities and topics in those answers, and mapped them to ideological positions (EU) or issue associations (US). In Europe, SE results disproportionately mentioned far-right entities beyond levels expected from polls, past elections, or media salience. In the US, Google strongly favored topics more important to Republican voters, while other search engines favored issues more relevant to Democrats. LLMs responses were more balanced, although there is evidence of overrepresentation of far-right (and Green) entities. These results show evidence of bias and open important discussions on how even small skews in widely used platforms may influence democratic processes, calling for systematic audits of their outputs.
    
\end{abstract}

\section{Introduction}

The vast majority of the world's population now access information online~\cite{pew2025newsplatform}, primarily through search engines (SEs) and, increasingly, through Large Language Models (LLMs)~\cite{robarts2025bbc}. SEs are used daily by billions worldwide ~\cite{semrush2026top} and have largely replaced traditional media as the first point of access for news and information~\cite{newman2025digital,pew2025newsplatform}. More recently, LLM-based chatbots such as ChatGPT have emerged as alternatives to SEs~\cite{bing_ai_usage}, while many SEs have integrated AI-generated summaries into their results~\cite{google_ai_search}.
While there is wide debate on the accuracy of LLMs, SEs are generally perceived as highly trustworthy ~\cite{in_google_we_trust, students_faculty_trust_google, misplaced_trust} and, unlike social media, are often seen as neutral~\cite{mont2022trust}. As a result, they have become central actors in the informational political ecosystem particularity ahead of elections~\cite{burman2024google}.

Despite their different technical foundations, both SEs and LLMs rely on complex, opaque algorithms developed through proprietary and largely non-transparent processes. While such systems can be unproblematic for everyday queries, their influence becomes consequential when responding to sensitive~\cite{Israel_Palestine} or political information needs. Empirical evidence shows that biased SE rankings--such as favoring one political candidate over another--can shift voting intentions among up to 20\% of undecided voters ~\cite{digital_personalization_effect, epstein2015search}, and affect issue salience and party visibility ~\cite{lee2016agenda, diakopoulos2018vote}, two dimensions that are well known to affect vote choices~\cite{hopmann2010effects, adams2024much}, particularly for radical right parties~\cite{vasilopoulou2024electoral}. Such algorithmic risks extend beyond ranking effects, and SEs have been shown to reproduce gender and racial biases in image and text ~\cite{otterbacher2017competent,Makhortykh2021DetectingRA,noble2018algorithms}, propagate erroneous information~\cite{Urman2021WhereTE}, and reinforce political bias following politically charged queries~\cite{self_imposed_filter_bubble,trielli2019partisan}. 

LLMs raise parallel concerns: they frequently generate false information (“hallucinations”)~\cite{apnews2023aihallucinations, tonmoy2024survey}, overlook factual inconsistencies even when authoritative sources are available~\cite{huang2023hallucination,venkit2024confidently}, exhibit sycophantic behavior by aligning outputs with perceived user views~\cite{laban2023are,sharma2024generative}, and may carry or amplify social and political biases inherited from inaccurate training data~\cite{silberg2019notes}.

Given the growing reliance on both SEs and LLMs for information seeking, and their potential to influence beliefs and decisions, this study investigates whether these algorithmic systems display political bias when responding to neutral election-related queries in the critical days leading up to an election -- when the growing numbers~\cite{late_voters} of late deciders are especially susceptible to campaign information~\cite{cameron2025electoral, willocq2019explaining}. 

This is non-trivial due to algorithmic opacity and to the difficulty of defining political neutrality across contexts. Since the influential work of Introna and Nissenbaum~\cite{introna2007shaping}, which first highlighted the societal implications of opaque, commercially driven algorithms, several approaches to measuring bias have been proposed. Early work by Mowshowitz and Kawaguchi~\cite{mowshowitz2002assessing} defined bias as a deviation from balanced representativeness of web documents for a given query or topic, constructing baselines by aggregating results across multiple SEs and quantifying statistically significant deviations from that average. More recent studies have focused on single contexts, primarily the US~\cite{shepherd2017method,auditing_partisan_bias,trielli2019partisan,urman2021matter,gezici2020evaluation,gezici2020evaluation}, with only a few exceptions~\cite{krafft2019search_engine_manipulation,makhortykh2025campaigning}. Most estimate bias by averaging the ideological leaning of sources on the results page~\cite{auditing_partisan_bias,gezici2020evaluation}, sometimes combined with sentiment analysis~\cite{gezici2020evaluation}. 

Findings remain mixed, reporting either dominance of mainstream left-leaning outlets~\cite{google_shaping_attention} or increased visibility of hyperpartisan and far-right sources~\cite{data_voids_far_right,hyperpartisan_sources}. Importantly, these studies rarely disentangle ranking criteria from contextual factors such as media popularity, nor do they provide a clear definition of neutrality. This is particularly relevant in the US context, where many widely trusted news outlets are classified as left-leaning~\cite{allsides2025mediabias}, raising ambiguity about whether observed patterns reflect algorithmic bias or the information ecosystem. Moreover, most rely on simple and explicitly political queries (e.g.,candidate or party names), which do not reflect typical user behavior and are known to skew results ~\cite{voter_centered_audits}.

For LLMs, political bias is often assessed through survey-style instruments, such as political compasses, with many studies finding a liberal or left-leaning tendency~\cite{potter-etal-2024-hidden,yang2024unpacking,political_bias_llms}. There is also evidence of sensitivity to language choice ~\cite{chatgpt_political_bias} and debiasing strategies~\cite{yang-etal-2025-rethinking-prompt,tran2022fairness}. Finally, model parameters -- such as temperature -- or prompts can affect results and even lead to selective answer refusals, in political contexts~\cite{urman2025silence}.

This work presents a novel approach to studying political bias in SEs and LLMs, using salience -- the frequency of mentions of political entities and polarized issues -- as a proxy. We measured how often different actors and agendas appeared in SE results and LLM outputs for neutral,  natural~\cite{voter_centered_audits} election-related prompts (e.g., “European elections 2024”, ``Who should I vote for''), with the intuition that systematic differences in mention frequency may reflect underlying bias toward certain political entities or agendas. Crucially, rather than treating overrepresentation as self-evident, we benchmark observed salience against external indicators that capture differential newsworthiness and user demand -- including past, expected electoral strength and legacy media coverage -- allowing us to ask: \textit{do SEs and LLMs overrepresent certain entities or agendas beyond what these factors would predict?}

\begin{figure*}[!ht]
    \centering
    \includegraphics[width=0.95\textwidth]{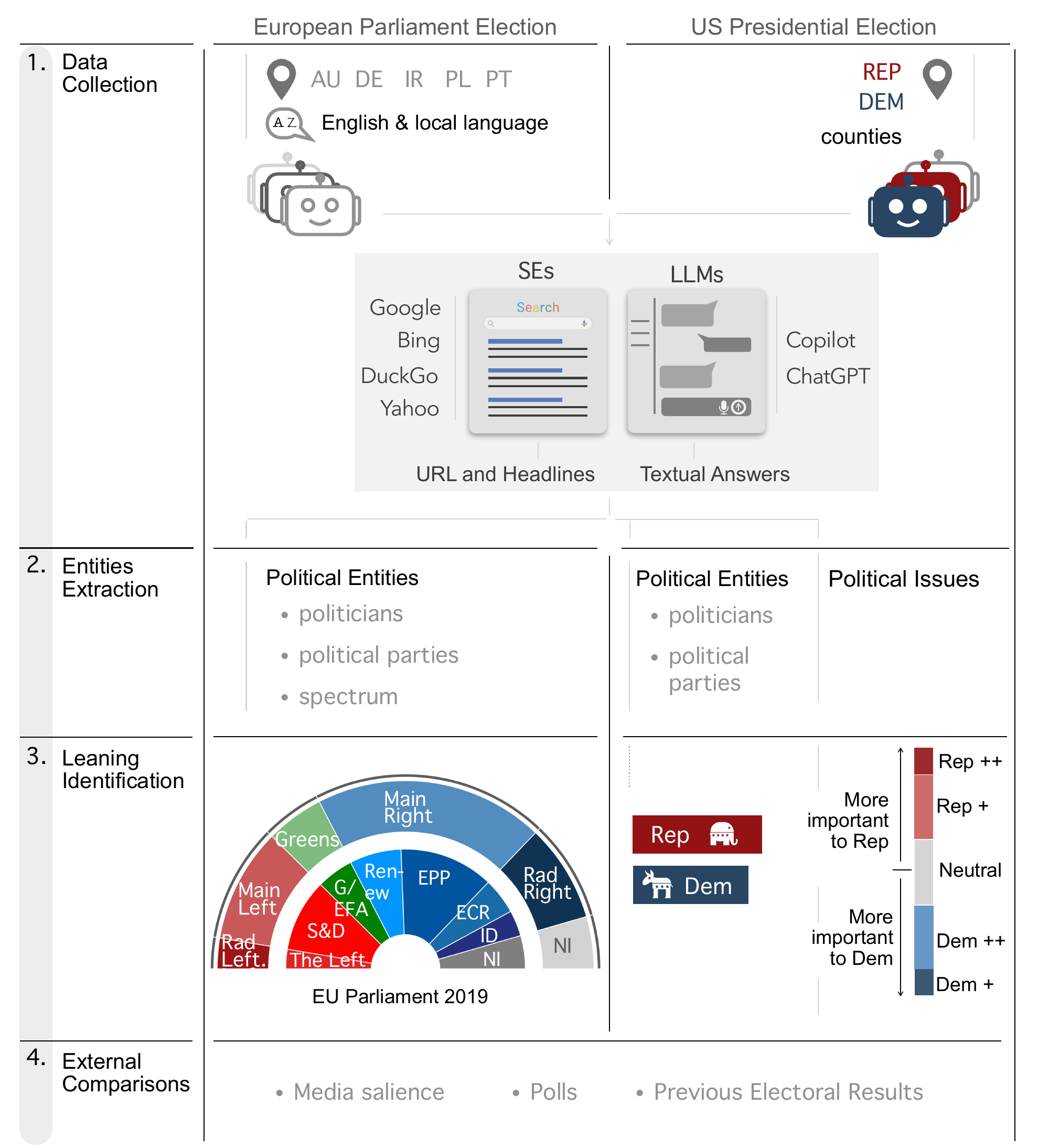}
    \caption{\textbf{Methodological scheme.} \textbf{(1)} Bots collected data from search engines (Google, Bing, DuckDuckGo, Yahoo) and LLMs (Copilot, ChatGPT), from different locations within the European Union (Austria - AU, Germany - DE, Ireland - IR, Poland - PL, and Portugal - PT) or from different counties in the United States of America. For LLMs, location was simulated by prompting the model in the most spoken language of each country (EU only), rather than through actual geolocation. \textbf{(2)} Political entities were extracted from the results collected in (1), i.e. from URLs and headlines for SEs and from textual answers for LLMs. \textbf{(3)} These entities were mapped to their ideological leanings using their placement on political spectra in the case of the EU, or according to the issues considered more important to Democratic or Republican voters in the case of the US. \textbf{(4)} The observed leanings were compared with external factors (media salience, polls, prior electoral results).}
    \label{fig:methodological_scheme}
\end{figure*}

We designed a privacy-preserving, bot-based tool and conducted a large-scale study of two major 2024 elections: to the European Parliament and the US Presidential elections. Bots simulated human behavior (clicking, scrolling, typing) and collected results from four SEs (Google, Bing, DuckDuckGo, Yahoo) and the two most popular LLMs at the time (Microsoft Copilot and ChatGPT). To capture geographic and political diversity, bots operated from five European countries and 15 US counties. From SE results (URLs and headlines) and LLM outputs, we quantified mentions of political figures, parties, ideological spectra, and politicized issues (US only) and compared their frequency against past election outcomes, polling, and media salience (see Figure~\ref{fig:methodological_scheme}).

Our findings, from close to 4,360 SE first-page results (each containing ~8 results per page), and 205 LLM answers show that SEs disproportionately mention far right entities, particularly in Europe, and that LLMs, while more cautious and prone to refusal, also demonstrate patterns of ideological preference – varying by model, language, and prompt wording. 
%While these findings are limited by the diversity of queries and dataset size, particularly in the case of LLMs, they reveal a concerning pattern: certain parties were systematically favored over others in the days preceding the election. Future research should expand the query set, dataset size, and range of political systems studied, to better understand what drives this behavior and whether it reflects training data, algorithmic design, or techniques implemented by such parties to increase their visibility (e.g. search engine optimization). 

\section{Results} 

In the days leading up to the 2024 European Parliament elections, we deployed 25 independent bots across five countries -- Austria, Germany, Ireland, Poland, and Portugal (Figure~\ref{fig:methodological_scheme} - Step 1). These countries were selected for political, geographical, linguistic, and polling diversity within the European Union: at the time of selection, Ireland’s polls showed a near tie between the radical left and the mainstream right parties; Portugal's a similar distribution between the two mainstream (left and right) parties; Germany a dominance of the mainstream right; Austria had the far-right nearly level with the main center-right party; and in Poland's polls radical right parties were leading, with almost no expected representation of the left (Supplementary Figure~\ref{fig:polls_eu}). The bots simultaneously queried four search engines (SEs) -- Google, Bing, DuckDuckGo, and Yahoo -- using neutral search terms (Table~\ref{tab:queries} and Methods). To capture language specific results and to simulate local and EU-wide users, queries were conducted in English and in each country’s main language. Similar prompts (Table~\ref{tab:queries}, Methods) were submitted to two popular large language models (LLMs): Copilot (with live web access) and ChatGPT-4o (without), from the same 5 locations, using only local languages. 

A similar procedure was applied for the 2024 US presidential elections. Bots were deployed from 15 counties with different historical leanings in states showing different polling intentions (pro-Republican, pro-Democratic, or highly divided; Supplementary Figures~\ref{fig:selected_states_and_counties} and \ref{fig:polls_us}). Given the predominantly bipartisan system of the US~\cite{britannica2025electionresults}, neutral queries such as those used in the EU audits were more likely to elicit mentions of both parties, limiting analytical richness. To address this, we adapted the queries to also include political issues (e.g., “most important issues US elections 2024”) and examined whether SEs and LLMs prioritized particular agendas (see Figure~\ref{fig:methodological_scheme} and Methods, section~\ref{leaning_identification}).

This resulted in a large dataset of 1,455 SE first pages (each containing ~8 results, totalling ~11,640 individual results) for the EU and 2,903 first-pages for the US elections (example in Figure~\ref{fig:screenshots_examples}-A, Supplementary Information, Figure S1 and Table S3), and in 128 and 77 LLM answers, respectively (example in Figure~\ref{fig:screenshots_examples} - B). 

\subsection{Search Engines and Large Language Models offer political content}
 
We first examined whether SE results and LLM responses to neutral political queries -- that is, queries that neither mentioned any party or politician nor revealed a political preference -- overrepresented particular political families. This involved two steps: (1) identifying political entities in the outputs, and (2) grouping them into broader political families -- European party groups for the EU elections, and the Republican/Democratic distinction for the US.

Entity identification combined automatic extraction with human validation (Methods~\ref{entities_identified}). For SEs, we focused on URLs and headlines (Supplementary Figure~\ref{fig:headline_url}); for the LLMs on the full answer text (Figure~\ref{fig:screenshots_examples}-B). For both elections and in the different languages, entities included political parties (e.g., `CDU' in Europe and `The Democratic Party' in the US), individual politicians (e.g., `Von der Leyen' or `Kamala Harris'), ideological references (e.g., `Radical Left'), and, specifically in the EU case study, international families (e.g., ``The Greens'' or `EPP').

For the US, we also analyzed issue salience by extracting mentions of electorally relevant topics (e.g.,"abortion', `immigration') based on~\cite{pew2024issues} and~\cite{yougov2025issues} (full list in Methods, Figure~\ref{fig:topics_importance}).

\begin{figure*}[!ht]
    \centering
    \includegraphics[width=0.93\textwidth]{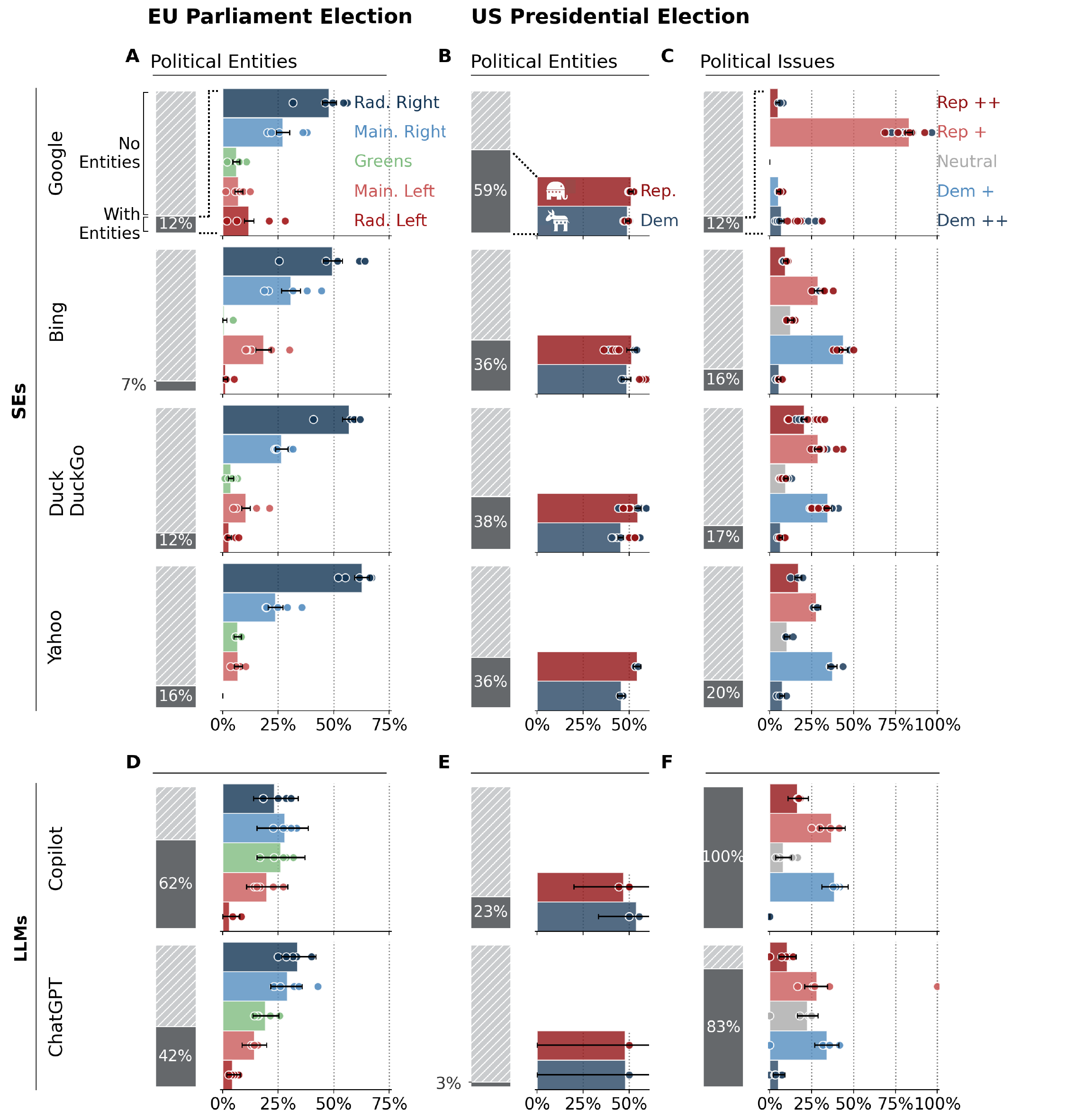}
    \caption{\textbf{Search Engine and LLM results classified by political mentions.} Dark gray bars show the proportion of SE results (A, B, C) or LLM textual answers (D, E, F) containing relevant mentions (political entities -- A, B, D and E -- or issues -- C and F). The plots on the right (color bars) show, for the results that mention at least one political entity, the average proportion of mentions by political leaning or issue categories. In A and D colors represent broad EU political families (Radical Left: dark red; Mainstream Left: red; Center/Greens: green; Mainstream Right: blue; Radical Right: dark blue). In B and D colors represent mentions to either Republican (red) or Democratic (blue) entities. In C and F colors correspond to the leaning of the mentioned issues (Rep ++ = much more important for Republicans, dark red; Rep + = slightly more important for Republicans, red; Dem + = slightly more important for Democrats, blue; Dem ++ = much more important for Democrats, dark blue, Neutral = equally important to both, gray). In A and D (EU) each dot represents a country; in B and C (US) each dot represents a county. 
    Error bars show 95\% confidence intervals computed via 1,000 bootstrap resamples (sampling with replacement from all mentions). Statistical details in Tables \ref{tab:agg_uni_eu} - \ref{tab:uni_global_llm} of Appendix.}
    \label{fig:leaning_results}
\end{figure*}

Grey bars in Figure~\ref{fig:leaning_results} show the proportion of query results with political mentions (politicians, parties and political groups) across four SEs (A–C) and two LLMs (D–F), for the EU (A, D) and US (B, E), with US issue mentions in panels C and F. While most SE results directed to general election related websites (e.g., europa.eu, usa.gov), even using the conservative approach of focusing only on URLs and headlines, results directly mentioning political entities were substantial: between 7\% Bing in the EU, and 59\% for Google in the US. Mentions were generally more frequent in the US, even for identical queries (e.g., “who should I vote for in [X] election”). Notably, a significant share appeared in the Top 3 results -- the positions most often clicked by users~\cite{Fay2024b} --, as well as in the prominent “Top News” section (Supplementary Figure~\ref{fig:category_websites_with_entities} and Figure~\ref{fig:proportion_leaning_top_3}).

Similarly, after excluding queries that led to refusals (Supplementary Table~\ref{tab:not_answer_llm}), LLM answers to EU-elections related questions often mentioned political entities. Even neutral queries such as ``What are the best European Parliament parties?'' elicited explicit references not only to parties, but also to individual politicians (Figure~\ref{sup_fig: us_llms_entities_proportion_query} and Table~\ref{tab:queries}). By contrast, in the US case, direct mentions of politicians or political parties were rare (Figure~\ref{fig:leaning_results}), but issue-framed queries such as ``What issues should I care the most about during American elections?'' always elicited an answer. 

\subsection{Search Engines and Large Language Models overrepresent certain political leanings}

We then analyzed the political orientation of the identified entities to assess whether SEs and LLMs mentioned particular sides of the political spectrum with increased frequency. For the EU audit, candidates, parties, and ideologies (national or international), were grouped into European political families, based on the 2019-2024 EU Parliament organization (Methods~\ref{entities_identified},  Figure~\ref{fig:methodological_scheme}). These families were further categorized into five broad ideological groups for political balance: Radical Left (always represented in dark red), Mainstream Left (red), Center (green), Mainstream Right (blue), and Radical Right (dark blue).

Figure~\ref{fig:leaning_results} (panel A, SE and panel D, LLMs, right bars) presents this leaning distribution for the European election extracted entities (see also Supplement~\ref{fig:distribution_by_eu_family}. Colored bars show the aggregated proportions for all countries, and dots show individual country distributions (detailed country analysis in Figure~\ref{fig:eu_results_by_country}). 
Being neutral queries, it is not obvious what the expected frequencies should be. If mentions were random, all families should be mentioned similarly (following a uniform distribution) or, if larger or historically larger parties (typically the so-called mainstream) were more represented in training data, these could be the ones more mentioned. However, we observed that, across all SEs and ChatGPT, entities associated with the radical right (dark blue) were much more mentioned and, in the case of Yahoo and DuckDuckGo, these counts exceed the sum of mentions to all other families. When compared to the uniform baseline, this overrepresentation was significant, with consistently positive standardized residuals (aggregated Stouffer z-test and Beta-Binomial likelihood ratio test, both $p<0.01$; Tables~\ref{tab:agg_uni_eu} and~\ref{tab:query_uni_eu}), while all other political families show significant underrepresentation (particularly the radical left - $p<0.01$ Stouffer z-test), with the exception of the mainstream right which is moderately overrepresented. 

%Entities associated with the mainstream right segment ranked second in frequency, and the radical left ranked last %(significantly underrepresented against the uniform distribution - $p<0.01$ Stouffer z-test). 
This pattern was consistent across queries (Supplementary Figure~\ref{sup_fig:eu_results_per_query}), and countries to the exception of Google in Germany and Austria (dots in Figure~\ref{fig:leaning_results} panel A and Supplementary Figure ~\ref{fig:eu_results_by_country}). 

When analyzing the nature of the SE suggested pages. We classified landing URLs according (see Methods-section~\ref{websites_classification}) and observed that most sites mentioning political entities had been classified as  ``News'', ``Political'' (e.g., party websites) or ``Reference'' (e.g., governmental) sites (Supplementary Table~\ref{tab:proportion_per_website_category_with_entities}, Figures~\ref{fig:screenshot_1}, ~\ref{fig:screenshot_2}). Radical right entities were mostly mentioned in the ``News'', especially on Google (Figures~\ref{sup_fig:type_results_per_leaning}, ~\ref{fig:category_websites_with_entities}), but queries such as “European parliament parties” could sometimes return "Reference" pages for far-right groups (e.g., Wikipedia page of Identity and Democracy) or more central and mainstream parties (ALDE, EPP), but never for left-leaning entities (Supplementary Figure~\ref{fig:screenshot_1}).

In the case of the US, the entities (e.g. ``Kamala Harris''), parties or orientations (e.g. ``Liberal'') were classified as either Democratic or Republican.
As expected, Figure~\ref{fig:leaning_results}, panel B, shows a comparable frequency of SE mentions to both parties across locations, and there were no notable differences in these frequencies based on the expected political leaning (``color'') of the state or county -- according to the 2024 electoral projections (blue and red dots in Figure~\ref{fig:leaning_results}). Still, while the magnitude of the differences was small, the direction was consistent: Democrats were systematically underrepresented and Republicans slightly overrepresented across queries, a pattern that proved statistically significant (Beta-Binomial likelihood ratio test, $p<0.01$; Table~\ref{tab:uni_us_parties}).

For the ``issue queries'' (e.g., ``most important issues in American elections''; Table~\ref{tab:queries}), we classified issues (e.g., ``Immigration'') based on pre-electoral surveys that compare the salience of topics across partisan groups ~\cite{pew2024issues,yougov2025issues}. Specifically, each issue was assigned to one of five categories: `much more important for Republicans' (Rep ++, dark red in Figure~\ref{fig:leaning_results}), 'slightly more important for Republicans´ (Rep+, red), 'similarly important for voters of both parties´ (Neutral, gray), 'slightly more important for Democrats´ (Dem+, blue), and 'much more important for Democrats´ (Dem ++, dark blue) (details in Methods~\ref{leaning_identification} and Figure~\ref{fig:topics_importance}).

Figure~\ref{fig:leaning_results}, panel C, shows distinct patterns across SEs. Bing, DuckDuckGo, and Yahoo displayed a relatively balanced distribution, with few mentions to highly polarized topics (dark red or dark blue), while 
Google strongly favored topics more important to Republicans (83\%). Excluding the broadly salient topic 'Economy' (Figure~\ref{fig:topics_importance}) reduced the skew, but Google still offered results on Republicans-preferred topics in 62\% of all classified URLs (Supplementary Figure~\ref{fig:no_economy}).  

Finally, we analyzed LLMs responses around the US Presidential election (Figure~\ref{fig:leaning_results}, panels E-F). Copilot's  distribution closely resembled Bing's, with a slight, non-significant preponderance of issues classified as much more relevant to Democrats (dark blue). ChatGPT more often mentioned non-neutral topics (less grey in Figure~\ref{fig:leaning_results}, Panel~F). Both LLMs suggested issues slightly more important to either party at similar rates (light red and blue), while topics that mostly interest Republicans (dark red) appeared almost twice as often than the ones that mostly interest Democrats (dark blue). 

In summary, in the US, both parties are similarly mentioned, except for Google, which strongly favors Republican-preferred issues. For the EU elections, all SEs and ChatGPT most frequently mentioned entities or topics associated with the Radical Right.

\subsection{The observed leaning corresponds to a systematic bias}

\begin{figure*}[htbp]
    \centering
    \includegraphics[width=0.91
    \textwidth]{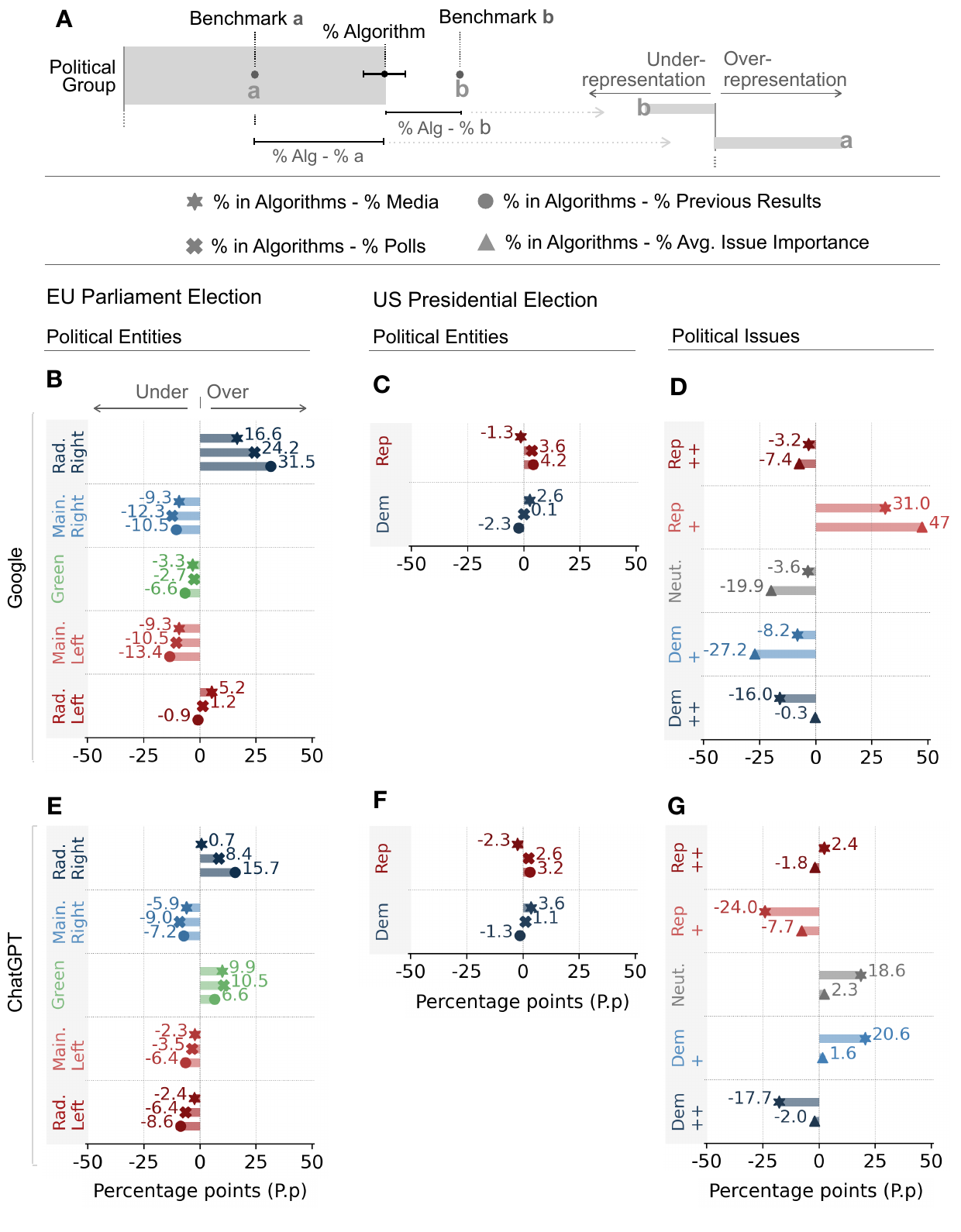}
    \caption{\textbf{Comparison with external factors. (A)} Explanatory scheme of the differences displayed in (B - G). Star = \% in Algorithms (SEs or LLMs) - \% Average attention in Media; Circle = \% in Algorithms - \% Previous Results; Cross = \% Algorithms - \% Polls; Triangle = \% Algorithms - \% Average importance of the Issue (US only), according to surveys~\cite{pew2024issues,yougov2025issues}. Positive values indicate overrepresentation by the algorithms, while negative values indicate underrepresentation. \textbf{(B - D)} Google attention to EU and US political leanings vs. Media, Previous results and polls. {(E - G)} ChatGPTattention to EU and US political leanings and issues compared with the same external factors. Z-scores for each comparison and political leaning are detailed in Appendix, Tables\ref{tab:agg_ext_eu}-\ref{tab:ext_global_llm}.}
    \label{fig:comparison_saliences}
\end{figure*}

The observed leanings are significantly different from random but, while it can be argued that SEs should not directly link to the official webpages of political parties when the queries are general, SEs and LLMs are trained on real data and tailored to offer answers that are aligned with the searchers needs. This raises the important question of what the expected party distribution should be. Should all parties receive equal attention (uniform distribution), even small and upcoming ones? Should the distribution reflect parliamentary representation, preserving the \textit{status quo}? Or should it follow more organic patterns, such as media coverage or user engagement? As mentioned, early studies on SE bias compared SEs to each other to detect outliers~\cite{mowshowitz2002assessing}, while more recent work measures bias using sentiment analysis or the political leaning of the suggested URL domains (e.g., right wing newspaper, left-leaning think tank)~\cite{gezici2020evaluation}. 

We used a different approach, comparing the extracted political mentions to external factors. In particular, we focused on the most popular SE and LLM at the time (Google and ChatGPT, respectively), and compared their answers with: 1) the official results of the 2019 EU parliamentary or the 2020 US Presidential elections -- under the rational that larger parties should appear more often); 2) aggregated polls for each of the audited countries -- as search activity is known to be correlated with votes~\cite{erokhin2025google}(details in Methods~\ref{methods_polls}); and 3) media salience, measured as the share of news from popular outlets mentioning each political entity or issue -- as SEs surface news content and LLMs are trained on it (details in Methods~\ref{media_salience}).

Figure~\ref{fig:comparison_saliences}, shows how political mentions deviate from media coverage (stars), previous election results (dots), and polling projections (crosses). Bars to the right indicate overrepresentation; bars to the left indicate underrepresentation (Panel A). 
We note that some countries already have a considerable radical right representation (to the exception of Ireland), and that media coverage disproportionately references Radical Right parties relative to their electoral importance in all analyzed EU countries except Poland~\cite{Damiao2026radicalright}. Therefore, these benchmarks are far from the uniform distribution tested before.

In the case of the European Parliamentary election, we found that Google (Panel~B) significantly under-mentions nearly all political groups while significantly over-mentioning the Radical Right (dark blue) which was overrepresented across all external benchmarks -- polling (crosses, +30 p.p.), previous electoral strength (circles, +32 p.p.), and media coverage (stars, +17 p.p.) -- (aggregated Stouffer $z-test$, $p<0.01$, Table~\ref{tab:agg_ext_eu}).
%, albeit inconsistently so across queries (Beta-Binomial likelihood ratio test, non-significant for Google only, Table~\ref{tab:que_ext_eu}). 
Conversely, all other parties were less mentioned than expected (bars to the left) when compared with the 2019 results, with the polls (except for the Mainstream Left), and even with the media (except for the Radical Left (see aggregated statistical results in Table~\ref{tab:agg_ext_eu} and by query in ~\ref{tab:que_ext_eu}). This pattern was also observed for the other SEs, suggesting a systemic, rather SE specific bias (Appendix Figure~\ref{fig:comparison_saliences_all}).

For the US presidential election (Panel C), Google gave similar visibility to both parties. However, when benchmarked against state-level polls and previous electoral results, mentions associated with the Republican party are significantly overrepresented while Democrats are consistently underrepresented across the states where bots were deployed (Figure~\ref{fig:selected_states_and_counties}; aggregated Stouffer $z-test$ and Beta-Binomial likelihood ratio test, $p<0.01$; Tables~\ref{tab:agg_us_parties} and~\ref{tab:que_us_parties}). This is not surprising, given that different parties won the 2020 and 2024 elections, and given the historically small vote-share difference between the two parties. Therefore, we complemented the analysis with an examination of issue salience -- specifically, how SE results weighted different issues relative to their importance in voter surveys \cite{pew2024issues,yougov2025issues} and with their coverage in legacy media (triangles and stars in Panel C, respectively). 

Google underrepresented issues prioritized by Democrats (light blue triangle bars to the left, 27 P.p.) and significantly overrepresented those more important to Republicans (light red triangles to the right, 47 P.p. - aggregated Stouffer $z-test$ and Beta-Binomial likelihood ratio test, $p<0.01$; Tables~\ref{tab:agg_us_issues} and~\ref{tab:que_us_issues}). In contrast, references to party-aligned issues were substantially more balanced across the other engines (Figure~\ref{fig:comparison_saliences_all} and Tables~\ref{tab:agg_us_issues},~\ref{tab:que_us_issues}).

Finally, we compared ChatGPT outputs with media attention, polls, and past electoral results (EU and US, Figure~\ref{fig:comparison_saliences}, panels E–F), and issue importance (US only; Panel G). In the EU, the Radical Right was again overrepresented relative to polls and past results, but not relative to media coverage, which was already quite high~\cite{Damiao2026radicalright}. Interestingly, the Greens algo get attention from LLMs beyond most external benchmarks. The Mainstream Right and Left, as well as the Radical Left were consistently below those benchmarks, although not significantly so.

In the United States (panel F), differences were again modest, reflecting the two-party system. For political issues (Panel G), ChatGPT showed a different behavior from Google's. Relative to media coverage (stars), ChatGPT underrepresented issues more important to Republicans (Rep +) and much more important to Democrats (Dem++), while overrepresenting both the topics slightly more important to Democrats (Dem +) and to voters from the two parties (Neutral, in gray). Moreover, in contrast to Google (Panel D) -- where the population-average importance markers (triangles) were displaced from the central line -- ChatGPT referenced issues in closer proportion to their overall importance among the population, with triangles clustering near the midline. Copilot showed a similar behavior (Figure~\ref{fig:leaning_results}) and Figure~\ref{fig:comparison_saliences_all}). 

Overall, these results indicate that political mentions in SE and LLM outputs cannot be fully explained by external factors such as media attention, polling trends, or past electoral performance. Instead, the two most popular platforms systematically favor or disadvantage certain political groups, exhibiting measurable bias. This bias is particularly pronounced in favor of the Radical Right in the EU (by all studied search engines and by ChatGPT) and of Republican Party issues in the US, in the case of Google only.

\section{Discussion}

In this study, we examined political bias of SEs and LLMs when answering neutral election-related queries. Bias was assessed relative to key benchmarks covering two major 2024 elections, across different countries, languages, and party systems. 

Unlike most previous studies, we focused on general, unbiased queries that better reflect how undecided voters search for information~\cite{voter_centered_audits}. These queries were done at slace, and avoided naming specific parties, politicians, or ideological orientations and instead mimicked natural user behavior -- short and direct for SEs, more conversational for LLMs.

Our approach also diverged in how we gauged political bias, focusing on political visibility -- i.e., direct mentions of parties, actors, and issues --, rather than more subjective source-based classifications~\cite{auditing_partisan_bias,gezici2020evaluation} or sentiment analysis~\cite{gezici2020evaluation}.

In the EU, our results reveal a strong and systematic overrepresentation of radical right actors, for all SEs and ChatGPT-4.0, which cannot be completely explained away by factors such as media salience and political support Although political mentions were relatively rare, they appeared prominently and were consistently skewed toward the radical right across countries and languages, including where these parties remained electorally weak (e.g. Ireland and Portugal). While it had already been observed that EU media attention favored the radical right~\cite{Damiao2026radicalright}, the bias of SE results was even stronger: in particular, SE News results mentioned radical-right parties and actors 80\%  more than its overall media salience. Moreover, SE results often included direct links to the Wikipedia pages or official websites of these parties, further amplifying their prominence online. This bias also cannot be explained by polling or prior results. 

In the US, the differences were smaller, whether we analyzed political entities or topics, reflecting the two party system. Both Google and ChatGPT emphasized Republican-prioritized topics, while the other SEs leaned slightly toward Democrats priorities. These effects were stronger among the top three SE results (Supplementary Figure\ref{fig:proportion_leaning_top_3}), which receive the most attention and trust~\cite{Fay2024b}. Still, (the slighter but significant) observed bias went beyond what could be predicted by differences in voting intentions or issue salience in the media (which already dedicated greater attention to topics preferred by Republican-voters).

These findings diverge from prior work analyzing both SEs and LLMs, which have attributed a ``liberal''/``left'' leaning or no explicit leaning at all~\cite{10.1145/3359231}. In the case of SEs, this difference can likely be explained by the methodology: by restricting analysis to neutral queries and using mentions of political families, instead of sentiment or source, we emphasize visibility. It is possible that many of these mentions stem from left-leaning sources or are negative. Indeed, one limitation of our study is that we did not evaluate valence, or sentiment. However, there is consistent evidence that visibility and issue salience influence vote choice regardless of tone~\cite{adams2024much,vasilopoulou2024electoral,hopmann2010effects,geers2017priming}. Moreover, many mentions appeared in News results or ``Reference'' pages, such as Wikipedia, which are often seen as neutral. Finally, given the scale of overrepresentation, there is a substantial visibility asymmetry that our approach detects and that source- and sentiment-based approaches may miss.

In the case of LLMs, other studies have classified them as predominantly left-leaning~\cite{potter-etal-2024-hidden,yang2024unpacking,political_bias_llms}. This difference likely stems from our task design: instead of eliciting opinions and comparing them to political surveys, we ran more neutral and arguably organic prompts, and examined which information LLMs prioritize when generating search-style answers -- also more closely mirroring their role in AI-generated summaries now integrated into SEs.

Because the analyzed SE and LLM algorithms are proprietary the mechanisms behind these patterns are unclear. In the case of the SEs, one possibility is user engagement (e.g. higher traffic to right-leaning web pages); other is higher search demand for radical-right entities. However, both raise a circular causality question: do these pages get more visits/interest because they appear more often in SE results, or are they displayed more because they get more engagement/queries? Other explanations include more effective use of search engine optimization techniques by some groups, or even deliberate algorithmic tweaking. 
Similarly, and although overall more balanced, both Copilot and ChatGPT produced results that are not straightforward to explain. Interestingly, both LLMs refused to answer a subset of election-related queries, with refusal rates varying substantially across countries (Tables~\ref{tab:not_answer_llm} and~\ref{tab:llms_answered_refused_per_country}), and whether the results were asked via API or not. Furthermore, both models mentioned political parties associated with the Greens more frequently than expected when benchmarked against media coverage, polls, and past electoral results (in this last case with the exception of the radical right, see Tables~\ref{tab:agg_ext_eu}~\ref{tab:que_ext_eu}) -- and more frequently than SEs did for comparable queries. This is in line with what had already been observed when asking ChatGPT to classify news headlines as pro or against a specific conflict ~\cite{damiao2025}. Together, this "central-leaning" and refusal behavior, raise questions about the role of fine-tuning and manual content moderation in shaping LLM outputs on politically sensitive topics.

While the pattern revealed by this study is worrying, several limitations should be acknowledged. The queries were formulated to ensure comparability, and certainly do not fully reflect the diversity of real-world search behavior; future studies should incorporate organically generated queries to better capture actual overall voter information-seeking patterns. Additionally, the number of collected LLM responses is relatively small and this is a very dynamical field. Larger longitudinal audits are necessary to assess the temporal stability of the observed patterns. In particular, it would be interesting to access if the observed biases are limited to pre-electoral periods or if they are consistent over time. Finally, while we found consistent bias across countries, it remains unclear how these results generalize to other political systems, languages, and cultural contexts. The observed differences in LLM refusal rates by country and language raise important questions about the effectiveness of safeguard mechanisms outside large countries or when queries are made in less widely used languages.

However, and regardless of the cause, our findings underscore a critical point: even neutral queries yielded non-neutral representations of the political landscape, and algorithms used daily by millions of citizens worldwide can systematically amplify biases in information exposure. While large scale, comparative audits such as this are essential for identifying distortions, they also highlight the urgent need for greater algorithmic transparency from the platform side. SEs and LLMs fall squarely within the scope of the EU Digital Services Act (DSA), which requires the assessment and mitigation of systemic risks to democratic processes~\cite{pierri2025dsa}. Independent auditing and accountability are fundamental to ensure that AI-driven systems do not distort information at scale.

\section{Methods}

This section provides the details of each step of the methodology employed, following the order in Figure~\ref{fig:methodological_scheme}: (1) web-crawlers (bots) performed simultaneous online election-related queries. (2) The query results were analyzed to identify mentions to political entities and/or political issues. (3) These mentions were classified and clustered according to their political families. (4) The mention's frequencies were compared with external sources to infer possible bias.

\subsection{Data Collection}

Results from four Search Engines (SEs) -- Google, Bing, DuckDuckGo and Yahoo -- and two Large Language Models (LLMs) -- Copilot and ChatGPT -- were collected during the week preceding each of the two case-study elections: the European Parliament elections (June 9, 2024) and the US presidential election (November 5, 2024). 

Google and Bing were selected as they are the two most widely used SEs worldwide~\cite{semrush_top_websites_2025}, DuckDuckGo due to its privacy-focused model~\cite{duckduckgo_privacy}, and Yahoo due to its relevance as a news and email platform~\cite{semrush_top_websites_2025}. Copilot and ChatGPT were chosen as the most widely used free-tier AI chat systems at the time, with and without internet access, respectively.

To ensure systematic large-scale collection, automated web crawlers (\hyperref[subsec:bots]{\textbf{bots}}) were employed in different \hyperref[subsec:audit]{\textbf{audits}} for all SEs (as in~\cite{Israel_Palestine}). For LLMs, data collection differed by case study. In the EU case study, outputs were partially collected manually, while ChatGPT outputs were additionally collected via the OpenAI API. In the US case study, Copilot results were collected using the same bot system employed for SEs, and ChatGPT outputs were also collected via the OpenAI API (model 4o,temperature = 1; prompt details in Appendix, Table~\ref{tab:prompts}).

\subsubsection{Webcrawlers}\label{subsec:bots} Each bot consisted of an automated browser (Firefox version 123) extending the functionalities of OpenWPM (version 0.28.0)~\cite{englehardt2016census}, and programmed to reproduce typical human online browsing behavior, which included: (1) visiting the website of one of the studied SEs or LLMs (e.g. \texttt{www.google.com}); (2) entering a query at a human-like typing pace; and (3) scrolling through the results page at a random and not immediate pace, ensuring all elements were fully loaded and could be collected using Selenium-based techniques~\cite{selenium}.

Each automated Firefox instance was also configured with a residential proxy to define geolocation, by specifically defining the Network settings of the automated Firefox browser ({\ttfamily network.proxy} settings), and with language settings adapted to the target country or region ({\ttfamily intl.accept\_languages}).  

\subsubsection{Audits}\label{subsec:audit} Each audit was carried out through the deployment of multiple pre-configured bots. These bots collected results from the four SEs and the two LLMs, queried about one of the two elections. The \hyperref[location]{\textbf{location}} and \hyperref[language]{\textbf{browser language}} parameters of each bot were adapted to the relevant case study (e.g., Poland, Polish - EU Parliament Election; New York City, English - US Presidential Election), and several replicas of each configuration were deployed within the same audit. All bots (with different locations and languages) were executed in parallel to minimize temporal variation; during each run, they accessed the same SE or LLM, issued the same query, and collected the corresponding results. For every query iteration, all bots were launched as fresh browser instances to eliminate possible carry-over effects from previous searches.

\vspace{0.2cm}
\mypara{Location:}\label{location} For the European Parliament elections, bots were deployed in five countries: Austria, Germany, Ireland, Poland, and Portugal, selected for their diversity of languages and polling intentions~\cite{austrian_polls,german_polls,irish_polls,polish_polls,portuguese_polls}, as detailed in Supplementary Figure~\ref{fig:polls_eu}. Per country five unique \href{https://brightdata.com/}{\textit{Bright Data}} residential IPs were used to collect search engine results, while for LLMs a single residential IP per country was used for manual data collection (accounting for approximately 45\% of the total textual responses in the EU case study).  

For the US elections, bots were deployed in counties with different voting leaning in the previous election (Presidential Elections of 2020), in states that showed different polling intentions (Democrat and Republican) in the week before the election, as detailed in Figure~\ref{fig:selected_states_and_counties} and ~\ref{fig:polls_us}. Since Bright Data offered limited coverage in Republican-leaning counties, \href{https://oxylabs.io/}{OxyLabs} proxies were also used. The final counties were:  

\begin{itemize}
    \item Leaning Republican (7): Wisconsin-Delafield, Georgia-Canton, Florida-Cape Coral, New York- Staten Island, Florida-Lakeland, Florida-Largo and Texas-Greenville.
    \item Leaning Democratic (8): New Jersey-Essex (Newark city), New York-New York City (varied), Massachusetts-Hampden (Springfield), Virginia-Reston, Wisconsin-Milwaukee, Georgia-Atlanta, New York- Bronx, and Texas- El Paso.
\end{itemize}  

In total, 8 leaning Democratic and 7 leaning Republican locations were used in the SE audit. For LLMs, because prior work provides no evidence that within-country geographic variation affects outputs, unlike its documented influence on search engine results~\cite{inconsistent_search_results}, we did not replicate the county-level design. Instead, we exclusively used four IP addresses for each of the four US locations available through Bright Data. 

\mypara{Browser Language:}\label{language} The browser language of the bots was set to English or to the most spoken language in each country: German (Germany, Austria), English (Ireland, USA), Portuguese (Portugal), and Polish (Poland). While English is not Ireland’s first official language~\cite{irish_constitution1937}, it is the most widely used~\cite{cso_ireland2022,ethnologue2023}. 

\vspace{0.2cm}

Supplementary Table~\ref{tab:audit_summary_rev} summarizes the final set of bot configurations applied in each audit type (SE or LLM) and case study. In total, 149 independent configuration bots were deployed and collected 4,358 result pages from SEs and 174 answers from LLMs. 

\subsubsection{Search Queries} 

In each audit, the bots executed multiple queries in English and/or in the language corresponding to their location. From the English text, queries were translated to each country's primary language using ChatGPT -- free tier -- and manually confirmed (see full list in Supplementary Table~\ref{tab:queries_eu_languages}). 

The queries were designed to simulate typical users seeking election-related information in the days leading up to the vote and were adapted to the platform type: longer, more conversational questions for LLMs, and shorter, more direct queries for SEs. See Table~\ref{tab:queries} for details.

For the European Parliament elections, queries were designed to capture explicit mentions of multiparty political entities (parties, politicians) (Table~\ref{tab:queries}). 

For the US presidential election (English only), queries focused on: entity-based queries (orange in Table~\ref{tab:queries}) to identify coverage biases in candidates, parties, and areas of the political spectrum, and topic-focused queries (blue in Table~\ref{tab:queries}) to assess issue prominence in Democrat/Republican contexts, compensating for being a mostly the bi-partisan system.

\begin{table*}[!ht]
\caption{Queries used for the European Parliament elections (left column) and the US Presidential elections (right column). Queries in the orange rows were used for analyzing political mentions; queries in the blue rows were used for political mentions and for studying topic salience.}
\label{tab:queries}
\vspace{0.4cm}
\centering
\small
\renewcommand{\arraystretch}{1.2}
\begin{tabular}{p{0.48\linewidth} p{0.48\linewidth}}
\toprule
\textbf{EU Parliament Election} & \textbf{US Presidential Election} \\
\midrule
\multicolumn{2}{c}{\textbf{SEs}} \\
\midrule
\cellcolor{orange!15} european parliament elections & \cellcolor{cyan!15} US election candidates policies \\
\cellcolor{orange!15} european parliament parties   & \cellcolor{cyan!15} US presidential elections 2024 \\
\cellcolor{orange!15} european elections choose party & \cellcolor{cyan!15} best president US 2024 \\
\cellcolor{orange!15} european elections top issues  & \cellcolor{cyan!15} who should I vote for US elections 2024 \\
\cellcolor{orange!15} who is going to win european elections 2024? & \cellcolor{cyan!15} most important issues US elections 2024 \\
\cellcolor{orange!15} who should I vote for european elections 2024? & \cellcolor{cyan!15} as an american most important issues US elections 2024 \\
\cellcolor{orange!15} & \cellcolor{orange!15} US election candidates economy \\
\cellcolor{orange!15} & \cellcolor{orange!15} US election candidates immigration \\
\cellcolor{orange!15} & \cellcolor{orange!15} US election candidates Israel Palestine \\
\cellcolor{orange!15} & \cellcolor{orange!15} US election candidates social security \\
\cellcolor{orange!15} & \cellcolor{orange!15} US election candidates racism \\
\cellcolor{orange!15} & \cellcolor{orange!15} US election candidates abortion \\
\midrule
\multicolumn{2}{c}{\textbf{LLMs}} \\
\midrule
\cellcolor{orange!15}
What are the European Parliament parties? & \cellcolor{cyan!15} most important issues in american elections \\
\cellcolor{orange!15}
What is the best European party? & \cellcolor{cyan!15} What issues should I care the most about during american elections? \\
\cellcolor{orange!15}
Who is going to win European elections 2024? & \cellcolor{cyan!15} us candidates policies \\
\cellcolor{orange!15}
Who should I vote for in European elections 2024? & \cellcolor{cyan!15} As an american, what issues should I care the most about during american elections? \\
\bottomrule
\end{tabular}
\end{table*}

\subsubsection{Collected Outputs} 

For each audit, individual query, and bot configuration, results were automatically collected.

SEs: URLs and respective headlines (Supplementary Figure~\ref{fig:headline_url}) displayed on the first page results (see example in Figure~\ref{fig:screenshots_examples} - A). This included the main results (green) and sections such as \textit{Top Stories}, in blue, and \textit{People Also Ask} (yellow). 

LLMs: complete textual responses, exemplified in Figure~\ref{fig:screenshots_examples}-B.

\subsubsection{SE results category classification}\label{websites_classification}

To identify the types of websites appearing on the first page of SE results, the unique set of websites retrieved across all searches in both elections was classified using ChatGPT-4o (accessed via the platform with internet access). The LLM was prompted to assign each website to one of the following predefined categories: News, Media Publications, Reference definition, Science/Academic, Political, Social Media, Forums/Discussion Boards, Entertainment Services, E-commerce/Retail Platforms, Corporate Websites, Educational Platforms, Search Engines/Aggregators, Utilities/Tools, Blogs, Adult/Gambling/Restricted and Fact-Checkers (see Table~\ref{tab:prompts}). 

The final taxonomy resulted from an iterative process that combined (i) an initial set of categories proposed by the model and (ii) multiple independent classification sessions (initiated in new chats) applied to a diverse subset of approximately 100 websites from the dataset. This procedure led to the inclusion of categories that ultimately did not appear in the final results (e.g., Adult/Gambling/Restricted, initially suggested by the model), as well as refinements designed to capture meaningful distinctions in the data, such as the separation of Political and Fact-Checker websites.

The resulting taxonomy was verified to adequately cover all relevant website types observed in the dataset. Providing detailed category definitions and a structured prompt yielded high classification accuracy (100/100 in validation tests). In cases where the model lacked prior knowledge of a website’s content, it was explicitly instructed to consult online sources before assigning a category.

\subsection{Entities Extraction}\label{entities_identified}

For each single SE output (URL and respective headline) and LLM textual response, all political entities (politicians, political parties/groups and political spectra) or political issues (US case study only) were extracted.  Extraction was performed using a LLM based approach, followed by manual validation. 

\vspace{0.5cm}

\mypara{LLM-based Approach:}\label{entities_llm_approach} Through OpenAI's API, ChatGPT-4o (with default temperature) was instructed to identify political entities (EU and US case studies) and political issues (US case only) in the URL and titles of the SE results, or in the full textual response in the case of the LLM outputs. 

In the EU Parliament election case, the model was instructed to detect any mentions of politicians (e.g. ``Manfred Weber''), political parties (e.g. ``CSU''), European Parliament groups (e.g. ``EPP''), or political leanings (e.g. ``right''). The model was also explicitly instructed to ignore official institutions, such as the European Parliament. When an entity was identified, ChatGPT was directed to assign the identified entity (politician, party or political group) to a national party and to the English version of the party's name (e.g. ``Hildegard Bentele'' $\rightarrow$ ``Christian Democratic Union - Germany''; ``CDU'' $\rightarrow$ ``Christian Democratic Union - Germany'') - see Table~\ref{tab:prompts}, Appendix). For each search result containing mentions to more than one entity, the output was a list with one party and country per entity. Mentions appearing in both the title and the URL were counted only once.

Cases involving mentions of political spectra (e.g. ``The Left in Portugal'') required an additional step to determine which specific political parties were being referenced. That is, when a spectrum term was detected in a headline or URL of a news item, the full article text was retrieved using the Python \texttt{Media Cloud Metadata Extractor}~\cite{mediacloud-metadata}. The URL, title, and article text were then resubmitted to ChatGPT-4o (default temperature), with instructions to identify any political parties mentioned (see Table~\ref{tab:prompts}, using only the information contained in the article. 

In the US context, ChatGPT-4o was applied both to the blue and orange queries in Table~\ref{tab:queries}. For the orange queries (political entities), the process mirrored that used in the EU case. The model was instructed to identify politicians, political parties and political spectrum references, and to map these to the one of the two main American political parties (e.g. ``Kamala Harris'' $\rightarrow$ ``Democratic Party''). Although the US political landscape includes parties beyond the Democratic and Republican parties, these two receive the vast majority of votes and therefore constituted the primary categories for classification. Additionally, in this case, political figures explicitly endorsing and connected to a political party at the time of the election were also considered when mentioned in connection to it (e.g., ``Elon Musk''). Mentions to political leanings (e.g., ``radical-right'') were very rare in this set of results and were not subjected to further verification. 

For the blue queries (topic-related queries), the model was instructed to assess whether each response referenced any of the following policy issues: \textit{Abortion, Economy, Healthcare, Supreme Court appointments, Foreign policy, Violent crime, Immigration, Gun policy, Terrorism, Taxes and government spending, Social Security, Climate change and the environment, Education, Racial and ethnic inequality, Civil rights and civil liberties} (see full prompt in Appendix Table~\ref{tab:prompts}). This set of political issues was derived from the intersection of two independent surveys identifying key voter concerns prior to the 2024 election ~\cite{pew2024issues,yougov2025issues} and was subsequently validated by one of the authors. The prompt instructed the model to identify relevant topics based on contextual meaning, synonyms, or closely related terms. For SE results, topic identification relied exclusively on headlines and URLs, whereas for LLM outputs it was based on the full generated text. Prompts were iteratively refined until a consecutive set of 100 SE topic evaluations was manually verified as correctly classified by ChatGPT.

\vspace{0.2cm}
\mypara{Manual Verification:} The previous step produced a list of unique URLs and headlines, along with the political entities and corresponding parties identified by ChatGPT-4o.

For the EU elections, manual verification was done on the entire set of unique search engine (URL/headline) pairs and LLM-generated answers, using \textit{Wikipedia} and the European Parliament official website \footnote{\label{fn:europarl}\href{https://www.europarl.europa.eu/meps/en/full-list}{Europarl}} as references. This full validation was necessary due to the diversity and complexity of the European political landscape, which spans multiple countries, parties, and candidate names. To ensure a fair comparison across the different types of political entities in this multi-country election, the analysis was restricted to national political parties that met all of the following criteria:

\begin{enumerate}
\item They were affiliated with an EU Member State at the time (excluding third countries and the United Kingdom).
\item They were running for the 2024 European Parliament election.
\item They were represented in the European Parliament during the 2019–2024 term~\cite{eu_political_groups}.
\end{enumerate}

For the US election, two random samples of 100 results each were manually checked for proper entity and topic identification, respectively. The first list was reviewed to identify false positives (entities classified as Democratic or Republican despite belonging to neither or to the other party), and two types of false negatives (entities not detected as political actors, and entities detected but not assigned to one of the two parties). False positives and misclassifications were extremely rare (no cases identified). In approximately 3\% of unique results, the model detected a political entity but did not assign a party affiliation (e.g., “Jason Miyares”). In these cases, affiliations were manually determined based on the most recent publicly documented information, consulting \textit{Wikipedia} where necessary.

Validation of topic identification followed two steps. First, the classification prompt had been iteratively refined using an initial set of results (as described in the previous sub-section), until 100 accurate classifications were made. Second, an independent random sample of 100 results was checked for the same type of errors (false positives and false negatives) as before. At this step, no errors were found.

Supplementary tables in section~\ref{app_sec:entities_n_results} show the final list of entities and their party/topic associations (for both SEs and LLMs).

\subsubsection{Mention counts}\label{mention_counts}

Unique mentions of political entities belonging to the same EU political family (EU case) and unique mentions of each political party and issue category (US case) were counted for each SE result and each collected news item. For the LLM outputs, unique entities mentioned in each response were counted. These counts were used directly (i.e., raw counts without additional normalization) to compute the proportion of mentions associated with each political leaning (see below) in each SE. 

Analysis were also computed at the location level (dots in Figure~\ref{fig:leaning_results}), using: 

\begin{equation}
P_{i,\ell} = \frac{M_{i,\ell}}{M_{i,\text{total}}} \times 100
\end{equation}

where $M_{i,\ell}$ denotes the number of mentions associated with political leaning $\ell$ in location $i$, and $M_{i,\text{total}}$ represents the total number of political mentions extracted for that location.

Finally, query-level analyses were performed; these results are reported in the Supplementary information in section~\ref{app_subsec:se_entities_per_query}.

\subsection{Political leaning identification}\label{leaning_identification}

Political entities (already linked to a political party) and political issues were assigned a political leaning in the EU or US contexts.

\subsubsection{EU Political Classification}

Using \textit{Wikipedia} and the \textit{Europarl} website as references, the final list of national parties was mapped to one of the established European Parliament groups during the 2019-2024 term: The Left; S\&D – Progressive Alliance of Socialists and Democrats; G/EFA – Greens/European Free Alliance; Renew – Renew Europe Group; EPP – European People’s Party (Christian Democrats); ECR – European Conservatives and Reformists; ID – Identity and Democracy; and NI – Non-attached (Figure 1, panel 3). Entities linked to newly formed alliances (not present in the 2019-2024 parliamentary configuration) were manually evaluated. This was the case for the Patriots for Europe and Europe of Sovereign Nations, which were classified under the Identity and Democracy (ID) group - as both derive from, and align ideologically with it ~\cite{FarRightReconfiguration}. 

In parallel, each national party was classified by political ideology using the 2024 Chapel Hill Expert Survey~\cite{ches2024}, which draws on assessments from 69 political scientists to position 279 parties across 31 countries, on dimensions including ideology, populism, democracy, EU integration, and specific policy stances. Based on this expert classification, parties were categorized into ideological groups such as Radical Right, Conservative, Liberal, Christian-Democratic, Socialist, Radical Left, Green, Regionalist, Confessional, and Agrarian/Centre. To ensure a balanced distribution of political leanings across the left–right spectrum, each classification was mapped to one of five broader ideological groups: \textbf{Radical Left}, represented in dark red, \textbf{Mainstream Left} (Socialist), in light red, \textbf{Greens}, in green, \textbf{Mainstream Right} (including Conservative, Liberal, and Christian-Democratic), in light blue, and \textbf{Radical Right} (Radical Right, Regionalist and Confessional), in dark blue in all figures. Therefore, each national party was classified both in terms of political ideology (according to the Chapel Hill survey) and as belonging to a EU parliamentary group during the 2019-2024 term. When the extracted entities included direct mentions to EU political groups or families (e.g., The European Conservatives and Reformists Party - ECR), the mode of the political leanings of all parties belonging to that family was computed (in the case of the ECR, it includes 23 parties, mostly classified as Radical Right, and it was categorized as Radical Right, see full association list in Supplementary Table ~\ref{tab:eu_parties_all}).

When multiple political entities belonging to the same EU party family appeared within a single URL or headline (e.g., Maximilian Krah, AfD, ID), they were counted as one single mention to that political group.

\subsubsection{US Political Classification} 
Two types of entities were collected: direct political entities and relevant political issues. Only unique party mentions were counted for each SE result and each LLM response.

Political entities were mapped as described in section~\ref{entities_identified}, to either the Democratic or Republican parties.

Issues were identified as described in~\ref{entities_llm_approach}, using two independent surveys~\cite{pew2024issues,yougov2025issues}: the Pew Research survey, asked approximately 1,147 respondents, in September 2024, “How important, if at all, are each of the following issues in making your decision about who to vote for in the 2024 presidential election?”, to which respondents could select one or more issues; the YouGov survey, asked about 5,742 respondents across four waves (weekly) in October 2024: “Which of these is the most important issue for you?”, with respondents selecting only one issue.

Both surveys reported the number of Republican N$^{\text{(R)}}$ and Democratic N$^{\text{(D)}}$ voters who selected each topic. To harmonize between surveys $s$ and topics $t$, the proportion of Republican (R$_{t,s}$) and Democratic (D$_{t,s}$) voters selecting that topic  - relative to the total number of respondents per party -, was computed:

\begin{equation}
R_{t,s} = \frac{N^{(R)}{t,s}}{N^{(R)}{\text{total},s}}
\end{equation}

\begin{equation}
D_{t,s} = \frac{N^{(D)}{t,s}}{N^{(D)}{\text{total},s}}
\end{equation}

Using these proportions, the leaning score ($L$) for each topic in survey $s$ was defined as:

\begin{equation}
L_{t,s} = R_{t,s} - D_{t,s}
\end{equation}

When a topic appeared in both surveys (e.g., Economy), the final leaning value was computed as the average of the two scores:

\begin{equation}
L_t = \frac{1}{2} \left( L_{t,1} + L_{t,2} \right)
\end{equation}

If a topic appeared in only one survey (Civil Rights), that value was used directly:

\begin{equation}
L_t = L_{t,s}
\end{equation}

A positive value of $L_t$ indicates greater importance for Republican voters;  a negative value indicates greater importance for Democratic voters.
Based on the value of $L_t$, issues were further classified according to the degree of importance. Rep ++ denotes issues that were substantially more important to Republicans than to Democrats; Rep + issues that were moderately more important to Republicans. Neutral corresponds to issues that were valued almost equally by both electorates. Dem + includes issues that were moderately more important to Democrats, and Dem ++ issues that were substantially more important to Democrats than to Republicans.

\begin{figure}[!ht]
    \centering
    \includegraphics[width=1\columnwidth]{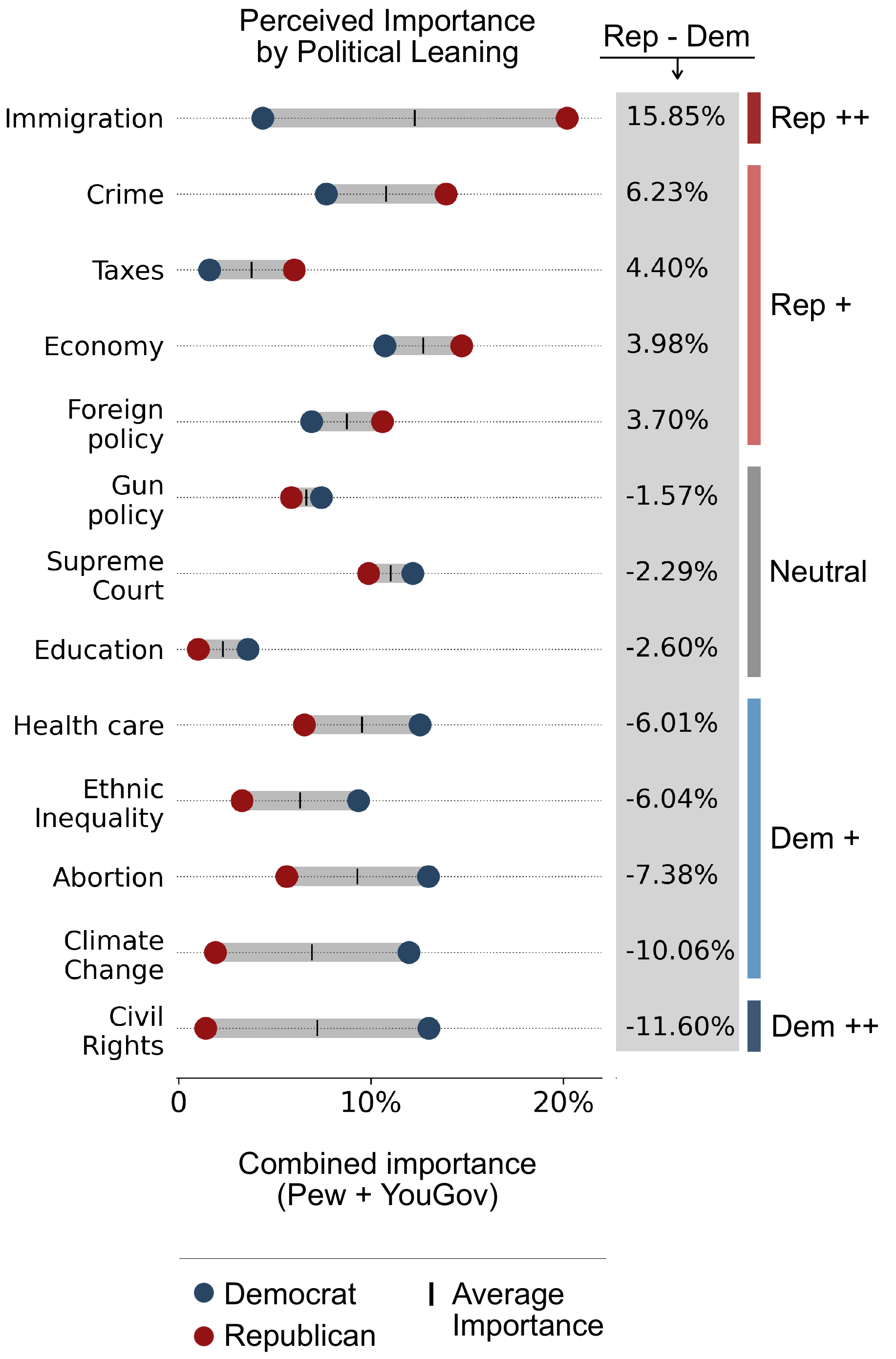}
    \caption{Leaning values for key topics, combining Pew Research~\cite{pew2024issues}and YouGov~\cite{yougov2025issues} data collected in October 2024.}
    \label{fig:topics_importance}
\end{figure}

Figure~\ref{fig:topics_importance} shows the computed leaning values and their importance classification. 

\subsection{Comparison with External Factors}\label{external_factors}

Attention (i.e., number of mentions) given to each political leaning and political issue by SEs and LLMs was compared with three external benchmarks: media salience, polls, and previous electoral results.

\subsubsection{Media Salience}\label{media_salience}

For the EU Parliamentary election, an existing large corpus of news articles ~\cite{Damiao2026radicalright} was used. This corpus was built using the open source aggregator \href{https://www.mediacloud.org/}{Media Cloud platform}, which returns the URL and headlines of online articles whose full HTML content contains user chosen terms. In this case, a pre-defined list of EU election-related keywords was built and used to query the database (see Appendix section~\ref{media_data_collection}, Table~\ref{tab:terms_media_EU_combined} for details). 

The dataset was limited to news from the most popular outlets with online presence in Austria, Germany, Ireland, Poland, and Portugal (see Table~\ref{tab:tab_media_sources_considered}). The news explicitly mentioned the 2024 EU elections and were published in the two months before the EU parliamentary elections (from April 9 to June 9, 2024). From their titles and URLs mentions to political entities or political leanings were extracted and classified as in section ~\ref{entities_identified}, combining automated extraction by ChatGPT with manual verification. In total, the EU media dataset includes close to 21,500 news items that explicitly mention a political entity and that could be mapped into one of the five broad political leanings (Radical Left, Mainstream Left, Greens, Mainstream Right and Radical Right).

A similar approach was used to select news items related with the US Presidential elections from the Top 20 Newspapers in the US, as in February 2025 (see Table~\ref{tab:tab_media_sources_considered}). Media Cloud was searched for news items containing election related terms (see Appendix section~\ref{media_data_collection}, Table~\ref{tab:en_us_media_terms}) between September 5 and November 5, 2024. From these items (N=18,368), mentions to political parties, politicians or political affiliation were extracted from their URLs and headlines, using OpenAI API, as detailed before (see Table~\ref{tab:prompts}). These mentions were further classified as Democratic or Republican. Political entities and political issues present in each article’s URL or headline were subsequently identified using the same procedure described section~\ref{entities_identified}.

For the EU elections, the identified political entities were mapped to broader political leaning categories using the Chapel Hill Expert Survey~\cite{ches2024}, following the criteria described in Section~\ref{leaning_identification}. For the US case, only entities associated with the Democratic or Republican parties were retained, as outlined previously. Political issues were mapped to issue categories using the same methodology applied to classify the search and LLM results.

\subsubsection{Polls and Previous Results}\label{methods_polls}

Poll data were collected at the country level to provide an external benchmark against which the attention given by SEs and LLMs to each political leaning or issue could be compared.

For the EU Parliament Election, polling estimates were obtained from \href{https://euobserver.com/eu-elections/areaba2f2f}{EuObserver}, a cross-national poll aggregator. The estimated number of seats per European party family and country was collected as of June 5, 2024 (three days before the election). For each of the five countries included in this study, the number of seats attributed to each political leaning was aggregated and converted into a percentage of the total seats contributed by those countries. Data on the previous number of seats held by each political family during the 2019–2024 term were retrieved from the record of the constitutive session~\cite{EuropeanParliament2024Results} and processed in the same way.

For the US Presidential Election, state-level polling intention data were collected from \href{https://www.270towin.com/2024-presidential-election-polls/#google_vignette}{270toWin}. These values were used directly as percentages. The most recent polling data available -- dated December 4, 2024 -- were used to select states with different polling intentions. At the national level, additional aggregated polling percentages for the two major parties were taken from The Economist’s cross-poll average to align the comparison with national media salience.

For previous results, the percentage of the popular vote and the vote distribution in the Electoral College obtained by each party in 2020 were used as historical baselines.

\subsubsection{Systematic Comparison and Statistical Analysis}

Differences between the proportions of mentions (described in Section~\ref{mention_counts}) and each of the three external reference values (media salience, poll data, and previous election results), were computed as:

\begin{equation}
    Diff = P_{c,\ell} - E_{c,\ell}
\end{equation}
where $E_{c,\ell}$ represents the external (“expected”) value for leaning $\ell$ in country $c$.
Positive values indicated an overrepresentation of a political category relative to the external baseline, whereas negative values indicated underrepresentation. For aggregated EU-level results, the overall difference for each political family was obtained by taking the mean of the country-level differences.

\paragraph{Statistical tests.} Two sets of tests were conducted, one against a uniform baseline and one against empirical external baselines:
\begin{enumerate}
    \item a \emph{uniform distribution}, assessing whether mentions were evenly distributed across political leanings ($1/K$ per category, where $K$ is the number of categories); and
    \item the \emph{empirical external reference values} (media salience, polling data, or previous election results), assessing deviations relative to real-world political distributions.
\end{enumerate}

\paragraph{Aggregated proportion test.} As a first test, we assessed whether the overall proportion of results mentioning each political leaning per country, differed from the country and state benchmark (e.g. Polls value for that leaning in that location) using a binomial proportion z-test:
\begin{equation}
    z = \frac{\hat{p} - p_0}{\sqrt{p_0(1-p_0)/N}}
\end{equation}
where $\hat{p}$ is the observed proportion of results mentioning leaning $\ell$, $p_0$ is the benchmark proportion, and $N$ is the total number of results containing at least one political entity mention. However, it assumes that all results are independent observations, which does not hold in practice: results returned for a single query, collected by bots with the same characteristics (location), at the same time, are very likely the product of the same algorithmic decision and are therefore possibly correlated (technical replicas). In addition, this test does not distinguish whether a single query, could be responsible for driving the significance of results.

\paragraph{Per-query test.} To address these concerns, we additionally treated the individual query as the unit of analysis. For each query,proportions were first averaged across technical replicas -- bots that submitted the same query simultaneously within the same collection round -- which we verified to constitute negligible sources of variance relative to between-query variance (median within/between ratio $< 0.05$). Queries where a given leaning category received no results contribute a proportion of zero, ensuring the denominator reflects all queries rather than only those where the leaning appeared.

For each leaning and country, a one-sample test assessed whether the mean per-query proportion differed significantly from the reference value. The test was selected adaptively based on the number of available queries $N$ and the type of data available:

\begin{itemize}
    \item \emph{Beta-Binomial Likelihood Ratio Test} when $N \geq 30$ and per-query article counts $(k_i, n_i)$ were available. This models each query's count as $k_i \sim \mathrm{BetaBinomial}(n_i, \alpha, \beta)$, where the mean proportion $\mu = \alpha/(\alpha+\beta)$ is estimated freely under $H_1$ and constrained to $p_0$ under $H_0$, while the dispersion parameter $M = \alpha + \beta$ remains free under both hypotheses. The likelihood ratio statistic $\Lambda = 2(\ell_{H_1} - \ell_{H_0}) \sim \chi^2(1)$ accounts for overdispersion arising from query-to-query variability in per-article mention rates, which a standard binomial test would ignore.

    \item \emph{Sign-flip permutation test} when $3 \leq N < 30$. The per-query deviations $d_q = p_q - p_0$ were computed, and the observed mean $\bar{d}$ was compared against a null distribution generated by randomly and independently flipping the sign of each $d_q$ over 9{,}999 permutations. The two-tailed p-value was computed as $p = (\#|\bar{d}_{\pi}| \geq |\bar{d}| + 1) / (9{,}999 + 1)$. This test makes no distributional assumptions and is exact under the null hypothesis that positive and negative deviations are equally likely.

\end{itemize}

Strata with fewer than three queries were excluded from significance testing and are reported as descriptive only.

Country-level test statistics were aggregated to an EU-level estimate using Stouffer's method~\cite{stouffer1949} weighted by $\sqrt{N_c}$, where $N_c$ is the number of queries in country $c$, giving greater influence to countries with larger and more reliable samples. For the US, the same state-level stratification and Stouffer aggregation were applied to search engine data where state-level polling is available; LLM results and issue coverage were tested on the national pool of queries.

\paragraph{Global distributional test.} Finally, to test whether the \emph{full distribution} of political leanings returned by an algorithm differed from a reference distribution, per-category Stouffer $z$-scores were combined into an omnibus statistic:
\begin{equation}
    \chi^2_{\text{approx}} = \sum_{\ell=1}^{K} z_\ell^2 \sim \chi^2(K-1)
\end{equation}
A significant result indicates that the algorithm's overall output distribution deviates from the benchmark; individual per-category tests identify which specific leanings drive that deviation.

All $p$-values were adjusted for multiple comparisons using the Holm--Bonferroni correction~\cite{holm1979}, applied separately within each analysis context (EU search engines, EU LLMs, US party search engines, US party LLMs, US issue search engines, US issue LLMs), with a significance threshold of $\alpha = 0.05$.

\section{Acknowledgments}
We thank members of the Social Physics and Complexity (SPAC) group at LIP for comments and critical reading of the manuscript. We thank HoneyComb for support in identifying meaningful topics for this study and BrightData for generously providing proxy services. This work has received funding from the European Research Council (ERC) under the European Union's (EU) Horizon 2020 and Horizon Europe research and innovation programs (Grant agreements 853566 and 101100653), granted to JGS, and from the Fundação para a Ciência e Tecnologia (FCT) PhD fellowship (2022.12547.BD), granted to ID.

\bibliographystyle{acm}
\bibliography{refs}

@techreport{pew2025newsplatform,
  author      = {{Pew Research Center}},
  title       = {News Platform Fact Sheet},
  institution = {Pew Research Center},
  year        = {2025},
  month       = {September},
  day         = {25},
  address     = {Washington, D.C.},
  url         = {https://www.pewresearch.org/journalism/fact-sheet/news-platform-fact-sheet/},
  note        = {Accessed: 2026-02-17}
}

@techreport{newman2025digital,
  author      = {Newman, Nic and Ross Arguedas, Amy and Robertson, Craig T. and Nielsen, Rasmus Kleis and Fletcher, Richard},
  title       = {Reuters Institute Digital News Report 2025},
  institution = {Reuters Institute for the Study of Journalism},
  year        = {2025},
  address     = {Oxford, UK},
  url         = {https://reutersinstitute.politics.ox.ac.uk/sites/default/files/2025-06/Digital_News-Report_2025.pdf},
  doi         = {10.60625/risj-8qqf-jt36},
  type        = {Report}
}

@misc{semrush2026top,
  author       = {{Semrush}},
  title        = {Top Websites in the World: {J}anuary 2026 {M}ost {V}isited and {P}opular {R}ankings},
  howpublished = {Semrush Traffic Analytics},
  year         = {2026},
  month        = {February},
  url          = {https://www.semrush.com/website/top/},
  note         = {Accessed: 2026-02-17}
}

@misc{robarts2025bbc,
  author       = {Stu Robarts},
  title        = {{AI} challenges the dominance of {G}oogle search},
  howpublished = {BBC News},
  year         = {2025},
  month        = {September},
  day          = {2},
  url          = {https://www.bbc.com/news/articles/c1dx9qy1eeno},
  note         = {Accessed: 2026-02-17}
}

@article{silberg2019notes,
  author = {Silberg, Jake and Manyika, James},
  title = {Notes from the AI Frontier: Tackling Bias in AI (and in Humans)},
  journal = {McKinsey Global Institute},
  year = {2019},
  month = {6},
  url = {https://www.mckinsey.com/featured-insights/artificial-intelligence/notes-from-the-ai-frontier-tackling-bias-in-ai-and-in-humans}
}

@article{apnews2023aihallucinations,
  author = {Associated Press},
  title = {Artificial Intelligence's Hallucinations: How Chatbots Like ChatGPT Generate Falsehoods},
  journal = {AP News},
  year = {2023},
  url = {https://apnews.com/article/artificial-intelligence-hallucination-chatbots-chatgpt-falsehoods-ac4672c5b06e6f91050aa46ee731bcf4},
  note = {Accessed: 2025-02-26}
}

@techreport{pew2024issues,
  author      = {{Pew Research Center}},
  title       = {Issues and the 2024 {Election}},
  institution = {Pew Research Center},
  year        = {2024},
  month       = sep,
  url         = {https://www.pewresearch.org/politics/2024/09/09/issues-and-the-2024-election/},
  urldate     = {November 10, 2025}
}

@misc{semrush_top_websites_2025,
  title        = {Most Visited Websites in the World, Updated May 2025},
  author       = {{Semrush}},
  howpublished = {\url{https://www.semrush.com/website/top/}},
  note         = {Accessed June 13, 2025},
  year         = {2025},
  month        = {May},
}

@misc{duckduckgo_privacy,
  author = {{DuckDuckGo}},
  title = {{DuckDuckGo Privacy Policy}},
  year = {2024},
  url = {https://duckduckgo.com/privacy},
  note = {Accessed: 2024-10-03}
}

@inproceedings{englehardt2016census,
author = {Englehardt, Steven and Narayanan, Arvind},
title = {Online Tracking: A 1-million-site Measurement and Analysis},
year = {2016},
isbn = {9781450341394},
publisher = {Association for Computing Machinery},
address = {New York, NY, USA},
url = {https://doi.org/10.1145/2976749.2978313},
doi = {10.1145/2976749.2978313},
booktitle = {Proceedings of the 2016 ACM SIGSAC Conference on Computer and Communications Security},
pages = {1388–1401},
numpages = {14},
location = {Vienna, Austria},
series = {CCS '16}
}

@misc{selenium,
  author       = {Selenium Project},
  title        = {Selenium WebDriver},
  howpublished = {\url{https://www.selenium.dev/documentation/}},
  note         = {Accessed: 2025-06-13}
}

@misc{irish_constitution1937,  
  title = {{Bunreacht na hÉireann} (Constitution of Ireland)},  
  author = {{Government of Ireland}},  
  year = {1937},  
  article = {Article 8},  
  url = {https://www.irishstatutebook.ie/eli/cons/en/html},  
  note = {Designates Irish as the "national and first official language" while recognizing English as a co-official language.}  
}

@techreport{cso_ireland2022,  
  title = {{Census of Population 2022 – Irish Language and Education}},  
  author = {{Central Statistics Office (CSO) Ireland}},  
  year = {2022},  
  institution = {CSO},  
  url = {https://www.cso.ie/en/releasesandpublications/ep/p-cp10esil/p10esil/ilg/},  
  note = {Reports that only 6\% of Ireland's population speaks Irish daily outside education.}  
}

@book{ethnologue2023,  
  title = {{Ethnologue: Languages of the World}},  
  author = {Eberhard, David M. and Simons, Gary F. and Fennig, Charles D.},  
  edition = {26th},  
  year = {2023},  
  publisher = {SIL International},  
  address = {Dallas, Texas},  
  url = {https://www.ethnologue.com/country/IE/},  
  note = {Documents English as "nearly universal" in Ireland, with Irish fluency below 2\%.}  
}

@misc{oxylabs,
  author       = {Oxylabs},
  title        = {Oxylabs – High Quality Proxy Service to Gather Data at Scale},
  howpublished = {\url{https://oxylabs.io/}},
  note         = {Accessed: 2025-06-20},
  year         = {2025},
}

@misc{mediacloud-metadata,
  author       = {Bhargava, Rahul and Gulley, Paige and Budne, Phil and Banos, Vangelis},
  title        = {{mediacloud‑metadata}: metadata extraction for online news articles},
  year         = {2024},
  howpublished = {\url{https://pypi.org/project/mediacloud-metadata/}},
  note         = {Version 1.4.1; Apache License; accessed 2025‑06‑24},
}

@misc{yougov2025issues,
  title = {Most Important Issues Facing the U.S.},
  author = {{YouGov}},
  year = {2025},
  note = {Accessed June 24, 2025. Available at: https://today.yougov.com/topics/politics/trackers/most-important-issues-facing-the-us},
  url = {}
}

@misc{german_polls,
    title = {Trendupdate vor der Europawahl},
    author = {Wahlkreisprognose},
    year = {2024},
    month = {6},
    note = {Accessed July 9, 2025},
    url = {https://www.wahlkreisprognose.de/2024/06/08/trendupdate-vor-der-europawahl/}
}

@misc{austrian_polls,
    title = {Umfrage: Mehrheit hält Migration für wichtigstes Thema bei der EU-Wahl},
    author = {Der Standard},
    year = {2024},
    month = {5},
    note = {Accessed July 9, 2025},
    url = {https://www.derstandard.at/story/3000000222257/umfrage-mehrheit-haelt-migration-fuer-wichtigstes-thema-bei-der-eu-wahl}
}

@misc{irish_polls,
    title = {New poll shows Sinn Féin out in front for upcoming European elections in June},
    author = {The Journal},
    year = {2024},
    month = {2},
    note = {Accessed July 9, 2025},
    url = {https://www.thejournal.ie/new-thejournalireland-thinks-poll-european-elections-6294904-Feb2024/}
}

@misc{portuguese_polls,
    title = {EU Parliamentary Projection: The End of the Rise of the Right?},
    author = {Europe Elects},
    year = {2024},
    month = {3},
    note = {Accessed July 9, 2025},
    url = {https://europeelects.eu/2024/03/31/march-2024/}
}

@misc{polish_polls,
    title = {Nowy sondaż i kłopot Trzeciej Drogi: małe zainteresowanie jej wyborców eurowyborami. Co z resztą?},
    author = {OKO.press},
    year = {2024},
    month = {5},
    note = {Accessed July 10, 2025},
    url = {https://oko.press/nowy-sondaz-i-klopot-trzeciej-drogi}
}

@misc{damiao2025,
      title={Digital Gatekeeping: An Audit of Search Engine Results shows tailoring of queries on the Israel-Palestine Conflict}, 
      author={Íris Damião and José M. Reis and Paulo Almeida and Nuno Santos and Joana Gonçalves-Sá},
      year={2025},
      eprint={2502.04266},
      archivePrefix={arXiv},
      primaryClass={cs.CY},
      url={https://arxiv.org/abs/2502.04266}, 
}

@misc{eu_political_groups,
    author = {European Parliament},
    title = {The Political groups of the European Parliament},
    year = {2025},
    url = {https://www.europarl.europa.eu/about-parliament/en/organisation-and-rules/organisation/political-groups},
    note = {Accessed: 2025-07-09}
}

@techreport{ches2024,
    author = {Jan Rovny and Ryan Bakker and Liesbet Hooghe and Seth Jolly and Gary Marks and Jonathan Polk and Marco Steenbergen and Milada Vachudova},
    title = {25 Years of Political Party Positions in Europe: The Chapel  Hill Expert Survey, 1999-2024},
    institution = {Chapel Hill Expert Survey},
    year = {2024},
    type = {Working Paper}
}

@article{in_google_we_trust,
  title={In Google we trust: Users’ decisions on rank, position, and relevance},
  author={Pan, Bing and Hembrooke, Helene and Joachims, Thorsten and Lorigo, Lori and Gay, Geri and Granka, Laura},
  journal={Journal of computer-mediated communication},
  volume={12},
  number={3},
  pages={801--823},
  year={2007},
  publisher={Oxford University Press Oxford, UK}
}

@article{students_faculty_trust_google,
  title={Search engine use behavior of students and faculty: User perceptions and implications for future research},
  author={Rieger, Oya Y},
  journal={First Monday},
  year={2009}
}

@article{misplaced_trust,
    author = {Sebastian Schultheiß and Dirk Lewandowski},
    title ={Misplaced trust? The relationship between trust, ability to identify commercially influenced results and search engine preference},
    journal = {Journal of Information Science},
    volume = {49},
    number = {3},
    pages = {609-623},
    year = {2023},
    doi = {10.1177/01655515211014157},
    url = {https://doi.org/10.1177/01655515211014157},
    eprint = {https://doi.org/10.1177/01655515211014157}
}

@misc{bing_ai_usage,
    author = {Yusuf Mehdi},
    title = {Announcing the next wave of AI innovation with Microsoft Bing and Edge},
    year = {2024},
    month = {5},
    url = {https://blogs.microsoft.com/blog/2023/05/04/announcing-the-next-wave-of-ai-innovation-with-microsoft-bing-and-edge/},
    note = {Accessed: 2025-07-18}
}

@misc{google_ai_search,
    author = {Elizabeth Reid},
    title = {Generative AI in Search: Let Google do the searching for you},
    year = {2024},
    month = {5},
    url = {https://blog.google/products/search/generative-ai-google-search-may-2024/},
    note = {Accessed: 2025-07-18}
}

@article{urman2025silence,
  author       = {Aleksandra Urman and Mykola Makhortykh},
  title        = {The Silence of the LLMs: Cross-Lingual Analysis of Political Bias and False Information Prevalence in ChatGPT, Google Bard, and Bing Chat},
  journal      = {Telematics and Informatics},
  year         = {2025},
  doi          = {10.1016/j.tele.2024.102211},
  url          = {https://scispace.com/pdf/the-silence-of-the-llms-cross-lingual-analysis-of-political-5cme3vqn4g.pdf},
  note         = {Accessed: 2025-07-07}
}

@article{auditing_partisan_bias,
author = {Robertson, Ronald E. and Jiang, Shan and Joseph, Kenneth and Friedland, Lisa and Lazer, David and Wilson, Christo},
title = {Auditing Partisan Audience Bias within Google Search},
year = {2018},
issue_date = {November 2018},
publisher = {Association for Computing Machinery},
address = {New York, NY, USA},
volume = {2},
number = {CSCW},
url = {https://doi.org/10.1145/3274417},
doi = {10.1145/3274417},
journal = {Proc. ACM Hum.-Comput. Interact.},
month = nov,
articleno = {148},
numpages = {22}
}

@article{data_voids_far_right,
    author = {Ov Cristian Norocel and Dirk Lewandowski},
    title ={Google, data voids, and the dynamics of the politics of exclusion},
    journal = {Big Data \& Society},
    volume = {10},
    number = {1},
    pages = {20539517221149099},
    year = {2023},
    doi = {10.1177/20539517221149099},
    URL = {https://doi.org/10.1177/20539517221149099},
    eprint = {https://doi.org/10.1177/20539517221149099}
}

@inbook{hyperpartisan_sources,
 URL = {http://www.jstor.org/stable/j.ctv1b0fvs5.5},
 author = {Guillén Torres and Richard Rogers},
 booktitle = {The Politics of Social Media Manipulation},
 pages = {97--122},
 publisher = {Amsterdam University Press},
 title = {Political news in search engines: Exploring Google’s susceptibility to hyperpartisan sources during the Dutch elections},
 urldate = {2025-07-11},
 year = {2020}
}

@article{self_imposed_filter_bubble,
    author = {A. G. Ekström and G. Madison and E. J. Olsson and M. Tsapos},
    title = {The search query filter bubble: effect of user ideology on political leaning of search results through query selection},
    journal = {Information, Communication \& Society},
    volume = {27},
    number = {5},
    pages = {878--894},
    year = {2024},
    publisher = {Routledge},
    doi = {10.1080/1369118X.2023.2230242},
    URL = {https://doi.org/10.1080/1369118X.2023.2230242},
    eprint = {https://doi.org/10.1080/1369118X.2023.2230242}    
}

@inproceedings{google_shaping_attention,
author = {Trielli, Daniel and Diakopoulos, Nicholas},
title = {Search as News Curator: The Role of Google in Shaping Attention to News Information},
year = {2019},
isbn = {9781450359702},
publisher = {Association for Computing Machinery},
address = {New York, NY, USA},
url = {https://doi.org/10.1145/3290605.3300683},
doi = {10.1145/3290605.3300683},
booktitle = {Proceedings of the 2019 CHI Conference on Human Factors in Computing Systems},
pages = {1–15},
numpages = {15},
location = {Glasgow, Scotland Uk},
series = {CHI '19}
}

@article{digital_personalization_effect,
    title = {The “digital personalization effect” (DPE): A quantification of the possible extent to which personalizing content can increase the impact of online manipulations},
    journal = {Computers in Human Behavior},
    volume = {166},
    pages = {108578},
    year = {2025},
    issn = {0747-5632},
    doi = {https://doi.org/10.1016/j.chb.2025.108578},
    url = {https://www.sciencedirect.com/science/article/pii/S0747563225000251},
    author = {Robert Epstein and Amanda Newland and Li Yu Tang}
}

@Article{political_bias_llms,
    author={Rettenberger, Luca
        and Reischl, Markus
        and Schutera, Mark},
    title={Assessing political bias in large language models},
    journal={Journal of Computational Social Science},
    year={2025},
    month={Feb},
    day={28},
    volume={8},
    number={2},
    pages={42},
    issn={2432-2725},
    doi={10.1007/s42001-025-00376-w},
    url={https://doi.org/10.1007/s42001-025-00376-w}
}

@Article{chatgpt_political_bias,
    author={Motoki, Fabio
        and Pinho Neto, Valdemar
        and Rodrigues, Victor},
    title={More human than human: measuring ChatGPT political bias},
    journal={Public Choice},
    year={2024},
    month={Jan},
    day={01},
    volume={198},
    number={1},
    pages={3-23},
    issn={1573-7101},
    doi={10.1007/s11127-023-01097-2},
    url={https://doi.org/10.1007/s11127-023-01097-2}
}

@Article{krafft2019search_engine_manipulation,
  author    = {Krafft, Tobias D. and Gamer, Michael and Zweig, Katharina A.},
  title     = {What did you see? A study to measure personalization in Google’s search engine},
  journal   = {EPJ Data Science},
  year      = {2019},
  volume    = {8},
  number    = {1},
  month     = dec,
  issn      = {2193-1127},
  doi       = {10.1140/epjds/s13688-019-0217-5},
  publisher = {Springer Science and Business Media LLC},
}

@misc{tonmoy2024survey,
  title        = {A Comprehensive Survey of Hallucination Mitigation Techniques in Large Language Models},
  author       = {S. M. Towhidul Islam Tonmoy and S. M. Mehedi Zaman and Vinija Jain and Anku Rani and Vipula Rawte and Aman Chadha and Amitava Das},
  year         = {2024},
  eprint       = {arXiv:2401.01313},
  primaryClass = {cs.CL},
  doi          = {10.48550/arXiv.2401.01313},
  url          = {https://arxiv.org/abs/2401.01313}
}

@inproceedings{potter-etal-2024-hidden,
  title     = {Hidden Persuaders: {LLM}s' Political Leaning and Their Influence on Voters},
  author    = {Potter, Yujin and Lai, Shiyang and Kim, Junsol and Evans, James and Song, Dawn},
  booktitle = {Proceedings of the 2024 Conference on Empirical Methods in Natural Language Processing (EMNLP)},
  pages     = {4244--4275},
  address   = {Miami, Florida, USA},
  month     = nov,
  year      = {2024},
  publisher = {Association for Computational Linguistics},
  doi       = {10.18653/v1/2024.emnlp-main.244},
  url       = {https://aclanthology.org/2024.emnlp-main.244/}
}

@misc{yang2024unpacking,
  title        = {Unpacking Political Bias in Large Language Models: A Cross‑Model Comparison on U.S. Politics},
  author       = {Yang, Kaiqi and Li, Hang and Chu, Yucheng and Lin, Yuping and Peng, Tai-Quan and Liu, Hui},
  year         = {2024},
  eprint       = {arXiv:2412.16746v3},
  primaryClass = {cs.CL},
  doi          = {10.48550/arXiv.2412.16746},
  url          = {https://arxiv.org/abs/2412.16746v3}
}

@inproceedings{tran2022fairness,
  title     = {Fairness Increases Adversarial Vulnerability},
  author    = {Tran, Cuong and Zhu, Keyu and Fioretto, Ferdinando and Van Hentenryck, Pascal},
  booktitle = {Proceedings of the 2022 NeurIPS Workshop on Fairness and ML},
  year      = {2022},
  url       = {https://arxiv.org/abs/2211.11835}
}

@inproceedings{yang-etal-2025-rethinking-prompt,
  title     = {Rethinking Prompt-based Debiasing in Large Language Models},
  author    = {Yang, Xinyi and Zhan, Runzhe and Yang, Shu and Wu, Junchao and Chao, Lidia S. and Wong, Derek F.},
  booktitle = {Findings of the Association for Computational Linguistics: ACL 2025},
  editor    = {Che, Wanxiang and Nabende, Joyce and Shutova, Ekaterina and Pilehvar, Mohammad Taher},
  pages     = {26538--26553},
  address   = {Vienna, Austria},
  month     = jul,
  year      = {2025},
  publisher = {Association for Computational Linguistics},
  doi       = {10.18653/v1/2025.findings-acl.1361},
  url       = {https://aclanthology.org/2025.findings-acl.1361/}
}

@misc{mont2022trust,
  title={The trust gap: how and why news on digital platforms is viewed more sceptically versus news in general},
  author={Mont'Alverne, Camila and Badrinathan, Sumitra and Ross Arguedas, Amy and Toff, Benjamin and Fletcher, Richard and Kleis Nielsen, Rasmus},
  year={2022},
  publisher={Reuters Institute for the Study of Journalism},
  doi = {10.60625/risj-skfk-h856}
}

@misc{burman2024google,
  author = {Burman, Theo},
  title = {Google Searches Spike for Harris, Trump Policies on Eve of Election Day},
  year = {2024},
  date = {2024-11-05},
  url = {https://www.newsweek.com/kamala-harris-donald-trump-policy-google-election-day-1980376},
  publisher = {Newsweek},
  note = {Accessed: 2025-09-27}
}

@inproceedings{otterbacher2017competent,
  title={Competent Men and Warm Women: Gender Stereotypes and Backlash in Image Search Results},
  author={Otterbacher, Jahna and Bates, Jo and Clough, Paul},
  booktitle={Proceedings of the 2017 CHI Conference on Human Factors in Computing Systems},
  pages={6620--6631},
  year={2017},
  publisher={ACM},
  doi={10.1145/3025453.3025727},
  url={https://eprints.whiterose.ac.uk/id/eprint/111419/7/Exploring_bias_FINAL_6_toshare.pdf}
}

@inproceedings{Makhortykh2021DetectingRA,
  title={Detecting race and gender bias in visual representation of {AI} on web search engines},
  author={Mykola Makhortykh and Aleksandra Urman and Roberto Ulloa},
  booktitle={Int. Workshop on Algorithmic Bias in Search and Recommendation},
  year={2021},
  url={https://api.semanticscholar.org/CorpusID:235659011}
}

@book{noble2018algorithms,
  title={Algorithms of Oppression: How Search Engines Reinforce Racism},
  author={Noble, Safiya Umoja},
  year={2018},
  publisher={New York University Press},
  address={New York},
  isbn={9781479837243},
  url={https://nyupress.org/books/9781479837243/}
}

@article{Urman2021WhereTE,
  title={Where the {Earth} is flat and 9/11 is an inside job: A comparative algorithm audit of conspiratorial information in web search results},
  author={Aleksandra Urman and Mykola Makhortykh and Roberto Ulloa and Juhi Kulshrestha},
  journal={ArXiv},
  year={2021},
  volume={abs/2112.01278},
  url={https://api.semanticscholar.org/CorpusID:244798677}
}

@article{huang2023hallucination,
  title={A Survey on Hallucination in Large Language Models: Principles, Taxonomy, Challenges, and Open Questions},
  author={Huang, Lei and Yu, Weijiang and Ma, Weitao and Zhong, Weihong and Feng, Zhangyin and Wang, Haotian and Chen, Qianglong and Peng, Weihua and Feng, Xiaocheng and Qin, Bing and others},
  journal={arXiv preprint arXiv:2311.05232},
  year={2023},
  url={https://arxiv.org/abs/2311.05232}
}

@article{venkit2024confidently,
  title={Confidently Nonsensical?: A Critical Survey on the Perspectives and Challenges of ‘Hallucinations’ in NLP},
  author={Venkit, Pranav Narayanan and Chakravorti, Tatiana and Gupta, Vipul and Biggs, Heidi and Srinath, Mukund and Goswami, Koustava and Rajtmajer, Sarah and Wilson, Shomir},
  journal={arXiv preprint arXiv:2404.07461},
  year={2024},
  url={https://arxiv.org/abs/2404.07461}
}

@article{laban2023are,
  title={Are You Sure? Challenging LLMs Leads to Performance Drops in the FlipFlop Experiment},
  author={Laban, Philippe and Murakhovs'ka, Lidiya and Xiong, Caiming and Wu, Chien-Sheng},
  journal={arXiv preprint arXiv:2311.08596},
  year={2023},
  url={https://arxiv.org/abs/2311.08596}
}

@inproceedings{sharma2024generative,
  title={Generative Echo Chamber? Effect of LLM-Powered Search Systems on Diverse Information Seeking},
  author={Sharma, Nikhil and Liao, Q. Vera and Xiao, Ziang},
  booktitle={Proceedings of the CHI Conference on Human Factors in Computing Systems},
  pages={1--17},
  year={2024},
  publisher={ACM},
  doi={10.1145/3613904.3642459},
  url={https://dl.acm.org/doi/10.1145/3613904.3642459}
}

@article{cameron2025electoral,
  title={Electoral context matters: why undecided voters in elections and referendums are different},
  author={Cameron, Sarah and McAllister, Ian},
  journal={Parliamentary Affairs},
  year={2025},
  doi={10.1093/pa/gsaf043},
  url={https://doi.org/10.1093/pa/gsaf043}
}

@incollection{introna2007shaping,
  title={Shaping the Web: Why the Politics of Search Engines Matters},
  author={Introna, Lucas D. and Nissenbaum, Helen},
  booktitle={Computer Ethics},
  editor={Himma, Kenneth Einar and Tavani, Herman T.},
  publisher={Routledge},
  year={2007},
  pages={275--289},
  doi={10.4324/9781315259697-19},
  url={https://www.taylorfrancis.com/chapters/edit/10.4324/9781315259697-19/shaping-web-politics-search-engines-matters-lucas-introna-helen-nissenbaum}
}

@article{mowshowitz2002assessing,
  title={Assessing Search Engine Bias},
  author={Mowshowitz, Abbe and Kawaguchi, Akira},
  journal={Communications of the ACM},
  volume={45},
  number={9},
  pages={56--60},
  year={2002},
  publisher={ACM},
  doi={10.1145/567498.567527},
  url={https://dl.acm.org/doi/10.1145/567498.567527}
}

@inproceedings{shepherd2017method,
  title={A Method for Detecting Bias in Search Rankings, with Evidence of Systematic Bias Related to the 2016 Presidential Election},
  author={Shepherd, Taylor and Zhang, Jun},
  booktitle={2017 IEEE International Conference on Big Data (Big Data)},
  pages={1486--1491},
  year={2017},
  publisher={IEEE},
  doi={10.1109/BigData.2017.8258083},
  url={https://ieeexplore.ieee.org/document/8258083}
}

@inproceedings{trielli2019partisan,
  title={Partisan Search Behavior and Google Results in the 2018 U.S. Midterm Elections},
  author={Trielli, Daniel and Diakopoulos, Nicholas},
  booktitle={Proceedings of the 2019 CHI Conference on Human Factors in Computing Systems},
  pages={1--15},
  year={2019},
  publisher={ACM},
  doi={10.1145/3290605.3300680},
  url={https://doi.org/10.1145/3290605.3300680}
}

@article{urman2021matter,
  title={The Matter of Chance: Auditing Web Search Results Related to the 2020 U.S. Presidential Primary Elections Across Six Search Engines},
  author={Urman, Aleksandra and Makhortykh, Mykola and Ulloa, Roberto},
  journal={arXiv preprint arXiv:2105.00756},
  year={2021},
  url={https://arxiv.org/abs/2105.00756}
}

@article{gezici2020evaluation,
  title={Evaluation Metrics for Measuring Bias in Search Engine Results},
  author={Gezici, Gizem and Lipani, Aldo and Saygin, Yucel and Yilmaz, Emine},
  journal={Information Retrieval Journal},
  volume={24},
  number={5},
  pages={400--432},
  year={2021},
  doi={10.1007/s10791-020-09386-w},
  url={https://link.springer.com/article/10.1007/s10791-020-09386-w}
}

@article{makhortykh2025campaigning,
  title={Campaigning through the Lens of Google: A Large-Scale Algorithm Audit of Google Searches in the Run-Up to the Swiss Federal Elections 2023},
  author={Makhortykh, Mykola},
  journal={arXiv preprint arXiv:2507.06018},
  year={2025},
  url={https://arxiv.org/abs/2507.06018}
}

@misc{allsides2025mediabias,
  title={AllSides Media Bias Chart},
  author={AllSides},
  year={2025},
  howpublished={\url{https://www.allsides.com/media-bias/media-bias-chart}},
  note={Accessed: 2025-10-15}
}

@inproceedings{voter_centered_audits,
author = {Mustafaraj, Eni and Lurie, Emma and Devine, Claire},
title = {The case for voter-centered audits of search engines during political elections},
year = {2020},
isbn = {9781450369367},
publisher = {Association for Computing Machinery},
address = {New York, NY, USA},
url = {https://doi.org/10.1145/3351095.3372835},
doi = {10.1145/3351095.3372835},
booktitle = {Proceedings of the 2020 Conference on Fairness, Accountability, and Transparency},
pages = {559–569},
numpages = {11},
location = {Barcelona, Spain},
series = {FAT* '20}
}

@misc{britannica2025electionresults,
  title={United States Presidential Election Results},
  author={Encyclopaedia Britannica},
  year={2025},
  howpublished={https://www.britannica.com/topic/United-States-Presidential-Election-Results-1788863},
  note={Accessed: 2025-10-15}
}

@article{willocq2019explaining,
  title={Explaining time of vote decision: The socio-structural, attitudinal, and contextual determinants of late deciding},
  author={Willocq, Simon},
  journal={Political Studies Review},
  volume={17},
  number={1},
  pages={53--64},
  year={2019},
  publisher={SAGE Publications Sage UK: London, England}
}

@article{epstein2015search,
  title={The search engine manipulation effect (SEME) and its possible impact on the outcomes of elections},
  author={Epstein, Robert and Robertson, Ronald E},
  journal={Proceedings of the National Academy of Sciences},
  volume={112},
  number={33},
  pages={E4512--E4521},
  year={2015},
  publisher={National Academy of Sciences}
}

@article{lee2016agenda,
  title={Agenda setting in the Internet Age: The reciprocity between online searches and issue salience},
  author={Lee, ByungGu and Kim, Jinha and Scheufele, Dietram A},
  journal={International Journal of Public Opinion Research},
  volume={28},
  number={3},
  pages={440--455},
  year={2016},
  publisher={Oxford University Press}
}

@incollection{diakopoulos2018vote,
  author    = {Diakopoulos, Nicholas and Trielli, Daniel and Stark, Jennifer and Mussenden, Sean},
  title     = {I Vote for—How Search Informs Our Choice of Candidate},
  booktitle = {Digital Dominance: The Power of Google, Amazon, Facebook, and Apple},
  editor    = {Moore, Martin and Tambini, Damian},
  publisher = {Oxford University Press},
  address   = {Oxford},
  pages     = {320--341},
  year      = {2018}
}

@article{hopmann2010effects,
  title={Effects of election news coverage: How visibility and tone influence party choice},
  author={Hopmann, David Nicolas and Vliegenthart, Rens and De Vreese, Claes and Alb{\ae}k, Erik},
  journal={Political communication},
  volume={27},
  number={4},
  pages={389--405},
  year={2010},
  publisher={Taylor \& Francis}
}

@article{adams2024much,
  title={How much does issue salience matter? A model with applications to the UK elections},
  author={Adams, James and Merrill III, Samuel and Zur, Roi},
  journal={European Journal of Political Research},
  volume={63},
  number={2},
  pages={798--809},
  year={2024},
  publisher={Wiley Online Library}
}

@article{vasilopoulou2024electoral,
  title={Electoral competition, the EU issue and far-right success in Western Europe},
  author={Vasilopoulou, Sofia and Zur, Roi},
  journal={Political Behavior},
  volume={46},
  number={1},
  pages={565--585},
  year={2024},
  publisher={Springer}
}

@article{geers2017priming,
  title={Priming issues, party visibility, and party evaluations: The impact on vote switching},
  author={Geers, Sabine and Bos, Linda},
  journal={Political Communication},
  volume={34},
  number={3},
  pages={344--366},
  year={2017},
  publisher={Taylor \& Francis}
}

@misc{EuropeanParliament2024Results,
  key          = {European Parliament},
  title        = {European Parliament 2024--2029: European Results},
  howpublished = {\url{https://results.elections.europa.eu/en/european-results/2024-2029/}},
  note         = {Official results of the 2024--2029 European Parliament elections},
  year         = {2024},
  organization = {European Union},
  url          = {https://results.elections.europa.eu/en/european-results/2024-2029/},
  urldate      = {November 10, 2025}
}

@misc{Fay2024b,
  author       = {Owen Fay},
  title        = {Value of \#1 position on {Google} - positional analysis study},
  howpublished = {Poll the People. Accessed: Sep. 5, 2024. [Online]. Available: \url{https://pollthepeople.app/the-value-of-google-result-positioning-3/}}
}

@article{Damiao2026radicalright,
  title        = {Cross-National Evidence of Disproportionate Media Visibility for the Radical Right in the 2024 European Elections},
  author       = {Dami{\~a}o, {\'I}ris and Franco, Jo{\~a}o and Silva, Mariana and Almeida, Paulo and Gon{\c c}alves-S{\'a}, Joana},
  journal      = {arXiv},
  year         = {2026},
  eprint       = {2601.05826},
  archivePrefix= {arXiv},
  primaryClass = {cs.CY},
  doi          = {10.48550/arXiv.2601.05826},
  url          = {https://arxiv.org/abs/2601.05826}
}

@article{Israel_Palestine,
  author={Dami{\~a}o, {\'I}ris and Reis, Jos{\'e} M. and Almeida, Paulo and Santos, Nuno and Gon{\c c}alves-S{\'a}, Joana},
  journal={IEEE Access}, 
  title={Unequal Content: Search Engine Personalization Influences Exposure to Information Regarding the Israel–Palestine Conflict}, 
  year={2026},
  volume={14},
  number={},
  pages={15347-15361},
  doi={10.1109/ACCESS.2025.3635871}}

@inproceedings{inconsistent_search_results,
author = {Kliman-Silver, Chloe and Hannak, Aniko and Lazer, David and Wilson, Christo and Mislove, Alan},
title = {Location, Location, Location: The Impact of Geolocation on Web Search Personalization},
year = {2015},
isbn = {9781450338486},
publisher = {Association for Computing Machinery},
address = {New York, NY, USA},
url = {https://doi.org/10.1145/2815675.2815714},
doi = {10.1145/2815675.2815714},
booktitle = "{Proc. 2015 Internet Meas. Conf. (IMC)}",
pages = {121–127},
numpages = {7},
location = {Tokyo, Japan},
series = {IMC '15}
}

@article{late_voters,
    author = {Irwin, Galen A. and Van Holsteyn, Joop J. M.},
    title = {What are they Waiting for? Strategic Information for Late Deciding Voters},
    journal = {International Journal of Public Opinion Research},
    volume = {20},
    number = {4},
    pages = {483-493},
    year = {2008},
    month = {11},
    issn = {0954-2892},
    doi = {10.1093/ijpor/edn040},
    url = {https://doi.org/10.1093/ijpor/edn040},
    eprint = {https://academic.oup.com/ijpor/article-pdf/20/4/483/2212130/edn040.pdf},
}

@book{stouffer1949,
  author    = {Stouffer, Samuel A. and others},
  title     = {The American Soldier: Adjustment During Army Life},
  year      = {1949},
  publisher = {Princeton University Press}
}

@article{holm1979,
  author  = {Holm, Sture},
  title   = {A Simple Sequentially Rejective Multiple Test Procedure},
  journal = {Scandinavian Journal of Statistics},
  year    = {1979},
  volume  = {6},
  number  = {2},
  pages   = {65--70}
}

@article{erokhin2025google,
  title={Applying Google Trends to analyze electoral outcomes: A 2024 cross-national perspective},
  author={Erokhin, Dmitry},
  journal={Social Sciences \& Humanities Open},
  volume={},
  number={},
  pages={101846},
  year={2025},
  publisher={Elsevier},
  doi={10.1016/j.ssaho.2025.101846}
}

@article{10.1145/3359231,
author = {Metaxa, Dana\"{e} and Park, Joon Sung and Landay, James A. and Hancock, Jeff},
title = {Search Media and Elections: A Longitudinal Investigation of Political Search Results},
year = {2019},
issue_date = {November 2019},
publisher = {Association for Computing Machinery},
address = {New York, NY, USA},
volume = {3},
number = {CSCW},
url = {https://doi.org/10.1145/3359231},
doi = {10.1145/3359231},
journal = {Proc. ACM Hum.-Comput. Interact.},
month = nov,
articleno = {129},
numpages = {17}
}

@misc{FarRightReconfiguration,
  key          = {Global Affairs},
  title        = {The Fragmentation of the {European Parliament} after the 2024 Elections},
  howpublished = {\url{https://en.unav.edu/web/global-affairs/the-fragmentation-of-the-european-parliament-after-the-2024-elections}},
  year         = {2024},
  organization = {Global Affairs and Strategic Studies, University of Navarra},
  urldate      = {November 10, 2025}
}

@article{pierri2025dsa,
  title={Research Opportunities and Challenges of the {EU}'s Digital Services Act},
  author={Pierri, Francesco and Araujo, Theo and Kruikemeier, Sanne and 
          Lorenz-Spreen, Philipp and Vanden Abeele, Mariek M. P. and 
          Vandenbosch, Laura and Gon\c{c}alves-S\'{a}, Joana and 
          Grabowicz, Przemyslaw A.},
  journal={Communications of the {ACM}},
  year={2025},
  eprint={2512.14223},
  archivePrefix={arXiv},
  primaryClass={cs.CY},
  doi={10.48550/arXiv.2512.14223}
}

\onecolumn
\clearpage
\appendix

\renewcommand{\thesection}{\Alph{section}.}
\renewcommand{\thesubsection}{\thesection\arabic{subsection}}

\setcounter{figure}{0} 
\renewcommand{\thefigure}{A\arabic{figure}}

\setcounter{table}{0}
\renewcommand{\thetable}{A\arabic{table}}

% \section*{Appendix Index}
% \addcontentsline{toc}{section}{Appendices}
% \begin{itemize}
%   \item \hyperref[app_sec:external_data]{External data sources: Polls, Previous and Current Results}
%   \item \hyperref[app_sec:data_collection]{Data Collection Supplementary Information}
%   \item \hyperref[app_sec:search_engine_results]{Supplementary Analysis of SE results}
% \end{itemize}

\section*{Supplementary Information Index}
\addcontentsline{toc}{section}{Appendix Index}

\begin{description}[leftmargin=2.5cm, style=nextline]
  \item[A. Data Sources and Benchmarking]
        \hyperref[app_sec:external_data]{A.1 Polls and electoral results (previous and current)} \dotfill \pageref{app_sec:external_data}

  \item[B. Data collection and Audit design]
        \hyperref[app_subsec:audits]{B.1 Audit configurations for SEs and LLMs} \dotfill \pageref{app_subsec:audits}

        \hyperref[app_subsec:queries]{B.2 Search and prompt queries used across languages} \dotfill \pageref{app_subsec:queries}

        \hyperref[app_sec:data_collection]{B.3 Illustrative SE results and LLM responses} \dotfill \pageref{app_subsec:illustrative_results}

        \hyperref[app_sec:data_collection]{B.4 Audit success rates (SEs and LLMs)} \dotfill \pageref{app_subsec:successes_se}

  \item[C. Political leaning classification frameworks]
    \hyperref[subsec:eu_mapping]{C.1 Mapping of EU political groups and leaning} \dotfill \pageref{subsec:eu_mapping}
    
    \hyperref[app_sec:entities_n_results]{C.2 Entities and their associated leaning across all collected results} \dotfill \pageref{app_sec:entities_n_results}

  \item[D. Statistical Analysis]
    \hyperref[app_subsec:uniform]{D.1 Testing deviations from a uniform distribution} \dotfill \pageref{app_subsec:uniform}

    \hyperref[app_subsec:external]{D.2 Testing comparisons with external benchmarks} \dotfill \pageref{app_subsec:external}

  \item[E. Results supplementary analysis]
    \hyperref[app_subsec:se_entities]{E.1 Proportion of results containing relevant entities (SEs and LLMs)} \dotfill \pageref{app_subsec:se_entities}

    \hyperref[app_sec:se_results_categories]{E.2 Categories of results retrieved by SEs} \dotfill \pageref{app_sec:se_results_categories}

    \hyperref[app_sec:se_differences]{E.3 Differences of SE results across bots} \dotfill \pageref{app_sec:se_differences}

    \hyperref[app_sec:top_3]{E.4 Top 3 Results Analysis} \dotfill \pageref{app_sec:top_3}

    \hyperref[app_sec:se_leaning]{E.5 Results Leaning} \dotfill \pageref{app_sec:se_leaning}

  \item[F. Robustness and Control Analyses]
  
    \hyperref[app_sec:control_1]{F.1 Control Analysis: Removing Economy from political issues} \dotfill \pageref{app_sec:control_1}

    \hyperref[app_sec:bias_external]{F.2 Bias of results against external sources} \dotfill \pageref{app_sec:bias_external}

   \item[G. Average of results bias against external benchmarks]

   \item[H. Prompts provided to ChatGPT]
    
  \item[H. Additional Information on the Media Data Collection]
  
    \hyperref[terms_media]{H.1 Terms used to collect MediaCloud News} \dotfill \pageref{terms_media}

    \hyperref[top_news]{H.2 Top 20 Newspapers per country} \dotfill \pageref{top_news}

\end{description}

\onecolumn
\section{Data Sources and Benchmarking}
\subsection{Polls and electoral results (previous and current)}\label{app_sec:external_data}

\begin{figure*}[htbp]
    \centering
    \includegraphics[width=0.9\textwidth]{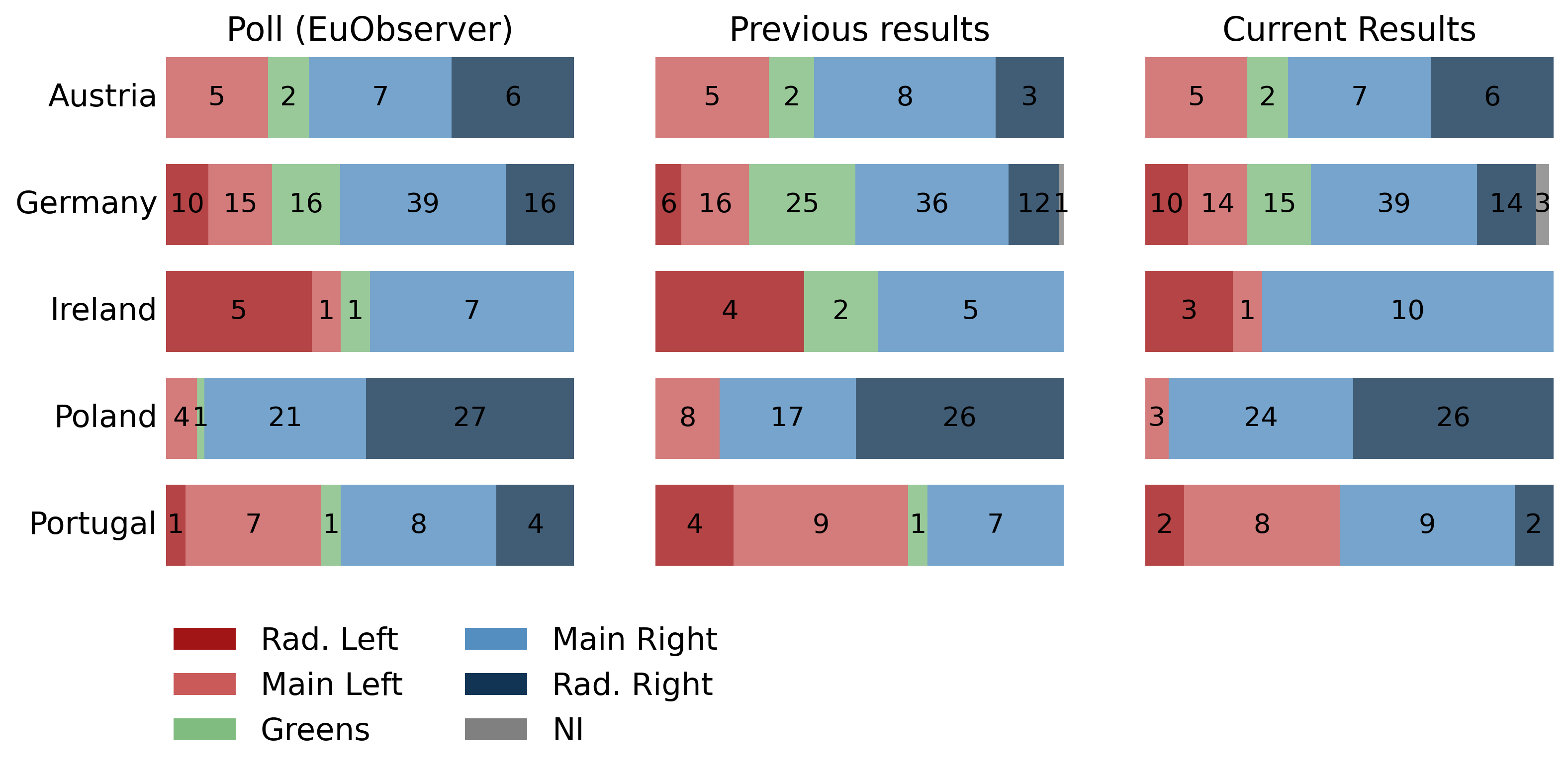}
    \caption{\textbf{EU Parliament Election external source data per country.} Column 1 reports the projected number of seats for each political leaning based on EuObserver polling during the week preceding the election. Column 2 presents the distribution of seats obtained by each political leaning in the 2019 constitutive session of the EU Parliament. Column 3 shows the actual 2024 election results for the five countries included in this study.}
    \label{fig:polls_eu}
\end{figure*}

\begin{figure*}[htbp]
    \centering
    \includegraphics[width=1\textwidth]{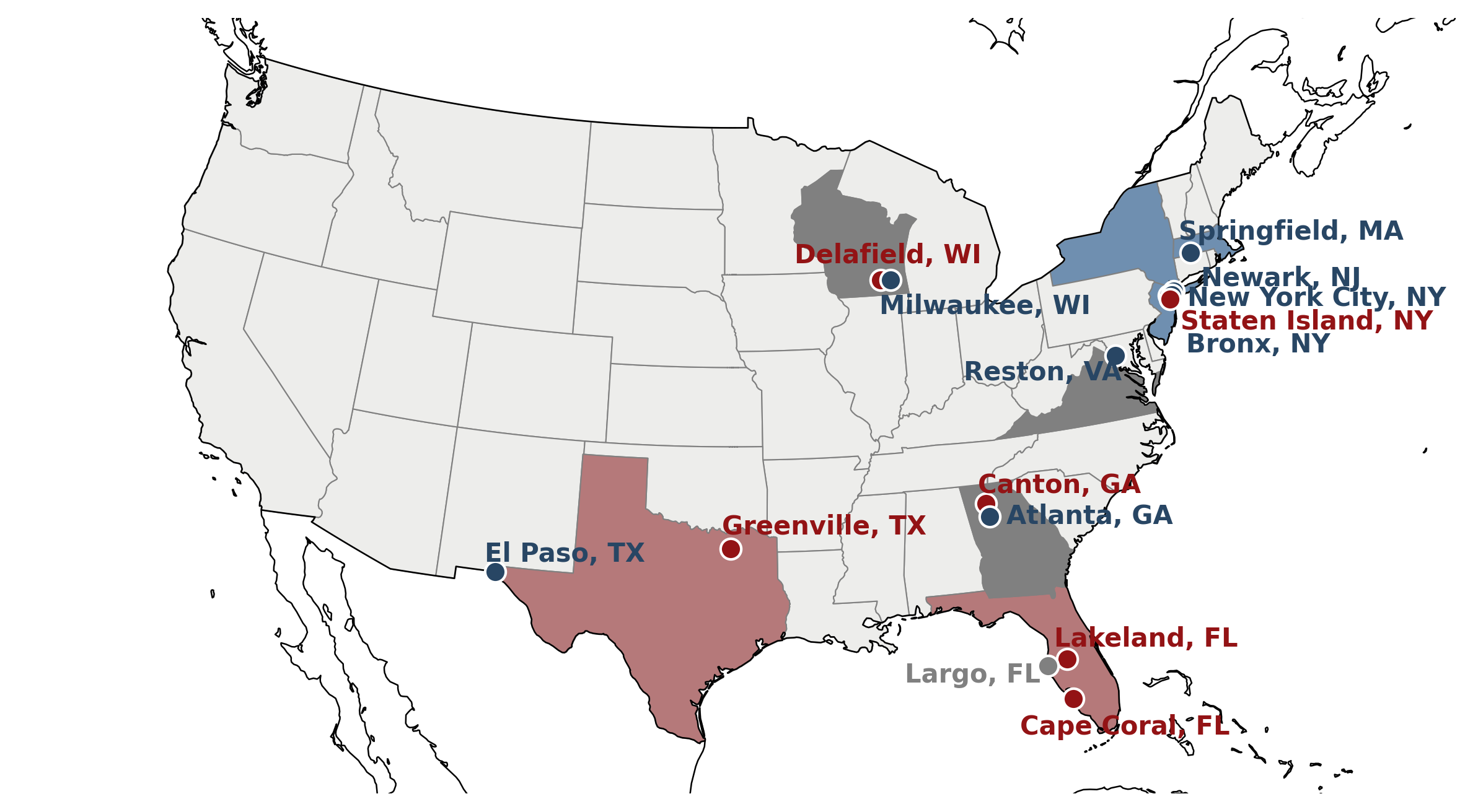}
    \caption{\textbf{Map of the United States showing the deployment locations of the bots.} Circles mark the counties selected within each state, colored according to the most voted party (red - Republican Party and blue - Democrat Party) in the 2020 US Presidential elections (prior to the one studied in the article). States containing selected counties are shaded according to state-level polling predictions for the 2024 election: grey indicates states where no party was projected to exceed 50\% of the vote (Wisconsin, Georgia, Virginia); blue indicates states where the Democratic Party was projected to receive more than 50\% (New York, New Jersey, Massachusetts); red indicates states where the Republican Party was projected to receive more than 50\% (Texas, Florida).}
    \label{fig:selected_states_and_counties}
\end{figure*}

\begin{figure*}[htbp]
    \centering
    \includegraphics[width=1\textwidth]{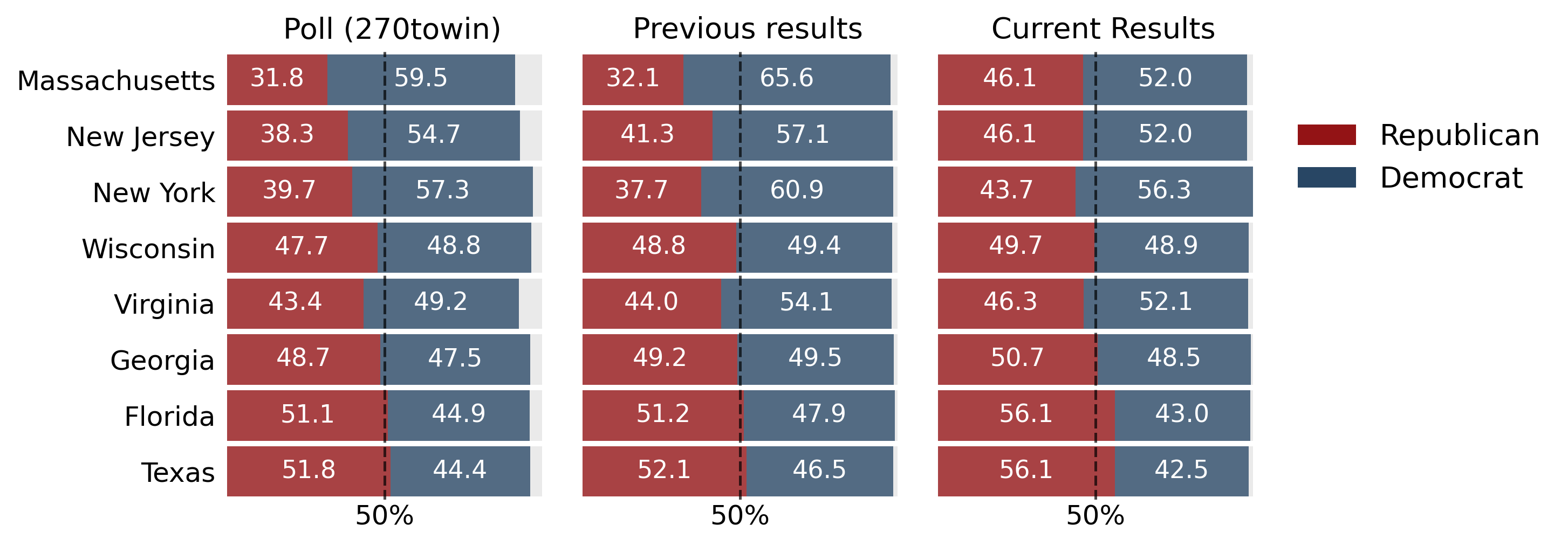}
    \caption{\textbf{US Presidential Election external source data by state.} Column 1 reports the projected vote share for each state based on pre-election polling in the week prior to the elections. Column 2 shows the actual vote share from the 2020 election. Column 3 displaus the official 2024 results per state. Red bars represent Republican vote share and blue bars represent Democratic vote share.}
    \label{fig:polls_us}
\end{figure*}

\clearpage
\section{Data collection and Audit design}\label{app_sec:data_collection}

\subsection{Audit Configurations for Search Engines and LLMs}\label{app_subsec:audits}
\begin{table*}[ht!]
\centering
\caption{\textbf{Configuration details of the web crawlers employed to collect data from search engines and large language models in the EU case study and the US Presidential Elections.
}}
\label{tab:audit_summary_rev}
\small
\begin{tabular}{
    >{\RaggedRight\arraybackslash}p{1.5cm}|
    >{\RaggedRight\arraybackslash}p{7cm}|
    >{\RaggedRight\arraybackslash}p{7cm}
}
\hline
\makecell[l]{\textbf{Audit}\\\textbf{Type}} & \makecell[l]{\textbf{EU}\\\textbf{Parliament Election}} & \makecell[l]{\textbf{US}\\\textbf{Presidential Election}} \\
\hline
\hline
\multirow{8}{*}{\textbf{SEs}} & & \\
& \textbf{Proxies:} Country-level (5 per loc) & \textbf{Proxies:} County-Level (3-4 per loc) \\
& \textbf{Loc:} AT, DE, IR, PL, PT & \textbf{Loc:} Democratic and Republican counties (15 total) \\
\cline{2-3} % Use \cline to separate audit configurations within the section
& \textbf{Audit 1 (English Lang)} & \textbf{Audit 1 (Base Configuration)} \\
& \textbf{Browser/Query Lang:} English & \textbf{Browser/Query Lang:} English \\
& \textbf{Audit Date:} Jun 6, 2024 & \textbf{Audit Date:} Nov 1, 2024 \\
\cline{2-2} % Separate Audit 1 and Audit 2 in the EU column
& \textbf{Audit 2 (Local Lang)} & \\
& \textbf{Browser/Query Lang:} Local language &  \\ % Placeholder row for alignment
& \textbf{Audit Date:} Jun 7, 2024 & \\
\hline
\hline
% --- LLMS SECTION ---
\multirow{9}{*}{\textbf{LLMs}} & & \\
& \textbf{Proxies:} One per country & \textbf{Proxies:} County-level (3 bots per location) \\
& \textbf{Loc:} AT, DE, IR, PL, PT & \textbf{Loc:} 4 different counties (BrightData proxies only) \\
& \textbf{Query Lang:} Local  language & \textbf{Query Lang:} English \\
\cline{2-3}
& \textbf{Audit 1 (Manual)} & \textbf{Audit 1 (Bots)} \\
& \textbf{Models:} Copilot, ChatGPT-3.5 (1 trial) & \textbf{Models:} Copilot \\
& \textbf{Date:} Jun 7, 2024 & \textbf{Date:} Nov 4, 2024 \\
\cline{2-3}
& \textbf{Audit 2 (API)} & \textbf{Audit 2 (API)} \\
& \textbf{Models:} ChatGPT-3.5 (3 trials) & \textbf{Models:} ChatGPT-4o \\
& \textbf{Date:} Jun 2, 2025 & \textbf{Date:} Nov 4, 2024 \\
\hline
\end{tabular}
\end{table*}

\clearpage
\subsection{Search and prompt queries used across languages}\label{app_subsec:queries}

\begin{table*}[ht!]
\centering
\small
\caption{Queries used for SE and LLM in EU election audits across languages and countries.}
\label{tab:queries_eu_languages}
\begin{tabular}{p{1.5cm}p{1.5cm}p{1.5cm}p{10cm}}
\toprule
\textbf{} & \textbf{Language} & \textbf{Country} & \textbf{Query} \\
\midrule

\multirow{24}{*}{\makecell[l]{Search\\Engines}}
& \multirow{6}{*}{English} & \multirow{6}{*}{Ireland}
& european parliament elections \\
& & & european parliament parties \\
& & & european elections choose party \\
& & & european elections top issues \\
& & & who should I vote for european elections 2024? \\
& & & who is going to win european elections 2024? \\
\cline{2-4}
& \multirow{6}{*}{German} & \multirow{6}{*}{\makecell[l]{Austria\\and \\Germany}}
& Europawahl \\
& & & Parteien im Europäischen Parlament \\
& & & Europawahl Partei wählen \\
& & & Europawahl wichtigste Themen \\
& & & Für wen sollte ich bei der Europawahl 2024 stimmen? \\
& & & Wer wird die Europawahl 2024 gewinnen? \\
\cline{2-4}

& \multirow{6}{*}{Polish} & \multirow{6}{*}{Poland}
& wybory do Parlamentu Europejskiego \\
& & & partie w Parlamencie Europejskim \\
& & & wybory europejskie wybierz partię \\
& & & wybory europejskie najważniejsze problemy \\
& & & na kogo powinienem głosować w wyborach europejskich 2024? \\
& & & kto wygra wybory europejskie 2024? \\
\cline{2-4}

& \multirow{6}{*}{Portuguese} & \multirow{6}{*}{Portugal}
& eleições do Parlamento Europeu \\
& & & partidos do Parlamento Europeu \\
& & & eleições europeias escolher partido \\
& & & eleições europeias principais questões \\
& & & em quem devo votar nas eleições europeias de 2024? \\
& & & quem vai ganhar as eleições europeias de 2024? \\

\midrule

\multirow{16}{*}{LLMs}
& \multirow{4}{*}{English} & \multirow{4}{*}{Ireland}
& What are the European Parliament parties? \\
& & & What is the best European party? \\
& & & Who is going to win European elections 2024? \\
& & & Who should I vote for in European elections 2024? \\
\cline{2-4}

& \multirow{4}{*}{German} & \multirow{4}{*}{\makecell[l]{Austria\\and\\Germany}}
& Was sind die Parteien des Europäischen Parlaments? \\
& & & Was ist die beste europäische Partei? \\
& & & Wer wird die Europawahlen 2024 gewinnen? \\
& & & Wen soll ich bei den Europawahlen 2024 wählen? \\
\cline{2-4}

& \multirow{4}{*}{Polish} & \multirow{4}{*}{Poland}
& Jakie są partie Parlamentu Europejskiego? \\
& & & Jaka jest najlepsza europejska partia? \\
& & & Kto wygra wybory europejskie w 2024 roku? \\
& & & Na kogo powinienem głosować w wyborach europejskich 2024? \\
\cline{2-4}

& \multirow{4}{*}{Portuguese} & \multirow{4}{*}{Portugal}
& Quais são os partidos do Parlamento Europeu? \\
& & & Qual é o melhor partido europeu? \\
& & & Quem vai ganhar as eleições europeias de 2024? \\
& & & Em quem devo votar nas eleições europeias de 2024? \\

\bottomrule
\end{tabular}
\end{table*}

\clearpage
\subsection{Illustrative SE results and LLM responses}\label{app_subsec:illustrative_results}

\begin{figure}[ht]
    \centering
    \includegraphics[width=1\textwidth]{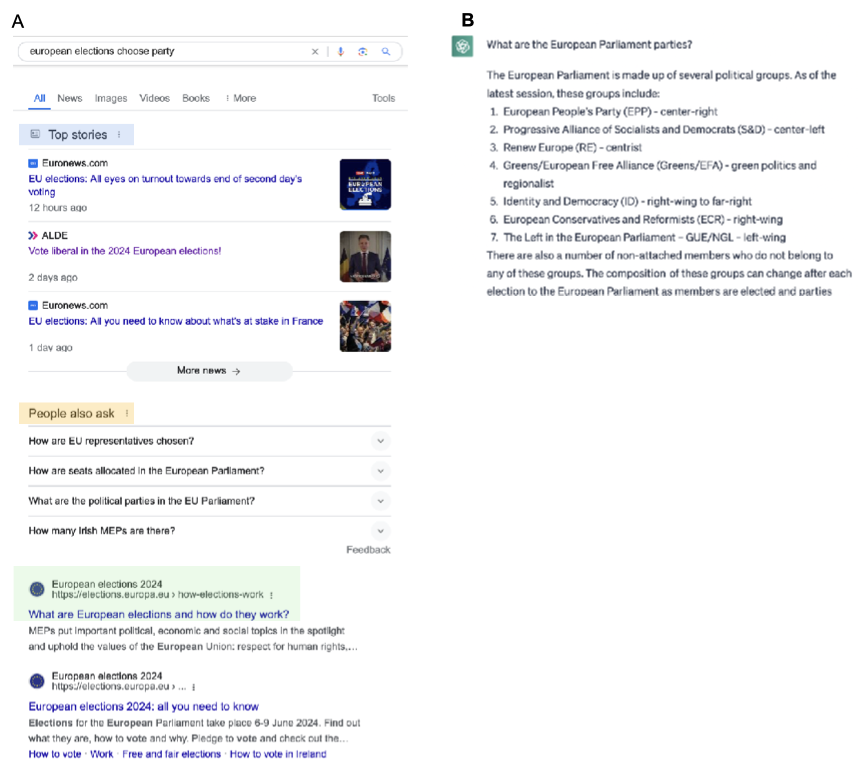}
    \caption{\textbf{Examples of Collected Results. A)} Screenshot of a SE results page showing all collected result types (Top Stories - blue; People Also Ask - yellow; Main Results - green). \textbf{B)} Screenshot illustrating typical responses generated by large language models (LLMs) for the queried prompts.}
    \label{fig:screenshots_examples}
\end{figure}

Figure~\ref{fig:screenshots_examples} illustrates the data collected from SEs (panel A) and LLMs (panel B). The figure also clarifies our definitions of a search engine result URL and its associated headline. Figures~\ref{fig:screenshot_1} and~\ref{fig:screenshot_2} exemplify some categories used to classify types of results.

\begin{figure*}[!ht]
    \centering
    \includegraphics[width=0.6\textwidth]{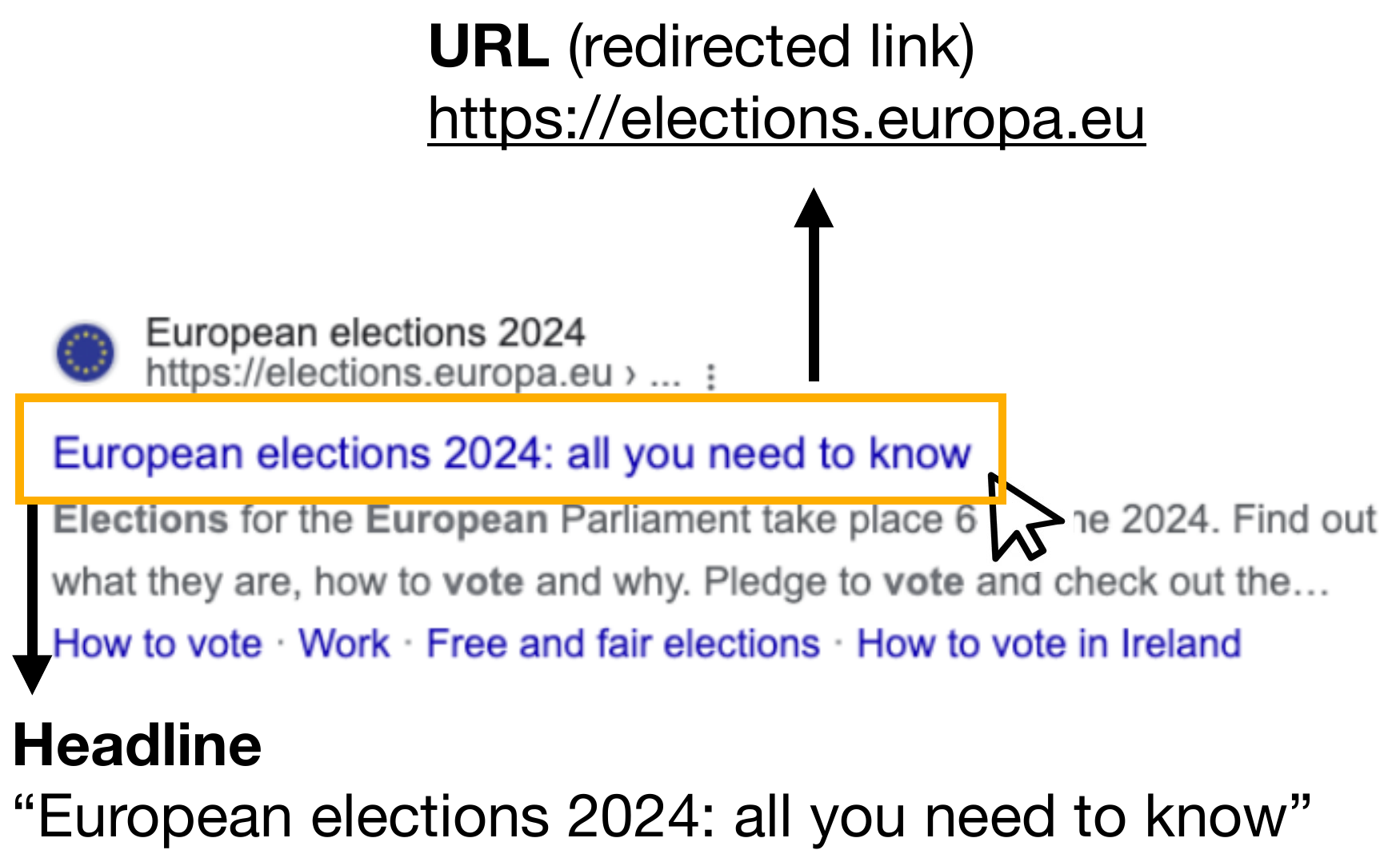}
    \caption{\textbf{URL and Headline definition.} Example of a result URL and respective headline.}
    \label{fig:headline_url}
\end{figure*}
\vspace{10cm}

All results from SEs -- top, ``people also ask'' and main results were collected and classified according to the type of website in question. Figure~\ref{fig:screenshot_1} and Figure~\ref{fig:screenshot_2} show examples of results pages collected by the bots. 

\begin{figure*}[!ht]
    \centering
    \includegraphics[width=0.6\textwidth]{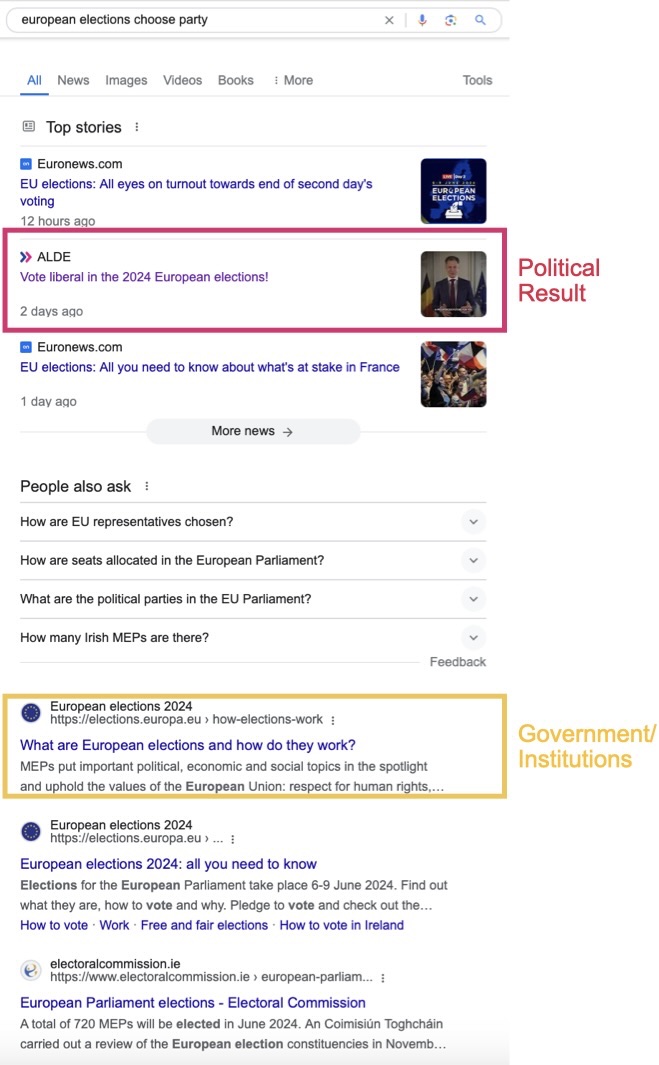}
    \caption{Screenshot of Google search results for the query ``european elections choose party'' collected by one of the bots. A political website (highlighted in pink) and a governmental website (highlighted in yellow) are marked, along with different versions of Google’s served results.}
    \label{fig:screenshot_1}
\end{figure*}

\begin{figure*}[!ht]
    \centering
    \includegraphics[width=0.6\textwidth]{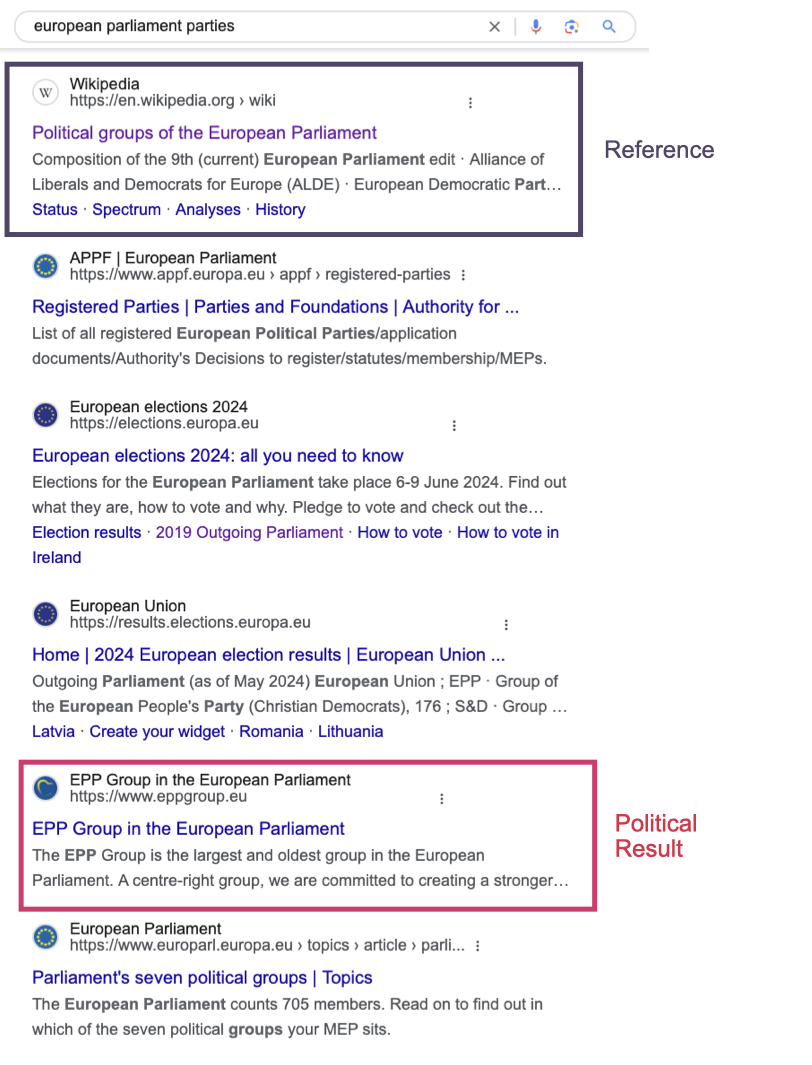}
    \caption{Screenshot of Google search results for the query ``european parliament parties'' collected by one of the bots. A reference website (highlighted in purple) and a political website are highlighted.}
    \label{fig:screenshot_2}
\end{figure*}

\clearpage
\subsection{Audit success rates (SEs and LLMs)}\label{app_subsec:successes_se}

\begin{itemize}
    \item \textbf{SE Audits}
\end{itemize}

\begin{figure*}[htbp]
    \centering
    \includegraphics[width=1\textwidth]{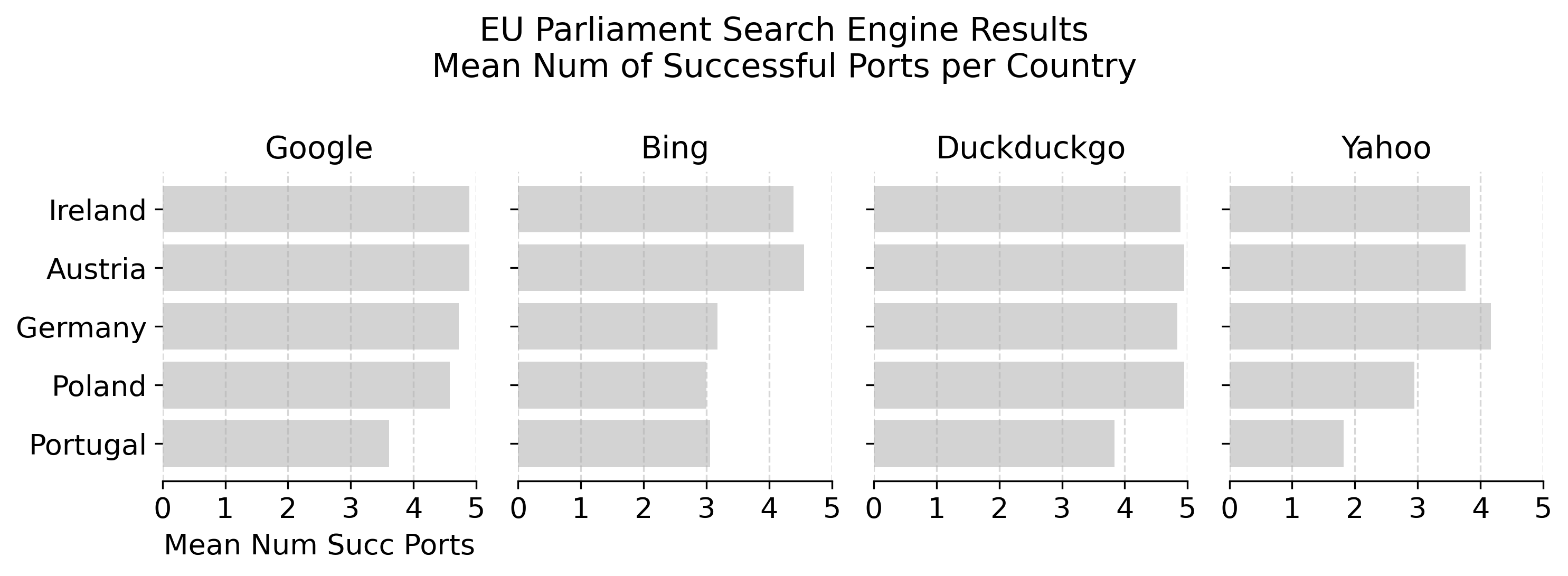}
    \caption{\textbf{EU Election - Mean number of Successful Ports per Country.} Average number of ports that successfully collected results per run of the European Parliament Union audit. Results are shown by country and search engine. Note that a total of 5 ports per country were deployed simultaneously.}
    \label{fig:eu_search_engines_average_ports}
\end{figure*}

\noindent{For the EU case study, a total of five independent bots were deployed per location. However, due to issues with the residential proxies or blocking by search engine platforms, not all bots successfully collected results in every run of the audit. Figure~\ref{fig:eu_search_engines_average_ports} displays the average number of bots (out of five) per location that successfully retrieved results.}

This number of results successfully collected resulted in the following number of first pages collected for analysis, as detailed in Table~\ref{sup_fig:eu_n_first_pages}. 

\begin{table*}[htbp]
\centering
\caption{\textbf{Results per SE for EU Case-Study:} Number of unique first pages of search results collected for each search engine on the EU Parliament Election.}
\begin{tabular}{p{3cm}p{7cm}}
\toprule
Search Engine & Number of SE page results per Location \\ 
\midrule
Google & 413\\ 
Bing & 328 \\ 
DuckDuckGo & 413 \\ 
Yahoo & 292 \\ 
\bottomrule
\end{tabular}
\label{sup_fig:eu_n_first_pages}
\end{table*}

\noindent{For the US case study, more replicas were performed per query, each one with a single exemplar per location. This change in approach was placed in order to avoid blockage from search engines, as we had less IPs available per location. Therefore, having multiple bots with the same IP address visiting the same search engine simultaneously, could cause the blockage of the IP. The total number of replicas, that is, the number of identical bots (same language) visited the same search engine and collected the results is shown in Table~\ref{sup_fig:us_n_replicas_bot}.}

\begin{table*}[ht!]
\centering
\caption{\textbf{Number of replicas per query across SEs:} Number of replicas (bots) which successfully collected results for the same query across the four approached SEs.}
\begin{tabular}{p{3cm}p{7cm}}
\toprule
Search Engine & Number query replicas per location \\ 
\midrule
Google & 3.07\\ 
Bing & 3.66 \\ 
DuckDuckGo & 3.66 \\ 
Yahoo & 3.10 \\
\bottomrule
\end{tabular}
\label{sup_fig:us_n_replicas_bot}
\end{table*}

This number of replicas resulted in the number of result pages detailed in Table~\ref{sup_fig:us_n_pages}, across the four search engines. Due to limited access to Oxylabs residential proxies, we did not use them to collect data from Yahoo. As a result, the total number of pages retrieved from Yahoo is lower compared to the other search engines.
\vspace{10cm}

\begin{table*}[ht!]
\centering
\caption{\textbf{Results per SE for US Case-Study:}Number of unique first pages of search results collected for each search engine about the US Presidential Election.}
\begin{tabular}{p{3cm}p{7cm}}
\toprule
Search Engine & N Replicas Query per Location \\ 
\midrule
Google & 973\\ 
Bing & 865 \\ 
DuckDuckGo & 1004 \\
Yahoo & 517 \\ 
\bottomrule
\end{tabular}
\label{sup_fig:us_n_pages}
\end{table*}

\clearpage
\begin{itemize}
    \item \textbf{LLMs response rates}
\end{itemize}

\begin{table*}[!ht]
\caption{Percentage of prompts answered and refused, for both LLMs (Copilot and ChatGPT) and both case-study elections.}
\label{tab:not_answer_llm}
\centering
\begin{tabular}{p{1.75cm} p{1.25cm} p{2.5cm} p{2.5cm} p{2.5cm} p{2.5cm}}

& & \multicolumn{2}{l}{\textbf{EU Parliament Election}}  & \multicolumn{2}{l}{\textbf{US Presidential Election}} \\ 
& & Answered & Refused & Answered & Refused \\ 
\midrule 
\multicolumn{2}{l}{\textbf{Copilot}}  & 80\% & 20\% &  100\%  & 0  \\ \hline
\multirow{3}{*}{\textbf{ChatGPT}} & Manual & 90\% & 10\% &   -  & -  \\ 
& & & &  & \\
& API & 100\%  & 0 &  100\%  & 0 \\
\end{tabular}
\end{table*}

\begin{table}[ht!]
\centering
\caption{Percentage of prompts answered and not answered (refused) by country for the EU Parliament Election case-study. For ChatGPT, we are considering only the manual responses, as the API ones were all answered, as shown in Table~\ref{tab:not_answer_llm}.}
\label{tab:llms_answered_refused_per_country}
\begin{tabular}{p{3cm} p{2.5cm} p{2.5cm} p{2.5cm} p{2.5cm}}
\toprule
& \multicolumn{2}{l}{\textbf{ChatGPT (manual)}} & \multicolumn{2}{l}{\textbf{Copilot}} \\
\textbf{Country} & Answered & Refused & Answered & Refused \\
\midrule
Austria   & 50\%  & 50\%  & 75\%  & 25\% \\
Germany   & 50\%  & 50\%  & 75\%  & 25\% \\
Ireland   & 83\%   & 17\% & 50\%  & 50\% \\
Poland    & 50\%  & 50\%  & 75\%  & 25\% \\
Portugal  & 50\%  & 50\%  & 75\%  & 25\% \\
\bottomrule
\end{tabular}
\end{table}

\clearpage
\section{Political leaning classification frameworks}

\subsection{Mapping of EU political groups and leaning}\label{subsec:eu_mapping}

\newcolumntype{L}[1]{>{\raggedright\arraybackslash}p{#1}}

\small
\begin{longtable}{|L{1.5cm}|p{1.5cm}|p{1.5cm}|p{4.5cm}|p{3.5cm}|p{2cm}|}
\caption{Classification of European Parliament Member Parties by CHES Political Family (2024 Wave). Note: Party family IDs and classifications are based on the 2024 Chapel Hill Expert Survey (CHES) dataset. Parties marked with (–) were not included in the 2024 expert survey cycle.}
\label{tab:eu_parties_all} \\

\hline
\textbf{EU Group} & \textbf{Country} & \textbf{Acronym} & \textbf{Full Party Name} & \textbf{Family Name} & \textbf{Family} \\
\hline \hline
\endfirsthead

\hline
\textbf{EU Group} & \textbf{Country} & \textbf{Acronym} & \textbf{Full Party Name} & \textbf{Family Name} & \textbf{Family}  \\
\hline \hline
\endhead
\hline

\hline
\multicolumn{3}{r}{\textit{Continued on next page}} \\
\endfoot
\endlastfoot
        \multirow{27}{*}{EPP} & Austria & ÖVP & Österreichische Volkspartei & Christian-Democratic & Main. Right \\ \cline{2-6}
        & Belgium & CD\&V & Christen-Democratisch en Vlaams & Christian-Democratic & Main. Right \\ \cline{2-6}
        & Belgium & Les Engagés & Les Engagés (formerly cdH) & Christian-Democratic & Main. Right \\ \cline{2-6}
        & Belgium & CSP & Christlich Soziale Partei & Christian-Democratic & Main. Right \\ \cline{2-6}
        & Bulgaria & GERB & Grazhdani za Evropeysko Razvitie na Balgariya & Conservatives & Main. Right \\ \cline{2-6}
        & Bulgaria & SDS & Sayuz na Demokratichnite Sili & Conservatives & Main. Right \\ \cline{2-6}
        & Bulgaria & DSB & Demokrati za Silna Balgariya & Conservatives & Main. Right \\ \cline{2-6}
        & Croatia & HDZ & Hrvatska Demokratska Zajednica & Christian-Democratic & Main. Right \\ \cline{2-6}
        & Cyprus & DISY & Dimokratikós Sinagermós & Conservatives & Main. Right \\ \cline{2-6}
        & Czechia & KDU-ČSL & Křesťanská a demokratická unie & Christian-Democratic & Main. Right \\ \cline{2-6}
        & Czechia & TOP 09 & Tradice Odpovědnost Prosperita 09 & Conservatives & Main. Right \\ \cline{2-6}
        & Czechia & STAN & Starostové a nezávislí & Agrarian/Centre & - \\ \cline{2-6}
        & Denmark & KF & Det Konservative Folkeparti & Conservatives & Main. Right \\ \cline{2-6}
        & Denmark & LA & Liberal Alliance & Liberal & Main. Right \\ \cline{2-6}
        & Estonia & Isamaa & Isamaa & Conservatives & Main. Right \\ \cline{2-6}
        & Finland & KOK & Kansallinen Kokoomus & Conservatives & Main. Right \\ \cline{2-6}
        & France & LR & Les Républicains & Conservatives & Main. Right \\ \cline{2-6}
        & Germany & CDU & Christlich Demokratische Union & Christian-Democratic & Main. Right \\ \cline{2-6}
        & Germany & CSU & Christlich-Soziale Union in Bayern & Christian-Democratic & Main. Right \\ \cline{2-6}
        & Germany & Familen & Familien-Partei Deutschlands & - & - \\ \cline{2-6}
        & Greece & ND & Nea Dimokratia & Conservatives & Main. Right \\ \cline{2-6}
        & Hungary & TISZA & Tisztelet és Szabadság Párt & - (New in 2024) & - \\ \cline{2-6}
        & Hungary & KDNP & Kereszténydemokrata Néppárt & Christian-Democratic & Main. Right \\ \cline{2-6}
        & Ireland & FG & Fine Gael & Christian-Democratic & Main. Right \\ \cline{2-6}
        & Italy & FI & Forza Italia & Conservatives & Main. Right \\ \cline{2-6}
        & Italy & SVP & Südtiroler Volkspartei & Regionalist & Rad. Right \\ \cline{2-6}
        & Latvia & JV & Jaunā Vienotība & Conservatives & Main. Right \\ \hline
        \multirow{18}{*}{EPP} & Lithuania & TS-LKD & Tėvynės sąjunga – Lietuvos krikščionys demokratai & Christian-Democratic & Main. Right \\ \cline{2-6}
        & Luxembourg & CSV & Chrëschtlech Sozial Vollekspartei & (Luxembourg excluded from CHES 2024) & - \\ \cline{2-6}
        & Malta & PN & Partit Nazzjonalista & Christian-Democratic & Main. Right \\ \cline{2-6}
        & Netherlands & CDA & Christen-Democratisch Appèl & Christian-Democratic & Main. Right \\ \cline{2-6}
        & Netherlands & BBB & BoerBurgerBeweging & Agrarian/Centre & Main. Right \\ \cline{2-6}
        & Netherlands & NSC & Nieuw Sociaal Contract & - (New in 2024) & - \\ \cline{2-6}
        & Poland & PO & Platforma Obywatelska & Christian-Democratic & Main. Right \\ \cline{2-6}
        & Poland & PSL & Polskie Stronnictwo Ludowe & Agrarian/Centre & Main. Right \\ \cline{2-6}
        & Portugal & PSD & Partido Social Democrata & Conservatives & Main. Right \\ \cline{2-6}
        & Portugal & CDS-PP & Centro Democrático e Social – Partido Popular & Christian-Democratic & Main. Right \\ \cline{2-6}
        & Romania & PNL & Partidul Național Liberal & Conservatives & Main. Right \\ \cline{2-6}
        & Romania & RMDSZ & Uniunea Democrată Maghiară din România & Regionalist & Rad. Right \\ \cline{2-6}
        & Slovakia & KDH & Kresťanskodemokratické hnutie & Christian-Democratic & Main. Right \\ \cline{2-6}
        & Slovenia & SDS & Slovenska Demokratska Stranka & Conservatives & Main. Right \\ \cline{2-6}
        & Slovenia & NSi & Nova Slovenija – Krščanski demokrati & Christian-Democratic & Main. Right \\ \cline{2-6}
        & Spain & PP & Partido Popular & Conservatives & Main. Right \\ \cline{2-6}
        & Sweden & M & Moderaterna & Conservatives & Main. Right \\ \cline{2-6}
        & Sweden & KD & Kristdemokraterna & Christian-Democratic & Main. Right \\ \hline
        \multirow{13}{*}{S\&D} & Austria & SPÖ & Sozialdemokratische Partei Österreichs & Socialist & Main. Left \\ \cline{2-6}
        & Belgium & PS & Parti Socialiste & Socialist & Main. Left \\ \cline{2-6}
        & Belgium & Vooruit & Vooruit & Socialist & Main. Left \\ \cline{2-6}
        & Bulgaria & BSP & Bulgarska Sotsialisticheska Partiya & Socialist & Main. Left \\ \cline{2-6}
        & Croatia & SDP & Socijaldemokratska Partija Hrvatske & Socialist & Main. Left \\ \cline{2-6}
        & Cyprus & EDEK & Ethnikí Dimokratikí Énosi Kéntrou & Socialist & Main. Left \\ \cline{2-6}
        & Czechia & SOCDEM & Sociální demokracie & Socialist & Main. Left \\ \cline{2-6}
        & Denmark & S & Socialdemokratiet & Socialist & Main. Left \\ \cline{2-6}
        & Estonia & SDE & Sotsiaaldemokraatlik Erakond & Socialist & Main. Left \\ \cline{2-6}
        & Finland & SDP & Suomen Sosiaalidemokraattinen Puolue & Socialist & Main. Left \\ \cline{2-6}
        & France & PS & Parti Socialiste & Socialist & Main. Left \\ \cline{2-6}
        & France & PP & Place Publique & - & Main. Left \\ \cline{2-6}
        & Germany & SPD & Sozialdemokratische Partei Deutschlands & Socialist & Main. Left \\ \cline{2-6}
        \multirow{16}{*}{S\&D}& Greece & PASOK & Panellinio Sosialistiko Kinima & Socialist & Main. Left \\ \cline{2-6}
        & Hungary & DK & Demokratikus Koalíció & Liberal & Main. Left \\ \cline{2-6}
        & Hungary & MSZP & Magyar Szocialista Párt & Socialist & Main. Left \\ \cline{2-6}
        & Ireland & Lab & Labour Party & Socialist & Main. Left \\ \cline{2-6}
        & Italy & PD & Partito Democratico & Socialist & Main. Left \\ \cline{2-6}
        & Latvia & SDPS & ""Saskaņa"" Sociāldemokrātiskā Partija & Socialist & Main. Left \\ \cline{2-6}
        & Lithuania & LSDP & Lietuvos Socialdemokratų Partija & Socialist & Main. Left \\ \cline{2-6}
        & Luxembourg & LSAP & Lëtzebuerger Sozialistesch Aarbechterpartei & (Luxembourg excluded from CHES) & Main. Left \\ \cline{2-6}
        & Malta & PL & Partit Laburista & Socialist & Main. Left \\ \cline{2-6}
        & Netherlands & PvdA & Partij van de Arbeid & Socialist & Main. Left \\ \cline{2-6}
        & Poland & NL & Nowa Lewica & Socialist & Main. Left \\ \cline{2-6}
        & Portugal & PS & Partido Socialista & Socialist & Main. Left \\ \cline{2-6}
        & Romania & PSD & Partidul Social Democrat & Socialist & Main. Left \\ \cline{2-6}
        & Slovenia & SD & Socialni Demokrati & Socialist & Main. Left \\ \cline{2-6}
        & Spain & PSOE & Partido Socialista Obrero Español & Socialist & Main. Left \\ \cline{2-6}
        & Sweden & S & Sveriges Socialdemokratiska Arbetareparti & Socialist & Main. Left \\ \hline
        \multirow{15}{*}{PfE (ID)} & Austria & FPÖ & Freiheitliche Partei Österreichs & Radical Right & Rad. Right \\ \cline{2-6}
        & Belgium & VB & Vlaams Belang & Radical Right & Rad. Right \\ \cline{2-6}
        & Czechia & ANO & ANO 2011 & Liberal* & Main. Right \\ \cline{2-6}
        & Czechia & Přísaha & Přísaha a Motoristé & - (New in 2024) & - \\ \cline{2-6}
        & Denmark & DF & Dansk Folkeparti & Radical Right & Rad. Right \\ \cline{2-6}
        & France & RN & Rassemblement National & Radical Right & Rad. Right \\ \cline{2-6}
        & Greece & FL & Foní Logikís (Voice of Reason) & Radical Right & Rad. Right \\ \cline{2-6}
        & Hungary & Fidesz & Fidesz – Magyar Polgári Szövetség & Radical Right & Rad. Right \\ \cline{2-6}
        & Hungary & KDNP & Kereszténydemokrata Néppárt & Christian-Democratic & Main. Right \\ \cline{2-6}
        & Italy & Lega & Lega per Salvini Premier & Radical Right & Rad. Right \\ \cline{2-6}
        & Latvia & LPV & Latvija pirmajā vietā (Latvia First) & Confessional & Rad. Right \\ \cline{2-6}
        & Netherlands & PVV & Partij voor de Vrijheid & Radical Right & Rad. Right \\ \cline{2-6}
        & Poland & RN & Ruch Narodowy (National Movement) & Radical Right & Rad. Right \\ \cline{2-6}
        & Portugal & CH & Chega & Radical Right & Rad. Right \\ \cline{2-6}
        & Spain & VOX & Vox & Radical Right & Rad. Right \\ \cline{2-6}
        ECR & Belgium & N-VA & Nieuw-Vlaamse Alliantie & Regionalist & Rad. Right \\ \hline
        \multirow{22}{*}{ECR}& Bulgaria & ITN & Ima Takav Narod (There is Such a People) & Radical Right & Rad. Right \\ \cline{2-6}
        & Croatia & DP & Domovinski pokret (Homeland Movement) & Radical Right & Rad. Right \\ \cline{2-6}
        & Cyprus & ELAM & Ethniko Laiko Metopo (National Popular Front) & Radical Right & Rad. Right \\ \cline{2-6}
        & Czechia & ODS & Občanská demokratická strana & Conservatives & Main. Right \\ \cline{2-6}
        & Denmark & DD & Danmarksdemokraterne (Denmark Democrats) & Radical Right & Rad. Right \\ \cline{2-6}
        & Estonia & ERK & Eesti Keskerakond (Center Party) & Liberal & Main. Right \\ \cline{2-6}
        & Finland & PS & Perussuomalaiset (Finns Party) & Radical Right & Rad. Right \\ \cline{2-6}
        & France & IDL & Identité-Libertés & - & - \\ \cline{2-6}
        & Greece & EL & Elliniki Lysi (Greek Solution) & Radical Right & Rad. Right \\ \cline{2-6}
        & Italy & FdI & Fratelli d'Italia (Brothers of Italy) & Radical Right & Rad. Right \\ \cline{2-6}
        & Latvia & NA & Nacionālā apvienība (National Alliance) & Radical Right & Rad. Right \\ \cline{2-6}
        & Latvia & AS & Apvienotais Saraksts (United List) & Agrarian/Centre & - \\ \cline{2-6}
        & Lithuania & LLRA-KŠS & Electoral Action of Poles in Lithuania & Confessional & Rad. Right \\ \cline{2-6}
        & Lithuania & LVŽS & Lithuanian Farmers and Greens Union & Agrarian/Centre & - \\ \cline{2-6}
        & Luxembourg & ADR & Alternativ Demokratesch Reformpartei & (Luxembourg excluded) & - \\ \cline{2-6}
        & Netherlands & SGP & Staatkundig Gereformeerde Partij & Confessional & Rad. Right \\ \cline{2-6}
        & Poland & PiS & Prawo i Sprawiedliwość & Conservatives & Rad. Right \\ \cline{2-6}
        & Poland & Suwerenna & Suwerenna Polska (Sovereign Poland) & Radical Right & Rad. Right \\ \cline{2-6}
        & Romania & AUR & Alianța pentru Unirea Românilor & Radical Right & Rad. Right \\ \cline{2-6}
        & Romania & PNCR & Partidul Național Conservator Român & - & - \\ \cline{2-6}
        & Spain & SALF & Se Acabó La Fiesta & - (New in 2024) & - \\ \cline{2-6}
        & Sweden & SD & Sverigedemokraterna (Sweden Democrats) & Radical Right & Rad. Right \\ \hline
        \multirow{3}{*}{Renew} & Austria & NEOS & Das Neue Österreich und Liberales Forum & Liberal & Main. Right \\ \cline{2-6}
        & Belgium & MR & Mouvement Réformateur & Liberal & Main. Right \\ \cline{2-6}
        & Belgium & Open Vld & Open Vlaamse Liberalen en Democraten & Liberal & Main. Right \\ \hline
        \multirow{32}{*}{Renew} & Belgium & Les Engagés & Les Engagés (transferred from EPP/EDP) & Christian-Democratic & Main. Right \\ \cline{2-6}
        & Bulgaria & DPS & Dvizhenie za prava i svobodi & Liberal & Main. Right \\ \cline{2-6}
        & Bulgaria & PP & Prodalzhavame promyanata & Liberal & Main. Right \\ \cline{2-6}
        & Denmark & V & Venstre, Danmarks Liberale Parti & Liberal & Main. Right \\ \cline{2-6}
        & Denmark & B & Radikale Venstre & Liberal & Main. Right \\ \cline{2-6}
        & Denmark & M & Moderaterne & Liberal & Main. Right \\ \cline{2-6}
        & Estonia & RE & Eesti Reformierakond & Liberal & Main. Right \\ \cline{2-6}
        & Estonia & K & Eesti Keskerakond & Liberal & Main. Right \\ \cline{2-6}
        & Finland & Kesk & Suomen Keskusta (Centre Party) & Agrarian/Centre & - \\ \cline{2-6}
        & Finland & SFP/RKP & Svenska folkpartiet i Finland & Liberal & Main. Right \\ \cline{2-6}
        & France & RE & Renaissance & Liberal & Main. Right \\ \cline{2-6}
        & France & MoDem & Mouvement démocrate & Liberal & Main. Right \\ \cline{2-6}
        & France & Horizons & Horizons & Liberal & Main. Right \\ \cline{2-6}
        & France & UDI & Union des démocrates et indépendants & Liberal & Main. Right \\ \cline{2-6}
        & Germany & FDP & Freie Demokratische Partei & Liberal & Main. Right \\ \cline{2-6}
        & Germany & FW & Freie Wähler & Agrarian/Centre & - \\ \cline{2-6}
        & Ireland & FF & Fianna Fáil & Agrarian/Centre & - \\ \cline{2-6}
        & Ireland & II & Independent Ireland & - (New in 2024) & - \\ \cline{2-6}
        & Latvia & LA & Latvijas attīstībai & Liberal & Main. Right \\ \cline{2-6}
        & Lithuania & LS & Liberalų sąjūdis & Liberal & Main. Right \\ \cline{2-6}
        & Lithuania & LP & Laisvės partija & Liberal & Main. Right \\ \cline{2-6}
        & Luxembourg & DP & Demokratesch Partei & (Luxembourg excluded) & - \\ \cline{2-6}
        & Netherlands & VVD & Volkspartij voor Vrijheid en Democratie & Liberal & Main. Right \\ \cline{2-6}
        & Netherlands & D66 & Democraten 66 & Liberal & Main. Right \\ \cline{2-6}
        & Poland & P2050 & Polska 2050 (Third Way) & Liberal & Main. Right \\ \cline{2-6}
        & Portugal & IL & Iniciativa Liberal & Liberal & Main. Right \\ \cline{2-6}
        & Romania & USR & Uniunea Salvați România & Liberal & Main. Right \\ \cline{2-6}
        & Slovakia & PS & Progresívne Slovensko & Liberal & Main. Right \\ \cline{2-6}
        & Slovenia & GS & Gibanje Svoboda & Liberal & Main. Right \\ \cline{2-6}
        & Spain & PNV/EAJ & Partido Nacionalista Vasco & Regionalist & - \\ \cline{2-6}
        & Sweden & C & Centerpartiet & Agrarian/Centre & - \\ \cline{2-6}
        & Sweden & L & Liberalerna & Liberal & Main. Right \\ \hline
        \multirow{5}{*}{G/EFA} & Austria & Die Grünen & Die Grünen – Die Grüne Alternative & Green & Greens \\ \cline{2-6}
        & Belgium & Ecolo & Écolo & Green & Greens \\ \cline{2-6}
        & Belgium & Groen & Groen & Green & Greens \\ \cline{2-6}
        & Croatia & Možemo! & Možemo! – politička platforma & Green & Greens \\ \cline{2-6}
        & Czechia & Piráti & Česká pirátská strana (Pirates) & Liberal* & Main. Right \\ \cline{2-6}
        \multirow{20}{*}{G/EFA} & Denmark & SF & Socialistisk Folkeparti (Green Left) & Green & Greens \\ \cline{2-6}
        & Finland & Vihr & Vihreä liitto (Green League) & Green & Greens \\ \cline{2-6}
        & France & LE & Les Écologistes (formerly EELV) & Green & Greens \\ \cline{2-6}
        & Germany & Grüne & Bündnis 90/Die Grünen & Green & Greens \\ \cline{2-6}
        & Germany & Volt & Volt Deutschland & - & - \\ \cline{2-6}
        & Germany & ÖDP & Ökologisch-Demokratische Partei & Green & Greens \\ \cline{2-6}
        & Greece & Kosmos & Kosmos & - (New in 2024) & - \\ \cline{2-6}
        & Italy & EV & Europa Verde & Green & Greens \\ \cline{2-6}
        & Latvia & PRO & Progresīvie & Green & Greens \\ \cline{2-6}
        & Lithuania & DSVL & Union of Democrats ""For Lithuania"" & Agrarian/Centre & - \\ \cline{2-6}
        & Luxembourg & DG & Déi Gréng & (Luxembourg excluded) & - \\ \cline{2-6}
        & Netherlands & GL & GroenLinks & Green & Greens \\ \cline{2-6}
        & Netherlands & Volt & Volt Nederland & - & - \\ \cline{2-6}
        & Romania & Ștefănuță & Independent (Nicu Ștefănuță) & - & - \\ \cline{2-6}
        & Slovenia & Vesna & Vesna – zelena stranka & Green & Greens \\ \cline{2-6}
        & Spain & Sumar & Movimiento Sumar & Radical Left & Rad. Left \\ \cline{2-6}
        & Spain & ERC & Esquerra Republicana de Catalunya & Regionalist & - \\ \cline{2-6}
        & Spain & BNG & Bloque Nacionalista Galego & Regionalist & - \\ \cline{2-6}
        & Spain & Compromís & Coalició Compromís & Regionalist & - \\ \cline{2-6}
        & Sweden & MP & Miljöpartiet de gröna & Green & Greens \\ \hline
        \multirow{13}{*}{The Left} & Austria & KPÖ & Kommunistische Partei Österreichs & Radical Left & Rad. Left \\ \cline{2-6}
        & Belgium & PTB-PVDA & Parti du Travail de Belgique & Radical Left & Rad. Left \\ \cline{2-6}
        & Cyprus & AKEL & Progressivós Laós tis Kýprou & Radical Left & Rad. Left \\ \cline{2-6}
        & Denmark & EL & Enhedslisten (Red-Green Alliance) & Radical Left & Rad. Left \\ \cline{2-6}
        & Finland & Vas & Vasemmistoliitto (Left Alliance) & Radical Left & Rad. Left \\ \cline{2-6}
        & France & LFI & La France Insoumise & Radical Left & Rad. Left \\ \cline{2-6}
        & France & PCF & Parti Communiste Français & Radical Left & Rad. Left \\ \cline{2-6}
        & Germany & Linke & Die Linke & Radical Left & Rad. Left \\ \cline{2-6}
        & Germany & Tierschutz. & Tierschutzpartei (Animal Welfare) & Green & Greens \\ \cline{2-6}
        & Greece & SYRIZA & Sinaspismós Rizospastikís Aristerás & Radical Left & Rad. Left \\ \cline{2-6}
        & Greece & PE & Plefsi Eleftherias (Course of Freedom) & Radical Left & Rad. Left \\ \cline{2-6}
        & Ireland & SF & Sinn Féin & Radical Left & Rad. Left \\ \cline{2-6}
        & Ireland & I4C & Independents 4 Change & - & - \\ \cline{2-6}
        \multirow{9}{*}{The Left}& Ireland & LP & Luke 'Ming' Flanagan (Independent) & - & - \\ \cline{2-6}
        & Italy & M5S & Movimento 5 Stelle (5 Star Movement) & No Family / Special & - \\ \cline{2-6}
        & Netherlands & PvdD & Partij voor de Dennen (Animal Rights) & Green & Greens \\ \cline{2-6}
        & Portugal & BE & Bloco de Esquerda & Radical Left & Rad. Left \\ \cline{2-6}
        & Portugal & PCP & Partido Comunista Português & Radical Left & Rad. Left \\ \cline{2-6}
        & Spain & Podemos & Podemos & Radical Left & Rad. Left \\ \cline{2-6}
        & Spain & EH Bildu & Euskal Herria Bildu & Regionalist & Rad. Left \\ \cline{2-6}
        & Spain & Sumar & Movimiento Sumar & Radical Left & Rad. Left \\ \cline{2-6}
        & Sweden & V & Vänsterpartiet & Radical Left & Rad. Left \\ \hline
        \multirow{8}{*}{ESN (ID)} & Bulgaria & Revival & Vazrazhdane & Radical Right & Rad. Right \\ \cline{2-6}
        & Czechia & SPD & Svoboda a přímá demokracie & Radical Right & Rad. Right \\ \cline{2-6}
        & France & R! & Reconquête! & Radical Right & Rad. Right \\ \cline{2-6}
        & Germany & AfD & Alternative für Deutschland & Radical Right & Rad. Right \\ \cline{2-6}
        & Hungary & MHM & Mi Hazánk Mozgalom (Our Homeland) & Radical Right & Rad. Right \\ \cline{2-6}
        & Lithuania & TTS & Tautos ir teisingumo sąjunga & Radical Right & Rad. Right \\ \cline{2-6}
        & Poland & NN & Nowa Nadzieja (New Hope) & Radical Right & Rad. Right \\ \cline{2-6}
        & Slovakia & Republika & Hnutie Republika & Radical Right & Rad. Right \\ \hline
        \multirow{15}{*}{NI} & Bulgaria & DPS-NN & Movement for Rights and Freedoms & Liberal & Main. Right \\ \cline{2-6}
        & Cyprus & IND & Fidias Panayiotou (Independent) & - & - \\\cline{2-6}
        & Czechia & Stačilo! & KSČM (Communist Party) & Radical Left & Rad. Left \\ \cline{2-6}
        & Czechia & Stačilo! & SD-SN (United Democrats) & - & - \\ \cline{2-6}
        & France & - & Malika Sorel (Independent) & - & - \\ \cline{2-6}
        & Germany & BSW & Bündnis Sahra Wagenknecht & Radical Left* & Rad. Left \\ \cline{2-6}
        & Germany & Die PARTEI & Die PARTEI & No Family / Special & - \\ \cline{2-6}
        & Germany & PdF & Partei des Fortschritts & - & - \\ \cline{2-6}
        & Greece & KKE & Kommounistikó Kómma Elládas & Radical Left & Rad. Left \\ \cline{2-6}
        & Greece & NIKI & Dimokratikó Patriotikó Kínima & Confessional & Rad. Right \\ \cline{2-6}
        & Italy & FN & Futuro Nazionale & - & - \\ \cline{2-6}
        & Netherlands & FvD & Forum voor Democratie & Radical Right & Rad. Right \\ \cline{2-6}
        & Poland & Konfederacja & Konfederacja Korony Polskiej & Radical Right & Rad. Right \\ \cline{2-6}
        & Romania & SOS RO & S.O.S. România & Radical Right & Rad. Right \\ \cline{2-6}
        & Slovakia & SMER-SD & Smer – sociálna demokracia & Socialist* & Main. Left \\ \hline
        \multirow{4}{*}{NI}& Slovakia & HLAS-SD & Hlas – sociálna demokracia & Socialist & Main. Left \\ \cline{2-6}
        & Slovakia & Republika & Hnutie Republika & Radical Right & Rad. Right \\ \cline{2-6}
        & Spain & SALF & Se Acabó La Fiesta & - & - \\ \cline{2-6}
        & Spain & Junts & Junts per Catalunya & Regionalist & - \\ \hline
\end{longtable}

\clearpage
\subsection{Entities and their associated leaning across all collected results}\label{app_sec:entities_n_results}

\begin{table}[h!]
\centering
\renewcommand{\arraystretch}{1.3}
\begin{tabular}{l l l p{8cm}}
\toprule
 & Leaning & EU family & Parties of entities extracted \\ 
\midrule
\multirow{10}{*}{\makecell[l]{SE\\Results}} 
& \multirow{2}{*}{Rad. Right} & ID & Identity and Democracy, Alternative for Germany - Germany, National Rally - France, Party for Freedom - Netherlands, Enough - Portugal, National Movement - Poland, Fidesz -- Hungarian Civic Union - Hungary, ID, patriots \\
&  & ECR & Brothers of Italy - Italy, Law and Justice - Poland, ECR \\
\cline{2-4}
&  \multirow{2}{*}{Main. Right} & EPP & Polish People's Party - Poland, Civic Platform - Poland, Democratic Alliance - Portugal, epp, Fine Gael (Family of the Irish) - Ireland, Nationalist Party - Malta, Christian Democratic Union - Germany \\
& & Renew & Poland 2050 - Poland, Renew, The Republic Onwards! | Renaissance - France, ALDE Party \\
\cline{2-4}
&  Greens & Greens & The Greens - Netherlands, G/EFA, Alliance 90 / Greens - Germany, Green Party - Ireland \\
\cline{2-4}
&  Main. Left & S\&D & Socialist Party - Portugal, S\&D, Social Democratic Party of Germany - Germany \\
\cline{2-4}
&  Rad. Left & The Left & Sinn Fein - Ireland, left\\
\midrule
\multirow{6}{*}{\makecell[l]{LLM\\Results}} & Rad. Right & ID/ECR & \makecell[l]{ID, ECR, AFD, PVV, RN, FDI, Fidesz} \\
\cline{2-4}
& Main. Right & EPP/Renew & \makecell[l]{EPP, Renew, AD, CDU/CSU, FDP} \\
\cline{2-4}
& Greens & G/EFA & \makecell[l]{Greens/EFA, Grünen, GL/PvdA, Zieloni, Volt} \\
\cline{2-4}
& Main. Left & S\&D & \makecell[l]{S\&D, SPD, SAP} \\
\cline{2-4}
& Rad. Left & The Left & \makecell[l]{The Left, Die Linke} \\
\cline{2-4}
&  - & - & \makecell[l]{NI, SD} \\
\bottomrule
\end{tabular}
\caption{SE results for EU Political Parties grouped by political family and EU family}
\label{tab:se_eu_parties}
\end{table}

\begin{table}[h!]
\centering
\renewcommand{\arraystretch}{1.3} % increase row height
\begin{tabular}{l l l p{7cm}}
\toprule
 & Analysis & Leaning & Entities extracted \\ 
\midrule
\multirow{2}{*}{\makecell[l]{SE\\Results}} 
 & \multirow{2}{*}{\makecell[l]{Political\\Entities}} & Republican Party - USA 
   & \makecell[l]{Donald Trump, JD Vance,\\ Virginia AG Miyares, Hung Cao, \\ Zach Nunn, Elon Musk} \\ 
   \cline{3-4}
 &  & Democratic Party - USA 
   & \makecell[l]{Kamala Harris, David Axelrod, \\ George Gascón, Tim Walz, \\ Mannion, Lanon Baccam, \\Rep. Jasmine Crockett} \\ 
\bottomrule
\end{tabular}
\caption{SE results for US Presidential Election by party}
\label{tab:se_us_presidential}
\end{table}

\clearpage
\section{Statistical Analysis}\label{app_sec:statistic}

\subsection{Testing deviations from a uniform distribution}\label{app_subsec:uniform}

\begin{table}[ht]
\centering
\small
\caption{\textbf{EU Case, Aggregated Test against Uniform distribution.} Deviation from uniform representation in EU search engine results and LLM responses ($K=5$ categories, uniform baseline = 20\% each). Each cell shows the difference in percentage points from a uniform distribution, testing whether the algorithm treats all political leanings equally. Statistical significance was assessed using a Binomial Z-test, where N= number of articles with political entities per search engine. The first four columns are search engines; the last two LLMs. Bold values significant after Holm--Bonferroni correction. \textsuperscript{*} $p<0.05$, \textsuperscript{**} $p<0.01$.}
\label{tab:agg_uni_eu}
\begin{tabular}{llrrrrrr}
\toprule
 & & \multicolumn{4}{c}{\textit{Search Engines}} & \multicolumn{2}{c}{\textit{LLMs}} \\
\cmidrule(lr){3-6}\cmidrule(lr){7-8}
Benchmark & Category & Google & Bing & \makecell[l]{Duck\\DuckGo} & Yahoo & ChatGPT & Copilot \\
\midrule
\multirow{5}{*}{\textbf{Uniform}} & R. Right & \textbf{+27.8pp\textsuperscript{**}} & \textbf{+29.3pp\textsuperscript{**}} & \textbf{+36.7pp\textsuperscript{**}} & \textbf{+42.7pp\textsuperscript{**}} & \textbf{+13.3pp\textsuperscript{**}} & +3.1pp \\
 & M. Right & \textbf{+7.2pp\textsuperscript{**}} & \textbf{+10.4pp\textsuperscript{**}} & \textbf{+6.4pp\textsuperscript{**}} & \textbf{+3.8pp\textsuperscript{*}} & \textbf{+9.0pp\textsuperscript{*}} & +7.7pp \\
 & Greens & \textbf{-13.9pp\textsuperscript{**}} & \textbf{-19.3pp\textsuperscript{**}} & \textbf{-16.3pp\textsuperscript{**}} & \textbf{-13.4pp\textsuperscript{**}} & -0.9pp & +6.2pp \\
 & M. Left & \textbf{-12.8pp\textsuperscript{**}} & -1.6pp & \textbf{-9.5pp\textsuperscript{**}} & \textbf{-13.1pp\textsuperscript{**}} & -5.8pp & +0.0pp \\
 & R. Left & \textbf{-8.3pp\textsuperscript{**}} & \textbf{-18.9pp\textsuperscript{**}} & \textbf{-17.2pp\textsuperscript{**}} & \textbf{-20.0pp\textsuperscript{**}} & \textbf{-15.7pp\textsuperscript{**}} & \textbf{-16.9pp\textsuperscript{**}} \\
\bottomrule
\end{tabular}
\end{table}
\begin{table}[ht]
\centering
\small
\caption{\textbf{EU Case, Query-Level Test against Uniform Distribution.} Deviation from uniform 
representation in EU search engine results and LLM responses ($K=5$ categories, uniform 
baseline = 20\% each). Each cell shows the difference in percentage points from a uniform 
distribution, testing whether the algorithm treats all political leanings equally across 
queries. Within each country, statistical significance was assessed using the Beta-Binomial 
likelihood ratio test where $N \geq 30$ queries were available, and a permutation test 
otherwise, where $N$ denotes the number of queries returning at least one result with a 
politically identifiable entity. Country-level z-scores were then pooled into a single 
EU-level estimate using Stouffer's method, weighting each country by $\sqrt{N}$. The first 
four columns are search engines; the last two LLMs. Bold values significant after 
Holm--Bonferroni correction. \textsuperscript{*} $p<0.05$, \textsuperscript{**} $p<0.01$.}
\label{tab:query_uni_eu}
\begin{tabular}{llrrrrr}
\toprule
 & & \multicolumn{4}{c}{\textit{Search Engines}} & \multicolumn{1}{c}{\textit{LLMs}} \\
\cmidrule(lr){3-6}\cmidrule(lr){7-7}
Benchmark & Category & Google & Bing & \makecell[l]{Duck\\DuckGo} & Yahoo & ChatGPT \\
\midrule
\multirow{5}{*}{\textbf{Uniform}} & R. Right & \textbf{+25.1pp\textsuperscript{*}} & \textbf{+25.0pp\textsuperscript{**}} & \textbf{+38.5pp\textsuperscript{**}} & \textbf{+39.7pp\textsuperscript{**}} & +15.1pp \\
 & M. Right & +13.5pp & \textbf{+16.6pp\textsuperscript{*}} & +6.1pp & +9.1pp & +5.9pp \\
 & Greens & \textbf{-17.4pp\textsuperscript{**}} & \textbf{-18.0pp\textsuperscript{**}} & \textbf{-18.0pp\textsuperscript{**}} & \textbf{-15.6pp\textsuperscript{**}} & -3.4pp \\
 & M. Left & \textbf{-13.5pp\textsuperscript{**}} & -7.1pp & \textbf{-11.0pp\textsuperscript{**}} & \textbf{-13.2pp\textsuperscript{**}} & -6.7pp \\
 & R. Left & -7.7pp & \textbf{-16.5pp\textsuperscript{**}} & \textbf{-15.6pp\textsuperscript{**}} & \textbf{-20.0pp\textsuperscript{**}} & \textbf{-15.3pp\textsuperscript{*}} \\
\bottomrule
\end{tabular}
\end{table}

\begin{table}[ht]
\centering
\small
\caption{\textbf{US Political Entities, Aggregated Test against Uniform Distribution.} Deviation 
from equal representation in US search engine results and LLM responses ($K=2$ categories, 
uniform baseline = 50\% each). Each cell shows the difference in percentage points from a 
uniform distribution, testing whether the algorithm treats Democratic and Republican entities 
equally. Statistical significance was assessed using a Binomial Z-test, where $N$ denotes 
the total number of articles containing at least one politically identifiable entity returned 
by the search engine across all states and queries. State-level results were aggregated into 
a single national estimate by summing article counts across states, with the expected 
proportion computed as the equal-weight average of state-level baselines. The first four 
columns are search engines; the last two LLMs. Bold values significant after 
Holm--Bonferroni correction. \textsuperscript{*} $p<0.05$, \textsuperscript{**} $p<0.01$.}
\label{tab:agg_uni_us_parties}
\begin{tabular}{llrrrrrr}
\toprule
 & & \multicolumn{4}{c}{\textit{Search Engines}} & \multicolumn{2}{c}{\textit{LLMs}} \\
\cmidrule(lr){3-6}\cmidrule(lr){7-8}
Benchmark & Category & Google & Bing & \makecell[l]{Duck\\DuckGo} & Yahoo & ChatGPT & Copilot \\
\midrule
\multirow{2}{*}{\textbf{Uniform}} & Dem. & -1.0pp & -1.3pp & \textbf{-4.6pp\textsuperscript{**}} & \textbf{-4.3pp\textsuperscript{**}} & +0.0pp & +3.3pp \\
 & Rep. & +1.0pp & +1.3pp & \textbf{+4.6pp\textsuperscript{**}} & \textbf{+4.3pp\textsuperscript{**}} & +0.0pp & -3.3pp \\
\bottomrule
\end{tabular}
\end{table}
\begin{table}[ht]
\centering
\small
\caption{\textbf{US Political Entities, Query-Level Test against Uniform distribution.} Deviation from equal representation in US search engine results and LLM responses ($K=2$ categories, uniform baseline = 50\% each). Each cell shows the difference in percentage points from a uniform distribution, testing whether the algorithm treats all political leanings equally. Statistical significance was assessed using the Beta-Binomial likelihood ratio test where $N \geq 30$ queries were available, and a permutation test otherwise. State-level z-scores were then pooled into a single estimate using Stouffer's method, weighting each state by $\sqrt{N}$.The first four columns are search engines; the last two LLMs. Bold values significant after Holm--Bonferroni correction. \textsuperscript{*} $p<0.05$, \textsuperscript{**} $p<0.01$.}
\label{tab:uni_us_parties}
\begin{tabular}{llrrrr}
\toprule
 & & \multicolumn{4}{c}{\textit{Search Engines}} \\
\cmidrule(lr){3-6}
Benchmark & Category & Google & Bing & \makecell[l]{Duck\\DuckGo} & Yahoo \\
\midrule
\multirow{2}{*}{\textbf{Uniform}} & Dem. & \textbf{-4.8pp\textsuperscript{**}} & \textbf{-1.2pp\textsuperscript{*}} & \textbf{-3.1pp\textsuperscript{**}} & \textbf{-6.5pp\textsuperscript{**}} \\
 & Rep. & \textbf{+4.8pp\textsuperscript{**}} & \textbf{+1.2pp\textsuperscript{*}} & \textbf{+3.1pp\textsuperscript{**}} & \textbf{+6.5pp\textsuperscript{**}} \\
\bottomrule
\end{tabular}
\end{table}

\begin{table}[ht]
\centering
\small
\caption{\textbf{US Political Issues, Aggregated Test against Uniform distribution.} Deviation from equal representation in US search engine results and LLM responses ($K=5$ categories, uniform baseline = 20\% each). Each cell shows the difference in percentage points from a uniform distribution, testing whether the algorithm treats all issue categorys equally. Statistical significance was assessed using a Binomial Z-test, where N= number of articles with political entities per search engine. State-level results were aggregated into 
a single national estimate by summing article counts across states, with the expected 
proportion computed as the equal-weight average of state-level baselines. The first four columns are search engines; the last two LLMs. Bold values significant after Holm--Bonferroni correction. \textsuperscript{*} $p<0.05$, \textsuperscript{**} $p<0.01$. }
\label{tab:agg_uni_us_issues}
\begin{tabular}{llrrrrrr}
\toprule
 & & \multicolumn{4}{c}{\textit{Search Engines}} & \multicolumn{2}{c}{\textit{LLMs}} \\
\cmidrule(lr){3-6}\cmidrule(lr){7-8}
Benchmark & Category & Google & Bing & \makecell[l]{Duck\\DuckGo} & Yahoo & ChatGPT & Copilot \\
\midrule
\multirow{5}{*}{\textbf{Uniform}} & Rep $++$ & \textbf{-15.2pp\textsuperscript{**}} & \textbf{-10.8pp\textsuperscript{**}} & +0.5pp & \textbf{-3.0pp\textsuperscript{*}} & \textbf{-9.5pp\textsuperscript{*}} & -3.5pp \\
 & Rep $+$ & \textbf{+63.1pp\textsuperscript{**}} & \textbf{+8.9pp\textsuperscript{**}} & \textbf{+8.8pp\textsuperscript{**}} & \textbf{+7.7pp\textsuperscript{**}} & \textbf{+8.1pp\textsuperscript{*}} & \textbf{+16.7pp\textsuperscript{**}} \\
 & Neutral & --- & \textbf{-7.6pp\textsuperscript{**}} & \textbf{-10.4pp\textsuperscript{**}} & \textbf{-9.8pp\textsuperscript{**}} & +2.2pp & \textbf{-12.1pp\textsuperscript{**}} \\
 & Dem $+$ & \textbf{-14.9pp\textsuperscript{**}} & \textbf{+24.0pp\textsuperscript{**}} & \textbf{+14.7pp\textsuperscript{**}} & \textbf{+17.5pp\textsuperscript{**}} & \textbf{+14.0pp\textsuperscript{**}} & \textbf{+18.8pp\textsuperscript{**}} \\
 & Dem $++$ & \textbf{-13.1pp\textsuperscript{**}} & \textbf{-14.5pp\textsuperscript{**}} & \textbf{-13.6pp\textsuperscript{**}} & \textbf{-12.5pp\textsuperscript{**}} & \textbf{-14.8pp\textsuperscript{**}} & --- \\
\bottomrule
\end{tabular}
\end{table}
\begin{table}[ht]
\centering
\small
\caption{\textbf{US Political Issues, Query-Level Test against Uniform Distribution.} Deviation from 
equal representation in US search engine results and LLM responses ($K=5$ categories, 
uniform baseline = 20\% each). Each cell shows the difference in percentage points from a 
uniform distribution, testing whether the algorithm treats all issue categories equally. 
Tests were conducted on a single national pool of queries using an adaptive one-sample 
test: a z-test for search engines (where $N \geq 30$ queries were available but raw article 
counts were not reliable) and a permutation test for LLMs (where $N = 4$--$5$ queries per 
engine). No cross-stratum aggregation was performed. The first four columns are search 
engines; the last two are LLMs. Bold values significant after Holm--Bonferroni correction. 
\textsuperscript{*} $p<0.05$, \textsuperscript{**} $p<0.01$.}
\label{tab:query_uni_us_issues}
\begin{tabular}{llrrrrrr}
\toprule
 & & \multicolumn{4}{c}{\textit{Search Engines}} & \multicolumn{2}{c}{\textit{LLMs}} \\
\cmidrule(lr){3-6}\cmidrule(lr){7-8}
Benchmark & Category & Google & Bing & \makecell[l]{Duck\\DuckGo} & Yahoo & ChatGPT & Copilot \\
\midrule
\multirow{5}{*}{\textbf{Uniform}} & Rep $++$ & \textbf{-19.2pp\textsuperscript{**}} & \textbf{-14.6pp\textsuperscript{**}} & +4.0pp & +2.6pp & -11.5pp & -2.8pp \\
 & Rep $+$ & \textbf{+52.3pp\textsuperscript{**}} & -1.6pp & +11.0pp & -3.8pp & +21.0pp & +13.2pp \\
 & Neutral & \textbf{-20.0pp\textsuperscript{**}} & -1.9pp & \textbf{-9.3pp\textsuperscript{**}} & \textbf{-10.7pp\textsuperscript{**}} & -1.9pp & -10.1pp \\
 & Dem $+$ & \textbf{-19.1pp\textsuperscript{**}} & \textbf{+27.2pp\textsuperscript{**}} & +7.0pp & +7.2pp & +8.3pp & +19.8pp \\
 & Dem $++$ & +6.0pp & -9.0pp & \textbf{-12.8pp\textsuperscript{**}} & +4.7pp & -15.9pp & -20.0pp \\
\bottomrule
\end{tabular}
\end{table}

\begin{table}[ht]
\centering
\small
\caption{\textbf{Global distributional test for search engines}. Each cell shows the approximate $\chi^2$ statistic ($\approx \sum_\ell z_\ell^2 \sim \chi^2(K-1)$), testing whether the full distribution differs from the benchmark. Bold values significant after Holm--Bonferroni correction. \textsuperscript{*} $p<0.05$, \textsuperscript{**} $p<0.01$.}
\label{tab:uni_global_se}
\begin{tabular}{llrrrr}
\toprule
Context & Benchmark & Google & Bing & \makecell[l]{Duck\\DuckGo} & Yahoo \\
\midrule
\multirow{1}{*}{\textbf{EU -- SE}} & Uniform & \textbf{53.77\textsuperscript{**}} & \textbf{51.23\textsuperscript{**}} & \textbf{81.21\textsuperscript{**}} & \textbf{70.54\textsuperscript{**}} \\
\midrule
\multirow{1}{*}{\textbf{US Parties -- SE}} & Uniform & \textbf{53.24\textsuperscript{**}} & \textbf{10.63\textsuperscript{**}} & \textbf{65.58\textsuperscript{**}} & \textbf{70.07\textsuperscript{**}} \\
\midrule
\multirow{1}{*}{\textbf{US Issues -- SE}} & Uniform & --- & \textbf{28.70\textsuperscript{**}} & --- & --- \\
\bottomrule
\end{tabular}
\end{table}
\begin{table}[ht]
\centering
\small
\caption{\textbf{Global distributional test for LLMs.} Each cell shows the approximate $\chi^2$ statistic ($\approx \sum_\ell z_\ell^2 \sim \chi^2(K-1)$), testing whether the full distribution differs from the benchmark. Bold values significant after Holm--Bonferroni correction. \textsuperscript{*} $p<0.05$, \textsuperscript{**} $p<0.01$.}
\label{tab:uni_global_llm}
\begin{tabular}{llrr}
\toprule
Context & Benchmark & ChatGPT & Copilot \\
\midrule
\multirow{1}{*}{\textbf{EU -- LLMs}} & Uniform & \textbf{20.24\textsuperscript{**}} & --- \\
\midrule
\multirow{1}{*}{\textbf{US Issues -- LLMs}} & Uniform & \textbf{10.41\textsuperscript{*}} & \textbf{12.07\textsuperscript{*}} \\
\bottomrule
\end{tabular}
\end{table}

\clearpage
\subsection{Testing comparisons with external values (Benchmarks)}\label{app_subsec:external}

\begin{table}[ht]
\centering
\small
\caption{\textbf{EU Case, Aggregated Test against external benchmarks.} Representation bias in EU 
search engine results and LLM responses by political leaning, measured against three external benchmarks: media consumption shares, pre-election poll averages, and 2019 election results. Each cell shows the difference in percentage points between the overall observed proportion and the benchmark, where $N$ denotes the total number of articles or responses containing at least one politically identifiable entity, summed across all countries. The expected benchmark proportion is computed as the unweighted average of country-level benchmark values, giving each country equal weight regardless of article volume. Statistical significance was assessed using a Binomial Z-test on the EU-aggregated counts. The first four columns are search engines; the last two are LLMs. Bold values significant after Holm--Bonferroni correction. \textsuperscript{*} $p<0.05$, 
\textsuperscript{**} $p<0.01$.}
\label{tab:agg_ext_eu}
\begin{tabular}{llrrrrrr}
\toprule
 & & \multicolumn{4}{c}{\textit{Search Engines}} & \multicolumn{2}{c}{\textit{LLMs}} \\
\cmidrule(lr){3-6}\cmidrule(lr){7-8}
Benchmark & Category & Google & Bing & \makecell[l]{Duck\\DuckGo} & Yahoo & ChatGPT & Copilot \\
\midrule
\multirow{5}{*}{\textbf{Media}} & R. Right & \textbf{+16.8pp\textsuperscript{**}} & \textbf{+18.3pp\textsuperscript{**}} & \textbf{+25.7pp\textsuperscript{**}} & \textbf{+31.7pp\textsuperscript{**}} & +2.3pp & -7.9pp \\
 & M. Right & \textbf{-10.4pp\textsuperscript{**}} & \textbf{-7.2pp\textsuperscript{*}} & \textbf{-11.2pp\textsuperscript{**}} & \textbf{-13.8pp\textsuperscript{**}} & -8.6pp & -9.9pp \\
 & Greens & -2.6pp & \textbf{-8.0pp\textsuperscript{**}} & \textbf{-5.0pp\textsuperscript{**}} & -2.1pp & \textbf{+10.4pp\textsuperscript{**}} & \textbf{+17.4pp\textsuperscript{**}} \\
 & M. Left & \textbf{-9.3pp\textsuperscript{**}} & +1.9pp & \textbf{-6.0pp\textsuperscript{**}} & \textbf{-9.6pp\textsuperscript{**}} & -2.3pp & +3.5pp \\
 & R. Left & \textbf{+5.5pp\textsuperscript{**}} & \textbf{-5.1pp\textsuperscript{**}} & \textbf{-3.4pp\textsuperscript{**}} & \textbf{-6.2pp\textsuperscript{**}} & -1.9pp & -3.1pp \\
\midrule
\multirow{5}{*}{\textbf{Polls}} & R. Right & \textbf{+24.5pp\textsuperscript{**}} & \textbf{+26.0pp\textsuperscript{**}} & \textbf{+33.4pp\textsuperscript{**}} & \textbf{+39.3pp\textsuperscript{**}} & +10.0pp & -0.3pp \\
 & M. Right & \textbf{-13.5pp\textsuperscript{**}} & \textbf{-10.2pp\textsuperscript{**}} & \textbf{-14.3pp\textsuperscript{**}} & \textbf{-16.9pp\textsuperscript{**}} & -11.7pp & -13.0pp \\
 & Greens & -2.0pp & \textbf{-7.4pp\textsuperscript{**}} & \textbf{-4.4pp\textsuperscript{**}} & -1.5pp & \textbf{+11.0pp\textsuperscript{**}} & \textbf{+18.1pp\textsuperscript{**}} \\
 & M. Left & \textbf{-10.6pp\textsuperscript{**}} & +0.7pp & \textbf{-7.2pp\textsuperscript{**}} & \textbf{-10.8pp\textsuperscript{**}} & -3.5pp & +2.3pp \\
 & R. Left & +1.5pp & \textbf{-9.1pp\textsuperscript{**}} & \textbf{-7.4pp\textsuperscript{**}} & \textbf{-10.2pp\textsuperscript{**}} & -5.9pp & -7.1pp \\
\midrule
\multirow{5}{*}{\textbf{Results 2019}} & R. Right & \textbf{+31.8pp\textsuperscript{**}} & \textbf{+33.3pp\textsuperscript{**}} & \textbf{+40.7pp\textsuperscript{**}} & \textbf{+46.6pp\textsuperscript{**}} & \textbf{+17.3pp\textsuperscript{**}} & +7.0pp \\
 & M. Right & \textbf{-11.7pp\textsuperscript{**}} & \textbf{-8.4pp\textsuperscript{**}} & \textbf{-12.5pp\textsuperscript{**}} & \textbf{-15.0pp\textsuperscript{**}} & -9.8pp & -11.1pp \\
 & Greens & \textbf{-5.9pp\textsuperscript{**}} & \textbf{-11.4pp\textsuperscript{**}} & \textbf{-8.3pp\textsuperscript{**}} & \textbf{-5.4pp\textsuperscript{**}} & +7.1pp & \textbf{+14.1pp\textsuperscript{*}} \\
 & M. Left & \textbf{-13.4pp\textsuperscript{**}} & -2.2pp & \textbf{-10.1pp\textsuperscript{**}} & \textbf{-13.7pp\textsuperscript{**}} & -6.4pp & -0.6pp \\
 & R. Left & -0.6pp & \textbf{-11.2pp\textsuperscript{**}} & \textbf{-9.6pp\textsuperscript{**}} & \textbf{-12.3pp\textsuperscript{**}} & -8.0pp & -9.3pp \\
\bottomrule
\end{tabular}
\end{table}
\begin{table}[ht]
\centering
\small
\caption{\textbf{EU Case, Query-Level Test against external benchmarks.} Representation bias in EU 
search engine results and LLM responses by political leaning, measured against three 
external benchmarks: media consumption shares, pre-election poll averages, and 2019 
election results. Each cell shows the mean difference in percentage points between the 
observed proportion and the benchmark, computed as the mean of per-query proportions, 
where queries returning no mentions of a category contribute zero. Tests were conducted 
separately for each country (Austria, Germany, Ireland, Poland, Portugal) using an 
adaptive one-sample test: the Beta-Binomial likelihood ratio test when $N \geq 30$ queries 
were available, and a permutation test otherwise, where $N$ denotes the number of queries 
returning at least one result with a politically identifiable entity. Country-level 
z-scores were pooled into a single EU-level estimate using Stouffer's method, weighting 
each country by $\sqrt{N}$. The first four columns are search engines; the last two are 
LLMs. Bold values significant after Holm--Bonferroni correction. \textsuperscript{*} 
$p<0.05$, \textsuperscript{**} $p<0.01$.}
\label{tab:que_ext_eu}
\begin{tabular}{llrrrrr}
\toprule
 & & \multicolumn{4}{c}{\textit{Search Engines}} & \multicolumn{1}{c}{\textit{LLMs}} \\
\cmidrule(lr){3-6}\cmidrule(lr){7-7}
Benchmark & Category & Google & Bing & \makecell[l]{Duck\\DuckGo} & Yahoo & ChatGPT \\
\midrule
\multirow{5}{*}{\textbf{Media}} & R. Right & +15.2pp & \textbf{+15.6pp\textsuperscript{**}} & \textbf{+26.9pp\textsuperscript{**}} & \textbf{+28.3pp\textsuperscript{**}} & +4.1pp \\
 & M. Right & -8.1pp & -4.1pp & \textbf{-11.1pp\textsuperscript{**}} & -8.8pp & -11.7pp \\
 & Greens & \textbf{-5.9pp\textsuperscript{**}} & \textbf{-6.5pp\textsuperscript{**}} & \textbf{-6.3pp\textsuperscript{**}} & -4.2pp & +7.9pp \\
 & M. Left & \textbf{-8.2pp\textsuperscript{**}} & -3.3pp & \textbf{-8.2pp\textsuperscript{**}} & \textbf{-9.1pp\textsuperscript{**}} & -3.2pp \\
 & R. Left & +7.0pp & \textbf{-1.8pp\textsuperscript{**}} & \textbf{-1.2pp\textsuperscript{*}} & \textbf{-6.2pp\textsuperscript{**}} & -1.5pp \\
\midrule
\multirow{5}{*}{\textbf{Polls}} & R. Right & +22.9pp & +23.3pp & \textbf{+34.6pp\textsuperscript{**}} & \textbf{+36.0pp\textsuperscript{**}} & +11.8pp \\
 & M. Right & -11.2pp & -7.1pp & \textbf{-14.2pp\textsuperscript{**}} & \textbf{-11.9pp\textsuperscript{**}} & -14.7pp \\
 & Greens & \textbf{-5.2pp\textsuperscript{**}} & \textbf{-5.9pp\textsuperscript{**}} & \textbf{-5.7pp\textsuperscript{**}} & -3.6pp & +8.5pp \\
 & M. Left & \textbf{-9.4pp\textsuperscript{**}} & -4.6pp & \textbf{-9.5pp\textsuperscript{**}} & \textbf{-10.3pp\textsuperscript{**}} & -4.5pp \\
 & R. Left & +3.0pp & \textbf{-5.7pp\textsuperscript{**}} & -5.2pp & \textbf{-10.2pp\textsuperscript{**}} & -5.4pp \\
\midrule
\multirow{5}{*}{\textbf{Results 2019}} & R. Right & +30.2pp & \textbf{+30.6pp\textsuperscript{*}} & \textbf{+41.9pp\textsuperscript{**}} & \textbf{+43.3pp\textsuperscript{**}} & +19.1pp \\
 & M. Right & -9.3pp & -5.3pp & \textbf{-12.3pp\textsuperscript{**}} & \textbf{-10.0pp\textsuperscript{**}} & -12.9pp \\
 & Greens & \textbf{-9.2pp\textsuperscript{**}} & \textbf{-9.8pp\textsuperscript{**}} & \textbf{-9.6pp\textsuperscript{**}} & \textbf{-7.5pp\textsuperscript{**}} & +4.5pp \\
 & M. Left & \textbf{-12.3pp\textsuperscript{**}} & \textbf{-7.4pp\textsuperscript{**}} & \textbf{-12.3pp\textsuperscript{**}} & \textbf{-13.2pp\textsuperscript{**}} & -7.3pp \\
 & R. Left & +0.8pp & \textbf{-7.9pp\textsuperscript{**}} & \textbf{-7.4pp\textsuperscript{**}} & \textbf{-12.3pp\textsuperscript{**}} & -7.6pp \\
\bottomrule
\end{tabular}
\end{table}

\begin{table}[ht]
\centering
\small
\caption{\textbf{US Political Entities, Aggregated Test against External Benchmarks.} Representation 
bias in US search engine results and LLM responses by political leaning, measured against three external benchmarks: media consumption shares, pre-election poll averages, and 2020 election results. Each cell shows the difference in percentage points between the overall observed proportion and the benchmark, where $N$ denotes the total number of articles or 
responses containing at least one politically identifiable entity, summed across all states and queries. The expected benchmark proportion is the unweighted average of state-level benchmark values, giving each state equal weight regardless of article volume. Statistical significance was assessed using a Binomial Z-test on the nationally aggregated counts. 
The first four columns are search engines; the last two are LLMs. Bold values significant after Holm--Bonferroni correction. \textsuperscript{*} $p<0.05$, \textsuperscript{**} $p<0.01$.}
\label{tab:agg_us_parties}
\begin{tabular}{llrrrrrr}
\toprule
 & & \multicolumn{4}{c}{\textit{Search Engines}} & \multicolumn{2}{c}{\textit{LLMs}} \\
\cmidrule(lr){3-6}\cmidrule(lr){7-8}
Benchmark & Category & Google & Bing & \makecell[l]{Duck\\DuckGo} & Yahoo & ChatGPT & Copilot \\
\midrule
\multirow{2}{*}{\textbf{Media}} & Dem. & \textbf{+2.6pp\textsuperscript{**}} & +2.3pp & -1.0pp & -0.6pp & +3.6pp & +7.0pp \\
 & Rep. & -1.3pp & -1.0pp & +2.3pp & +1.9pp & -2.3pp & -5.7pp \\
\midrule
\multirow{2}{*}{\textbf{Polls}} & Dem. & +0.1pp & -0.2pp & \textbf{-3.5pp\textsuperscript{**}} & \textbf{-3.2pp\textsuperscript{*}} & +1.1pp & +4.4pp \\
 & Rep. & \textbf{+3.6pp\textsuperscript{**}} & \textbf{+3.9pp\textsuperscript{*}} & \textbf{+7.2pp\textsuperscript{**}} & \textbf{+6.9pp\textsuperscript{**}} & +2.6pp & -0.7pp \\
\midrule
\multirow{2}{*}{\textbf{Last Elections}} & Dem. & \textbf{-2.3pp\textsuperscript{**}} & -2.6pp & \textbf{-5.9pp\textsuperscript{**}} & \textbf{-5.6pp\textsuperscript{**}} & -1.3pp & +2.0pp \\
 & Rep. & \textbf{+4.2pp\textsuperscript{**}} & \textbf{+4.5pp\textsuperscript{**}} & \textbf{+7.8pp\textsuperscript{**}} & \textbf{+7.5pp\textsuperscript{**}} & +3.2pp & -0.1pp \\
\bottomrule
\end{tabular}
\end{table}
\begin{table}[ht]
\centering
\small
\caption{\textbf{US Political Entities, Query-Level Test against external benchmarks.} Each cell shows the mean difference in percentage points between the observed proportion and the benchmark (media consumption shares, pre-election poll averages, and 2020 election results) computed as the mean of per-query proportions (queries returning no mentions of a category contribute zero). Statistical significance was assessed using the Beta-Binomial likelihood ratio test where $N \geq 30$ queries were available, and a permutation test otherwise. State-level z-scores were then pooled into a single estimate using Stouffer's method, weighting each state by $\sqrt{N}$.The first four columns are search engines; the last two LLMs. Bold values significant after Holm--Bonferroni correction. \textsuperscript{*} $p<0.05$, \textsuperscript{**} $p<0.01$.}
\label{tab:que_us_parties}
\begin{tabular}{llrrrr}
\toprule
 & & \multicolumn{4}{c}{\textit{Search Engines}} \\
\cmidrule(lr){3-6}
Benchmark & Category & Google & Bing & \makecell[l]{Duck\\DuckGo} & Yahoo \\
\midrule
\multirow{2}{*}{\textbf{Media}} & Dem. & -1.6pp & \textbf{+3.0pp\textsuperscript{*}} & +1.4pp & -2.2pp \\
 & Rep. & +2.9pp & -1.7pp & -0.1pp & +3.5pp \\
\midrule
\multirow{2}{*}{\textbf{Polls}} & Dem. & \textbf{-4.8pp\textsuperscript{**}} & -0.2pp & \textbf{-1.8pp\textsuperscript{**}} & \textbf{-11.8pp\textsuperscript{**}} \\
 & Rep. & \textbf{+9.4pp\textsuperscript{**}} & +4.8pp & \textbf{+6.5pp\textsuperscript{**}} & \textbf{+16.8pp\textsuperscript{**}} \\
\midrule
\multirow{2}{*}{\textbf{Last Elections}} & Dem. & \textbf{-7.4pp\textsuperscript{**}} & -2.8pp & -4.5pp & \textbf{-14.8pp\textsuperscript{**}} \\
 & Rep. & \textbf{+8.9pp\textsuperscript{**}} & +4.3pp & \textbf{+5.9pp\textsuperscript{**}} & \textbf{+16.3pp\textsuperscript{**}} \\
\bottomrule
\end{tabular}
\end{table}

\begin{table}[ht]
\centering
\small
\caption{\textbf{US Political Issues, Aggregated Test against external benchmarks.} Representation 
bias in US search engine results and LLM responses by political leaning, measured against three external benchmarks: media consumption shares, pre-election poll averages, and 2020 election results. Each cell shows the difference in percentage points between the overall observed proportion and the benchmark, where $N$ denotes the total number of articles or 
responses containing at least one politically identifiable entity, summed across all states and queries. The expected benchmark proportion is the unweighted average of state-level benchmark values, giving each state equal weight regardless of article volume. Statistical significance was assessed using a Binomial Z-test on the nationally aggregated counts. 
The first four columns are search engines; the last two are LLMs. Bold values significant after Holm--Bonferroni correction. \textsuperscript{*} $p<0.05$, \textsuperscript{**} $p<0.01$.}
\label{tab:agg_us_issues}
\begin{tabular}{llrrrrrr}
\toprule
 & & \multicolumn{4}{c}{\textit{Search Engines}} & \multicolumn{2}{c}{\textit{LLMs}} \\
\cmidrule(lr){3-6}\cmidrule(lr){7-8}
Benchmark & Category & Google & Bing & \makecell[l]{Duck\\DuckGo} & Yahoo & ChatGPT & Copilot \\
\midrule
\multirow{5}{*}{\textbf{Media}} & Rep $++$ & \textbf{-3.2pp\textsuperscript{**}} & +1.2pp & \textbf{+12.4pp\textsuperscript{**}} & \textbf{+9.0pp\textsuperscript{**}} & +2.4pp & \textbf{+8.5pp\textsuperscript{**}} \\
 & Rep $+$ & \textbf{+31.0pp\textsuperscript{**}} & \textbf{-23.2pp\textsuperscript{**}} & \textbf{-23.3pp\textsuperscript{**}} & \textbf{-24.4pp\textsuperscript{**}} & \textbf{-24.0pp\textsuperscript{**}} & \textbf{-15.4pp\textsuperscript{**}} \\
 & Neutral & --- & \textbf{+8.8pp\textsuperscript{**}} & \textbf{+6.0pp\textsuperscript{**}} & \textbf{+6.6pp\textsuperscript{**}} & \textbf{+18.6pp\textsuperscript{**}} & +4.3pp \\
 & Dem $+$ & \textbf{-8.2pp\textsuperscript{**}} & \textbf{+30.7pp\textsuperscript{**}} & \textbf{+21.4pp\textsuperscript{**}} & \textbf{+24.1pp\textsuperscript{**}} & \textbf{+20.6pp\textsuperscript{**}} & \textbf{+25.5pp\textsuperscript{**}} \\
 & Dem $++$ & \textbf{-16.0pp\textsuperscript{**}} & \textbf{-17.4pp\textsuperscript{**}} & \textbf{-16.5pp\textsuperscript{**}} & \textbf{-15.4pp\textsuperscript{**}} & \textbf{-17.7pp\textsuperscript{**}} & --- \\
\midrule
\multirow{5}{*}{\makecell[l]{\textbf{Issue}\\ \textbf{Importance}}} & Rep $++$ & \textbf{-7.4pp\textsuperscript{**}} & \textbf{-3.0pp\textsuperscript{**}} & \textbf{+8.3pp\textsuperscript{**}} & \textbf{+4.8pp\textsuperscript{**}} & -1.8pp & +4.3pp \\
 & Rep $+$ & \textbf{+47.3pp\textsuperscript{**}} & \textbf{-6.9pp\textsuperscript{**}} & \textbf{-7.0pp\textsuperscript{**}} & \textbf{-8.1pp\textsuperscript{**}} & -7.7pp & +0.9pp \\
 & Neutral & --- & \textbf{-7.5pp\textsuperscript{**}} & \textbf{-10.3pp\textsuperscript{**}} & \textbf{-9.7pp\textsuperscript{**}} & +2.3pp & \textbf{-12.0pp\textsuperscript{**}} \\
 & Dem $+$ & \textbf{-27.2pp\textsuperscript{**}} & \textbf{+11.7pp\textsuperscript{**}} & +2.4pp & \textbf{+5.1pp\textsuperscript{**}} & +1.6pp & +6.5pp \\
 & Dem $++$ & -0.3pp & -1.7pp & -0.8pp & +0.3pp & -2.0pp & --- \\
\bottomrule
\end{tabular}
\end{table}
\begin{table}[ht]
\centering
\small
\caption{\textbf{US Political Issues, Query-level test agains external benchmarks.} Each cell shows the mean difference in percentage points between the observed proportion and the benchmark (media consumption shares, pre-election poll averages, and 2020 election results) computed as the mean of per-query proportions (queries returning no mentions of a category contribute zero). Statistical significance was assessed using the Beta-Binomial likelihood ratio test where $N \geq 30$ queries were available, and a permutation test otherwise. State-level z-scores were then pooled into a single estimate using Stouffer's method, weighting each state by $\sqrt{N}$.The first four columns are search engines; the last two LLMs. Bold values significant after Holm--Bonferroni correction. \textsuperscript{*} $p<0.05$, \textsuperscript{**} $p<0.01$.}
\label{tab:que_us_issues}
\begin{tabular}{llrrrrrr}
\toprule
 & & \multicolumn{4}{c}{\textit{Search Engines}} & \multicolumn{2}{c}{\textit{LLMs}} \\
\cmidrule(lr){3-6}\cmidrule(lr){7-8}
Benchmark & Category & Google & Bing & \makecell[l]{Duck\\DuckGo} & Yahoo & ChatGPT & Copilot \\
\midrule
\multirow{5}{*}{\textbf{Media}} & Rep $++$ & \textbf{-7.2pp\textsuperscript{**}} & \textbf{-2.6pp\textsuperscript{*}} & \textbf{+16.0pp\textsuperscript{**}} & +14.6pp & +0.4pp & +9.1pp \\
 & Rep $+$ & \textbf{+20.2pp\textsuperscript{*}} & \textbf{-33.7pp\textsuperscript{**}} & \textbf{-21.1pp\textsuperscript{**}} & \textbf{-35.9pp\textsuperscript{**}} & -11.2pp & -18.9pp \\
 & Neutral & -3.6pp & \textbf{+14.5pp\textsuperscript{**}} & \textbf{+7.1pp\textsuperscript{**}} & \textbf{+5.8pp\textsuperscript{*}} & +14.5pp & +6.3pp \\
 & Dem $+$ & \textbf{-12.4pp\textsuperscript{**}} & \textbf{+33.8pp\textsuperscript{**}} & \textbf{+13.7pp\textsuperscript{**}} & \textbf{+13.8pp\textsuperscript{**}} & +15.0pp & +26.4pp \\
 & Dem $++$ & +3.1pp & -12.0pp & \textbf{-15.7pp\textsuperscript{**}} & +1.8pp & -18.8pp & -22.9pp \\
\midrule
\multirow{5}{*}{\textbf{Issue Importance}} & Rep $++$ & \textbf{-11.4pp\textsuperscript{**}} & \textbf{-6.8pp\textsuperscript{**}} & +11.8pp & +10.4pp & -3.7pp & +5.0pp \\
 & Rep $+$ & \textbf{+36.4pp\textsuperscript{**}} & \textbf{-17.4pp\textsuperscript{**}} & -4.8pp & \textbf{-19.7pp\textsuperscript{**}} & +5.1pp & -2.6pp \\
 & Neutral & \textbf{-19.9pp\textsuperscript{**}} & -1.8pp & \textbf{-9.2pp\textsuperscript{**}} & \textbf{-10.6pp\textsuperscript{**}} & -1.8pp & -10.0pp \\
 & Dem $+$ & \textbf{-31.5pp\textsuperscript{**}} & \textbf{+14.8pp\textsuperscript{**}} & -5.3pp & -5.2pp & -4.0pp & +7.4pp \\
 & Dem $++$ & \textbf{+18.8pp\textsuperscript{*}} & +3.8pp & +0.0pp & +17.5pp & -3.1pp & -7.2pp \\
\bottomrule
\end{tabular}
\end{table}

\begin{table}[ht!]
\centering
\small
\caption{\textbf{Global distributional test for search engines.} Each cell shows the approximate $\chi^2$ statistic ($\approx \sum_\ell z_\ell^2 \sim \chi^2(K-1)$), testing whether the full distribution differs from the benchmark. Bold values significant after Holm--Bonferroni correction. \textsuperscript{*} $p<0.05$, \textsuperscript{**} $p<0.01$.}
\label{tab:ext_global_se}
\begin{tabular}{llrrrr}
\toprule
Context & Benchmark & Google & Bing & \makecell[l]{Duck\\DuckGo} & Yahoo \\
\midrule
\multirow{4}{*}{\textbf{EU -- SE}} & Media & \textbf{63.60\textsuperscript{**}} & \textbf{77.22\textsuperscript{**}} & \textbf{118.82\textsuperscript{**}} & \textbf{137.60\textsuperscript{**}} \\
 & Polls & \textbf{71.93\textsuperscript{**}} & \textbf{71.65\textsuperscript{**}} & \textbf{94.06\textsuperscript{**}} & \textbf{97.11\textsuperscript{**}} \\
 & Results 2019 & \textbf{78.60\textsuperscript{**}} & \textbf{82.41\textsuperscript{**}} & \textbf{134.90\textsuperscript{**}} & \textbf{125.24\textsuperscript{**}} \\
 & Results 2024 & \textbf{62.08\textsuperscript{**}} & \textbf{49.88\textsuperscript{**}} & \textbf{96.83\textsuperscript{**}} & \textbf{108.16\textsuperscript{**}} \\
\midrule
\multirow{3}{*}{\textbf{US Parties -- SE}} & Last Elections & \textbf{209.47\textsuperscript{**}} & 4.79 & \textbf{56.27\textsuperscript{**}} & \textbf{181.57\textsuperscript{**}} \\
 & Media & 2.44 & \textbf{14.59\textsuperscript{**}} & 6.28 & 0.05 \\
 & Polls & \textbf{217.63\textsuperscript{**}} & \textbf{8.22\textsuperscript{*}} & \textbf{79.95\textsuperscript{**}} & \textbf{152.78\textsuperscript{**}} \\
\midrule
\multirow{2}{*}{\textbf{US Issues -- SE}} & Issue Importance & \textbf{3491.66\textsuperscript{**}} & \textbf{145.03\textsuperscript{**}} & \textbf{42.13\textsuperscript{**}} & \textbf{94.59\textsuperscript{**}} \\
 & Media & \textbf{648.83\textsuperscript{**}} & \textbf{429.35\textsuperscript{**}} & \textbf{111.97\textsuperscript{**}} & \textbf{193.01\textsuperscript{**}} \\
\bottomrule
\end{tabular}
\end{table}
\begin{table}[ht]
\centering
\small
\caption{\textbf{Global distributional test for LLMs.} Each cell shows the approximate $\chi^2$ statistic ($\approx \sum_\ell z_\ell^2 \sim \chi^2(K-1)$), testing whether the full distribution differs from the benchmark. Bold values significant after Holm--Bonferroni correction. \textsuperscript{*} $p<0.05$, \textsuperscript{**} $p<0.01$.}
\label{tab:ext_global_llm}
\begin{tabular}{llrr}
\toprule
Context & Benchmark & ChatGPT & Copilot \\
\midrule
\multirow{4}{*}{\textbf{EU -- LLMs}} & Media & 10.33 & --- \\
 & Polls & 14.64 & --- \\
 & Results 2019 & 10.45 & --- \\
 & Results 2024 & \textbf{16.08\textsuperscript{*}} & --- \\
\midrule
\multirow{2}{*}{\textbf{US Issues -- LLMs}} & Issue Importance & 4.14 & 9.89 \\
 & Media & 8.70 & 10.99 \\
\bottomrule
\end{tabular}
\end{table}

\clearpage

\clearpage
\section{Results supplementary analysis}\label{app_sec:search_engine_results}

\subsection{Proportion of results containing relevant entities (SEs and LLMs)}\label{app_subsec:se_entities}

\subsubsection{Proportions per location and algorithm}

\begin{figure*}[ht!]
    \centering
    \includegraphics[width=0.9\textwidth]{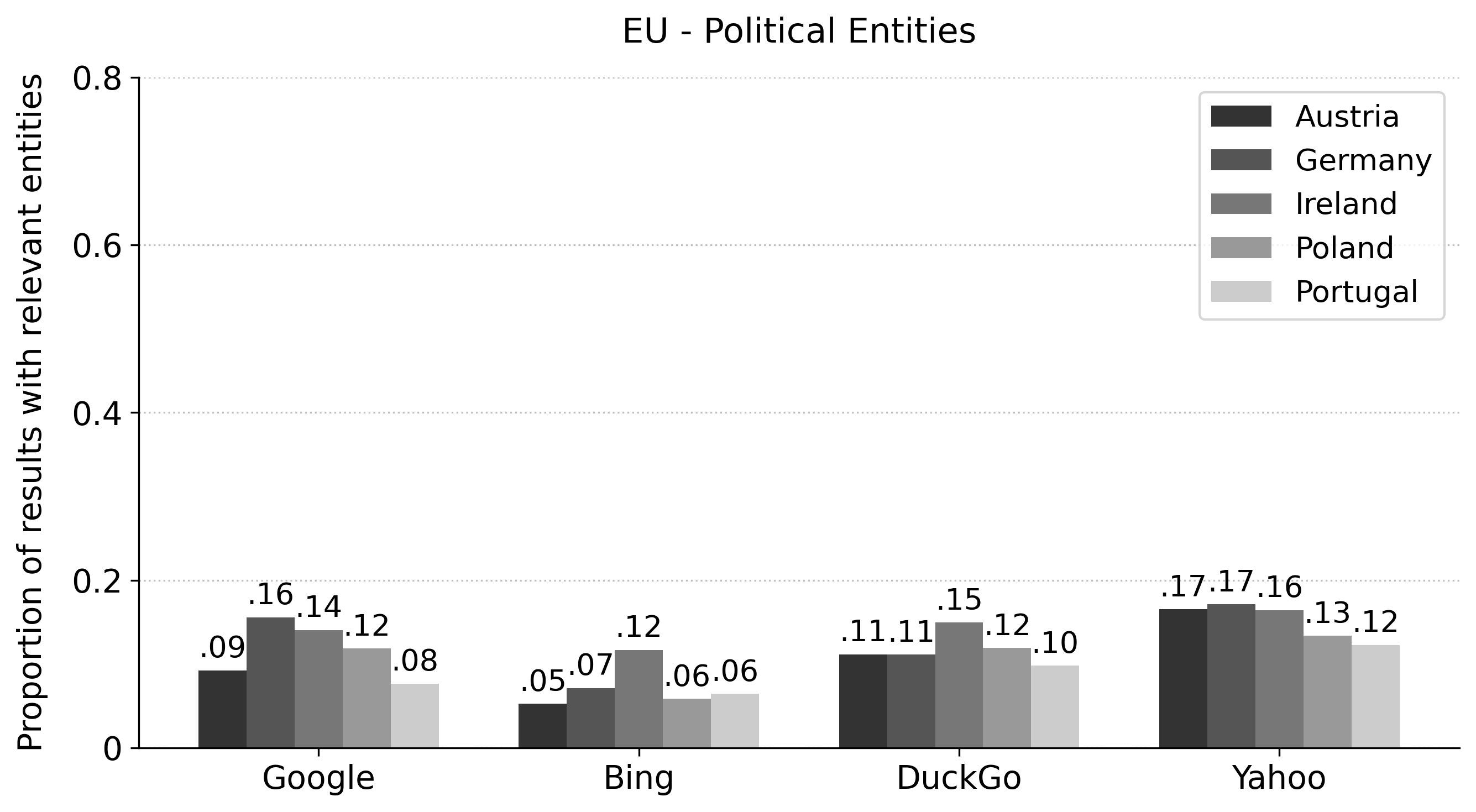}
    \caption{Proportion of results containing political entities, by country and search engine, in the European elections case study.}
    \label{sup_fig: eu_proportion_entities_per_loc_engine}
\end{figure*}

\begin{table}[ht]
\centering
\begin{tabular}{llc}
\hline
LLM  & Country & \% results with entities \\
\hline
\multirow{5}{*}{ChatGPT} 
 & Austria & 40.00 \\
 & Germany & 40.00 \\
 & Ireland & 44.44 \\
 & Poland & 40.00 \\
 & Portugal & 46.67 \\
\hline
\multirow{5}{*}{Copilot} 
 & Austria & 66.67 \\
 & Germany & 75.00 \\
 & Ireland & 33.33 \\
 & Poland & 66.67 \\
 & Portugal & 66.67 \\
\hline
\end{tabular}
\caption{Percentage of prompts with answers containing relevant political entities by country and LLM model. It includes data from the EU elections case-study only.
}
\end{table}

\begin{figure*}[ht!]
    \centering
    \includegraphics[width=0.7\textwidth]{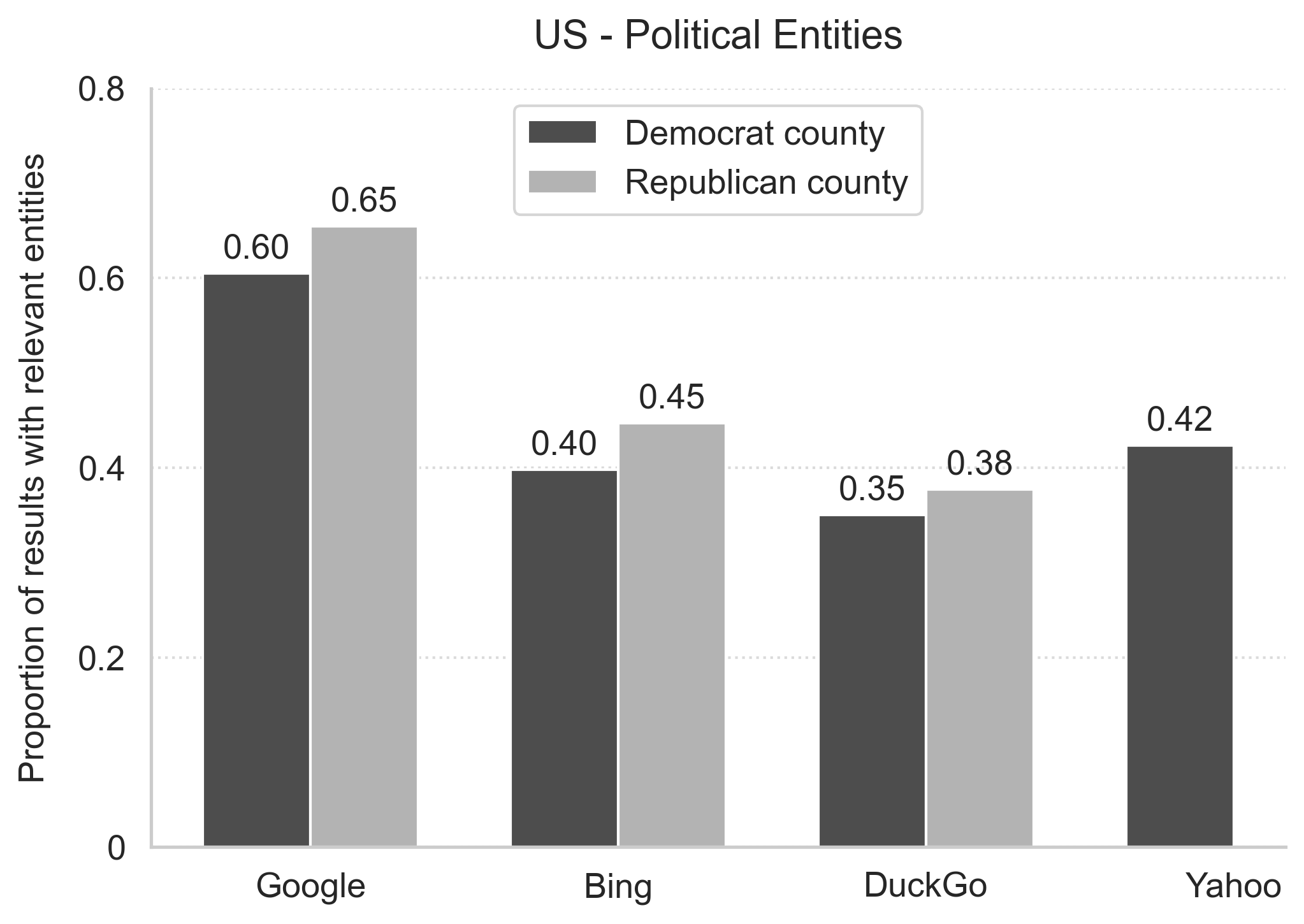}
    \caption{Proportion of results containing political entities, by county color (leaning in the polls) and search engine, in the US presidential elections case study. Bots in Republican counties did not search on Yahoo.}
    \label{sup_fig: us_proportion_entities_per_loc_engine}
\end{figure*}

\begin{figure*}[ht!]
    \centering
    \includegraphics[width=0.7\textwidth]{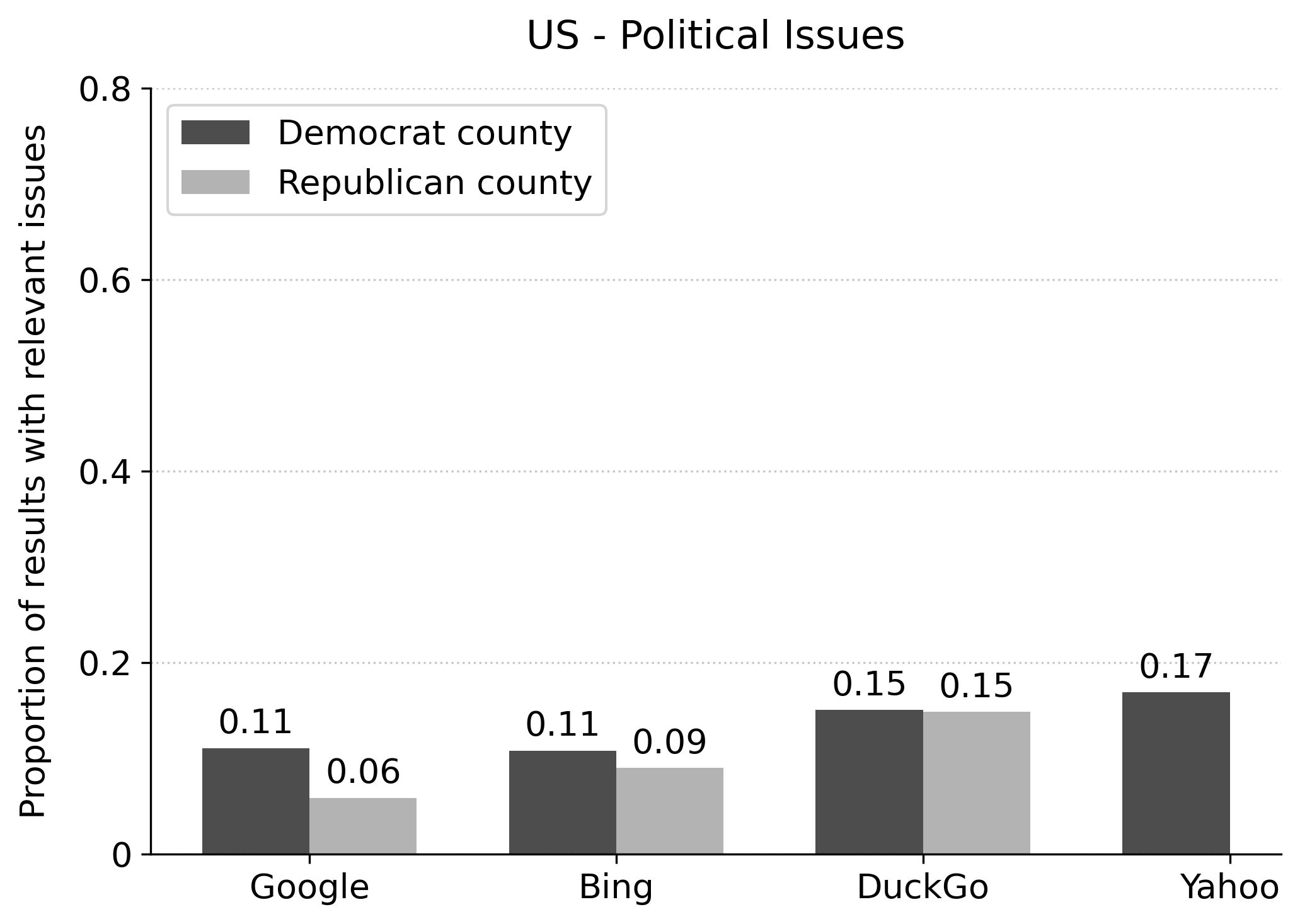}
    \caption{Proportion of results containing political issues (e.g., 'Economy´, 'Abortion´), by county color (leaning in the polls) and search engine, in the US presidential elections case study. Bots in Republican counties did not search on Yahoo.}
    \label{sup_fig: us_proportion_issues_per_loc_engine}
\end{figure*}

\clearpage
\subsubsection{Proportions per query}\label{app_subsec:se_entities_per_query}

\begin{figure*}[htbp]
    \centering
    \includegraphics[width=1\textwidth]{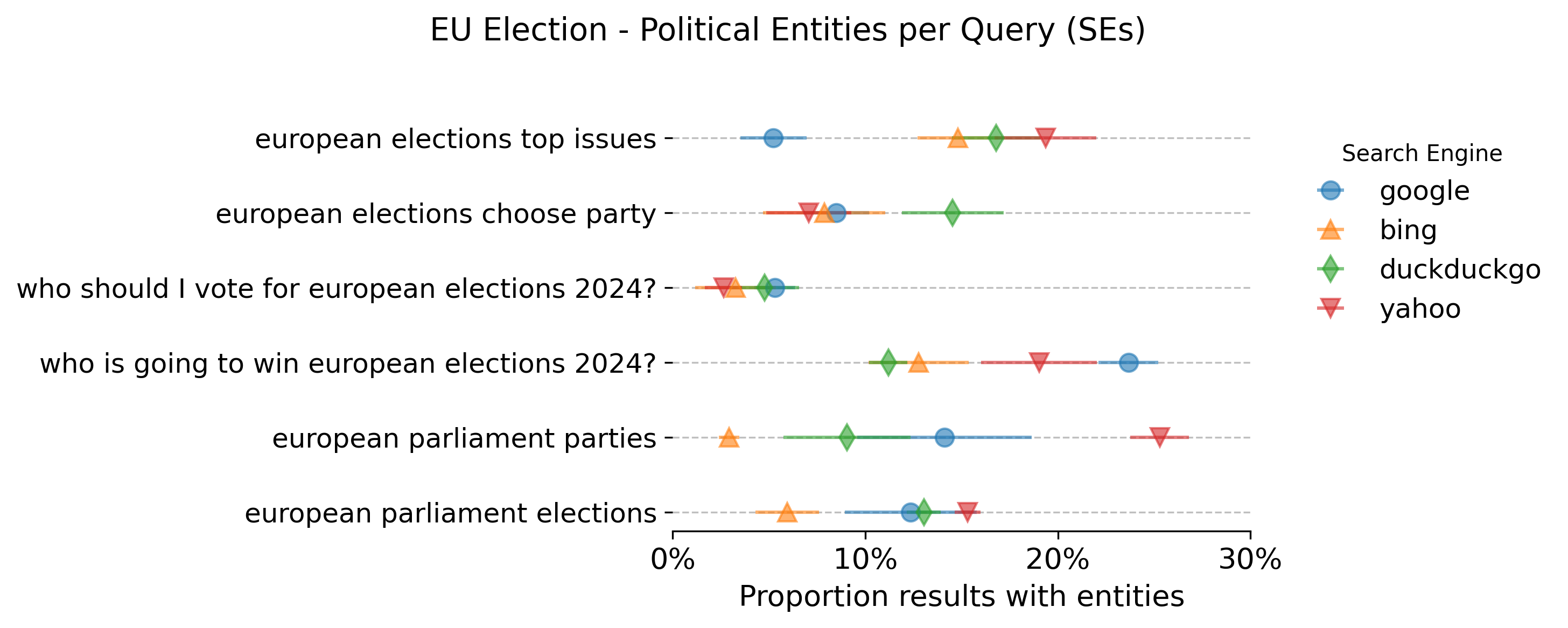}
    \caption{Proportions of SE results containing political entities, by query (EU Parliament Election case study).}
    \label{sup_fig: eu_proportion_query}
\end{figure*}

\begin{figure*}[htbp]
    \centering
    \includegraphics[width=1\textwidth]{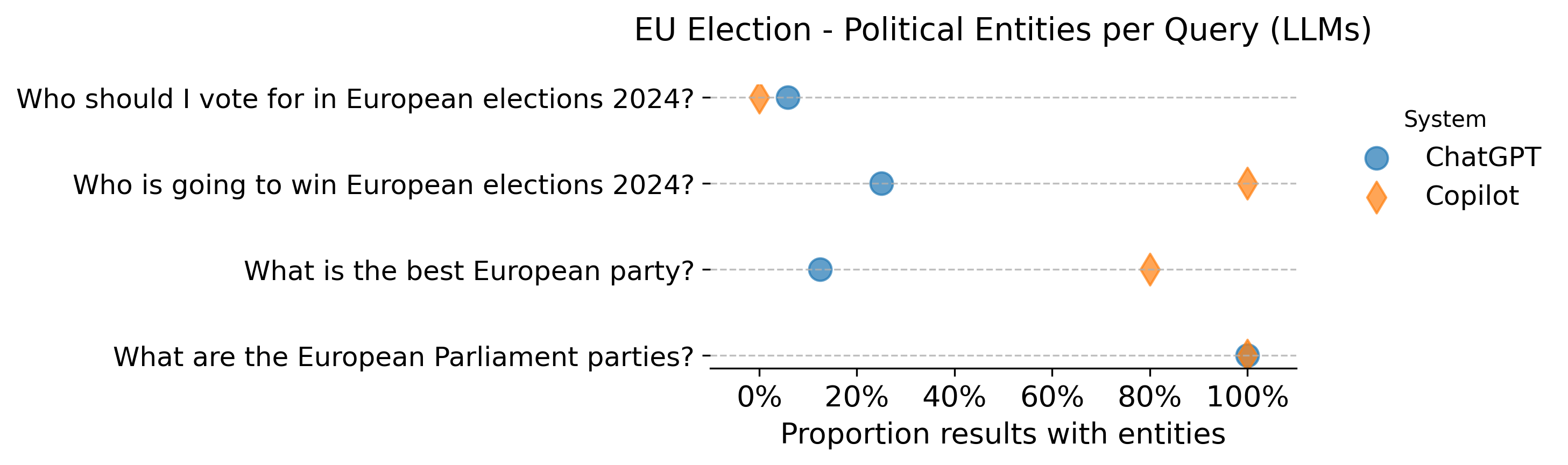}
    \caption{Proportions of LLM answers containing political entities, by query (EU Parliament Election case study).}
    \label{sup_fig: eu_llm_proportion_query}
\end{figure*}

\begin{figure*}[htbp]
    \centering
    \includegraphics[width=1\textwidth]{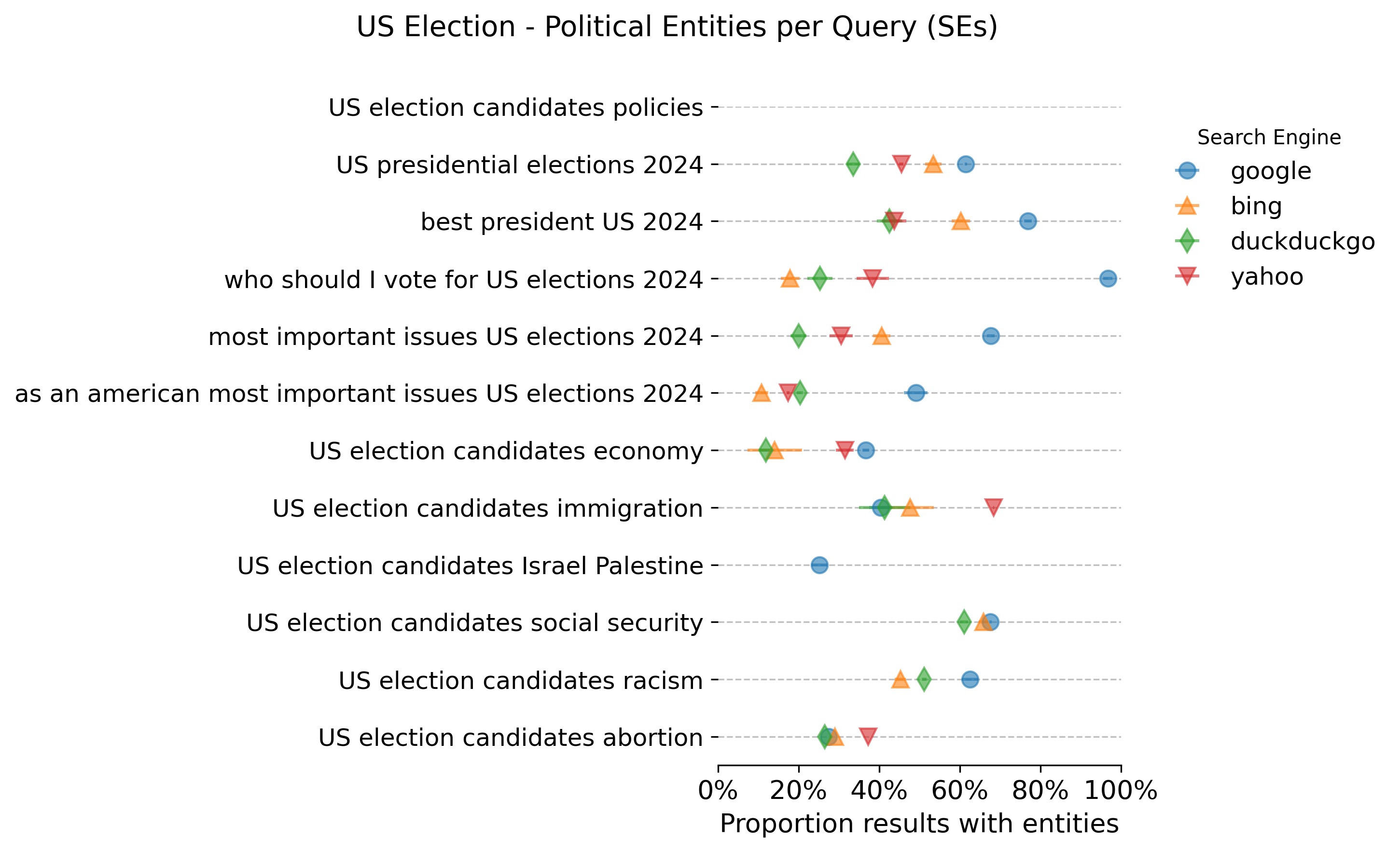}
    \caption{Proportion of SE results containing political entities, by query (US Presidential Election case study).}
    \label{sup_fig: us_entities_proportion_query}
\end{figure*}

\begin{figure*}[htbp]
    \centering
    \includegraphics[width=1\textwidth]{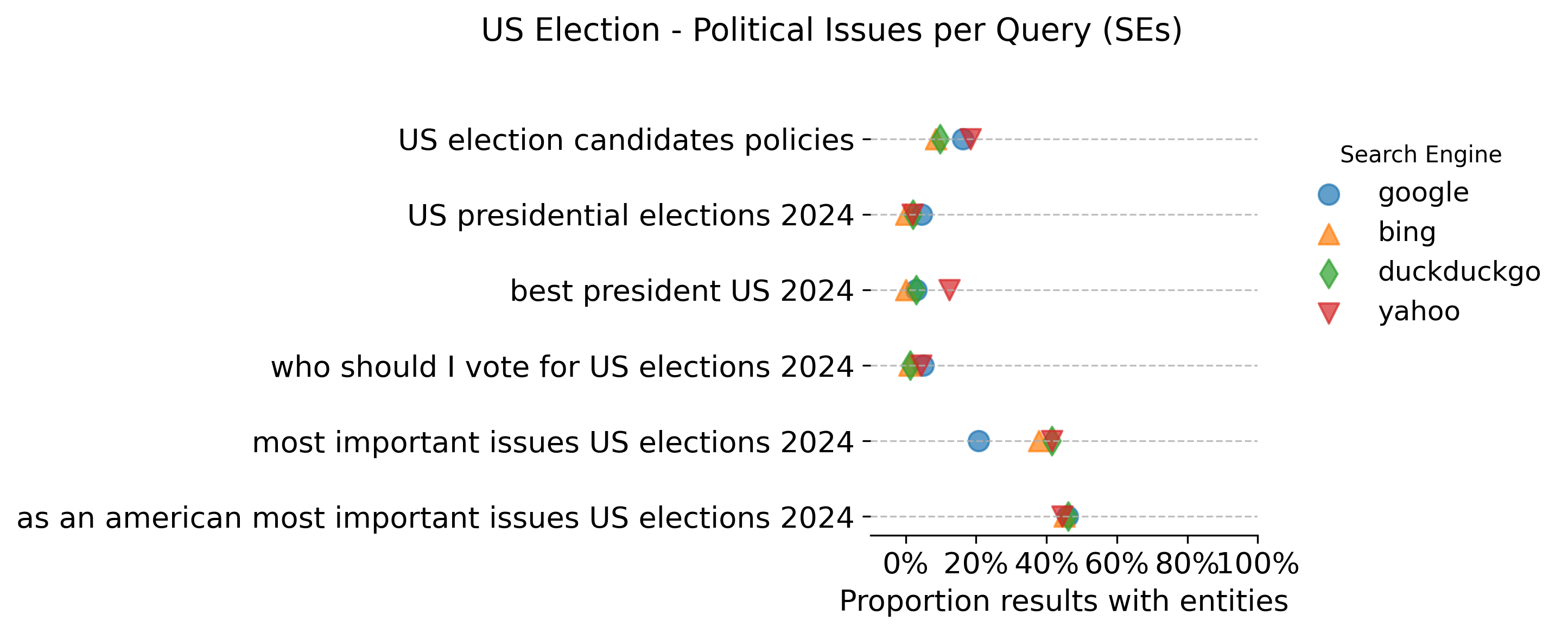}
    \caption{Proportion of SE results containing mentions to relevant political issues, by query (US Presidential Election case study).}
    \label{sup_fig: us_issues_proportion_query}
\end{figure*}

\begin{figure*}[ht!]
    \centering
    \includegraphics[width=1\textwidth]{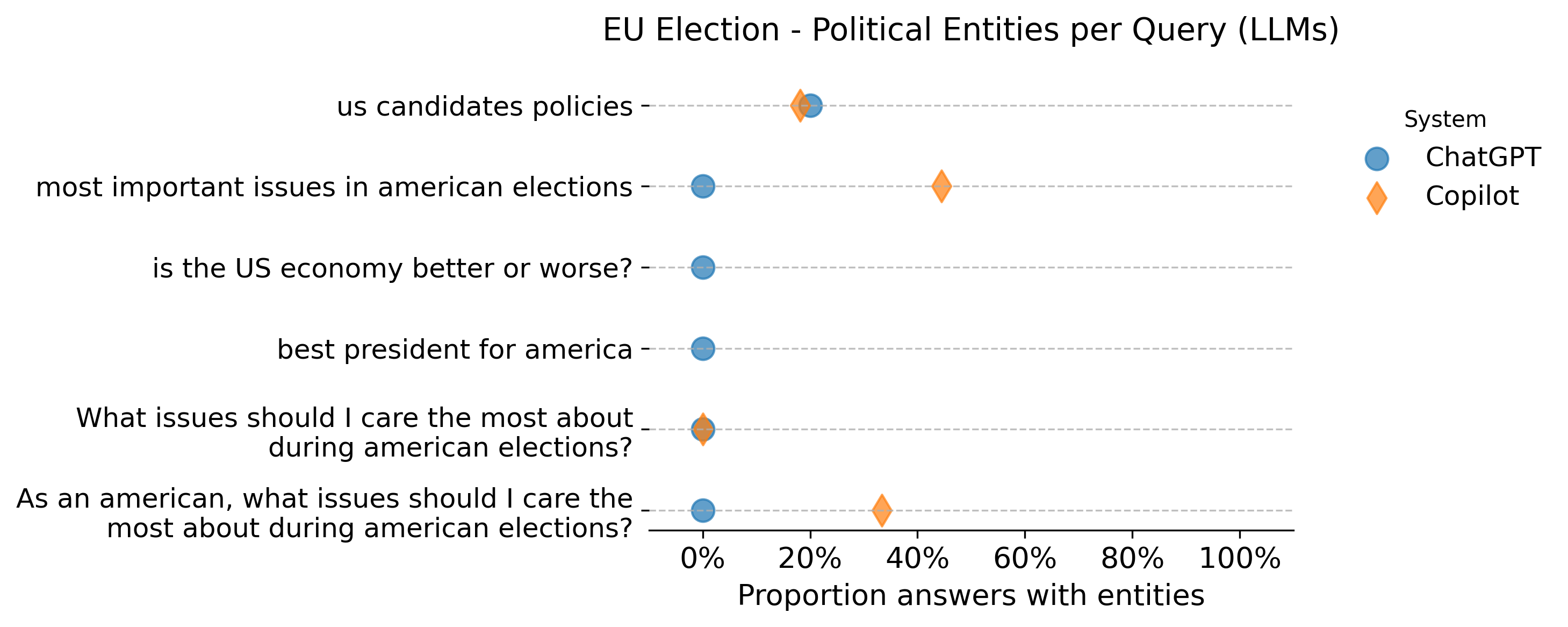}
    \caption{Proportion of LLMs answers containing mentions to relevant political entities, by query (US Presidential Election case study).}
    \label{sup_fig: us_llms_entities_proportion_query}
\end{figure*}

\begin{figure*}[ht!]
    \centering
    \includegraphics[width=1\textwidth]{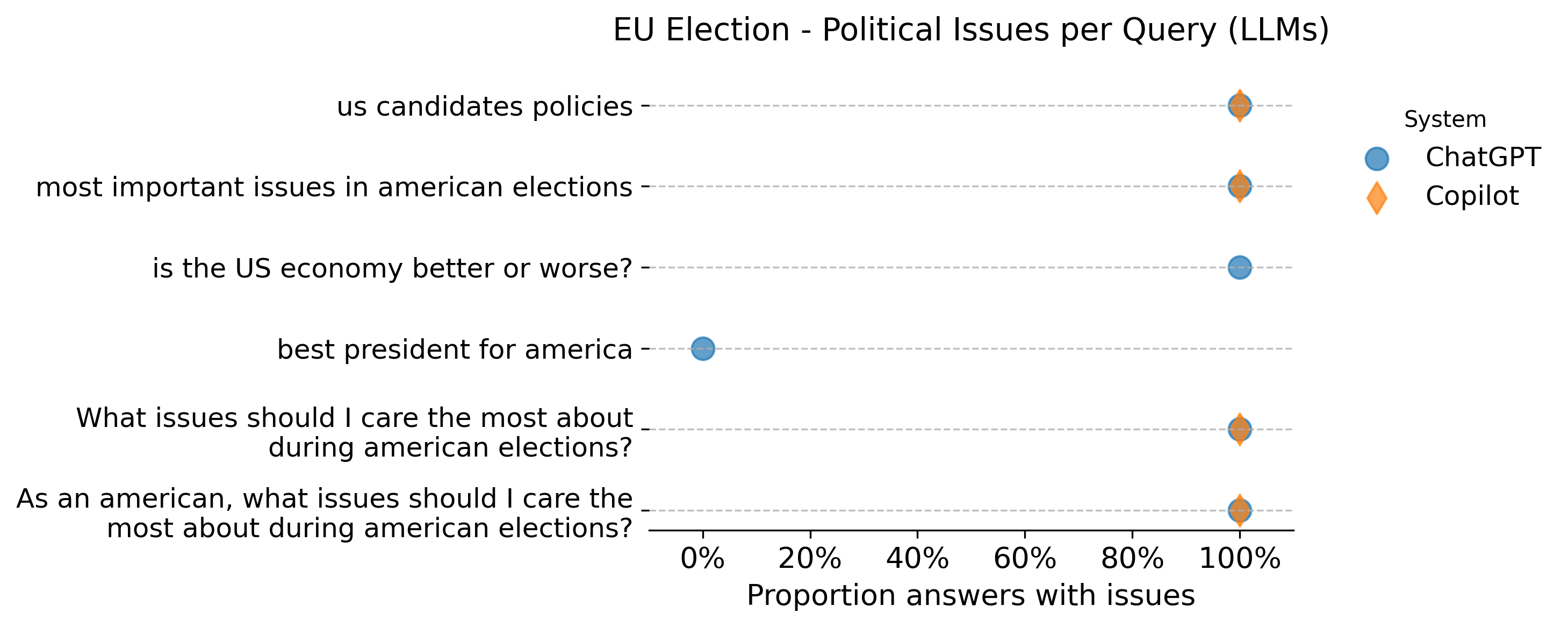}
    \caption{Proportion of LLMs answers containing mentions to relevant political issues, by query (US Presidential Election case study).}
    \label{sup_fig: us_llms_issues_proportion_query}
\end{figure*}
\clearpage

\subsection{Categories of results retrieved by SEs}\label{app_sec:se_results_categories}
\begin{table}[ht!]
\caption{\textbf{Percentage of entities per website category:} Percentage of results by website category containing political entities and political issues (US only) for all search engines and both elections.}
\label{tab:proportion_per_website_category_with_entities}
\small
\centering
\begin{tabular}{l l r l r l r}
\toprule
& \multicolumn{2}{c}{\cellcolor{gray!20}EU Parliament Election} & 
\multicolumn{4}{c}{\cellcolor{gray!20}US Presidential Election} \\
\cmidrule(lr){2-3} \cmidrule(lr){4-7}
& Website Category & \% Entities &
Website Category & \% Entities &
Website Category & \% Issues \\
\midrule
\multirow{7}{*}{Google}
& Corporate & 64.7 & Entertainment & 59.5 & - & - \\
& \makecell[l]{Government/\\Institutions} & 0.2 & \makecell[l]{Media\\Publications} & 60.3 & \makecell[l]{Government/\\Institutions} & 71.0 \\
& News & 30.1 & News & 74.8 & News & 10.5 \\
& Political & 92.3 & \makecell[l]{Non-Profit/\\NGOs} & 3.8 & \makecell[l]{Non-Profit/\\NGOs} & 1.1 \\
& Reference & 9.3 & Reference & 72.0 & - & - \\
& \makecell[l]{Science/\\Academic} & 76.4 & \makecell[l]{Science/\\Academic} & 8.7 & \makecell[l]{Science/\\Academic} & 45.9 \\
& Top News & 42.2 & Top News & 72.5 & Top News & 10.7 \\
\midrule
\multirow{5}{*}{Bing}
& \makecell[l]{Government/\\Institutions} & 0.3 & \makecell[l]{Media\\Publications} & 11.5 & \makecell[l]{Media\\Publications} & 1.9 \\
& News & 29.6 & News & 78.2 & News & 14.4 \\
& Political & 100.0 & \makecell[l]{Non-Profit/\\NGOs} & 4.4 & \makecell[l]{Non-Profit/\\NGOs} & 17.5 \\
& Top News & 12.7 & Top News & 25.6 & Top News & 4.4 \\
& - & - & Reference & 70.2 & Reference & 0.3 \\
& - & - & - & - & \makecell[l]{Science/\\Academic} & 71.7 \\
\midrule
\multirow{6}{*}{DuckGo}
& \makecell[l]{Government/\\Institutions} & 0.1 & \makecell[l]{Media\\Publications} & 10.3 & \makecell[l]{Media\\Publications} & 2.6 \\
& News & 29.7 & News & 77.4 & News & 14.6 \\
& Political & 100.0 & \makecell[l]{Non-Profit/\\NGOs} & 3.5 & \makecell[l]{Non-Profit/\\NGOs} & 18.2 \\
& Reference & 0.4 & Reference & 73.2 & Reference & 0.4 \\
& Top News & 31.7 & Top News & 31.4 & Top News & 35.5 \\
& - & - & - & - & \makecell[l]{Science/\\Academic} & 66.0 \\
\midrule
\multirow{4}{*}{Yahoo}
& News & 21.5 & Reference & 100.0 & News & 20.3 \\
& Top News & 49.2 & Top News & 44.1 & Top News & 26.0 \\
& - & - & - & - & \makecell[l]{Non-Profit/\\NGOs} & 10.4 \\
& - & - & - & - & \makecell[l]{Science/\\Academic} & 52.2 \\
\bottomrule
\end{tabular}
\end{table}

\begin{figure*}[ht!]
    \centering
    \includegraphics[width=1\textwidth]{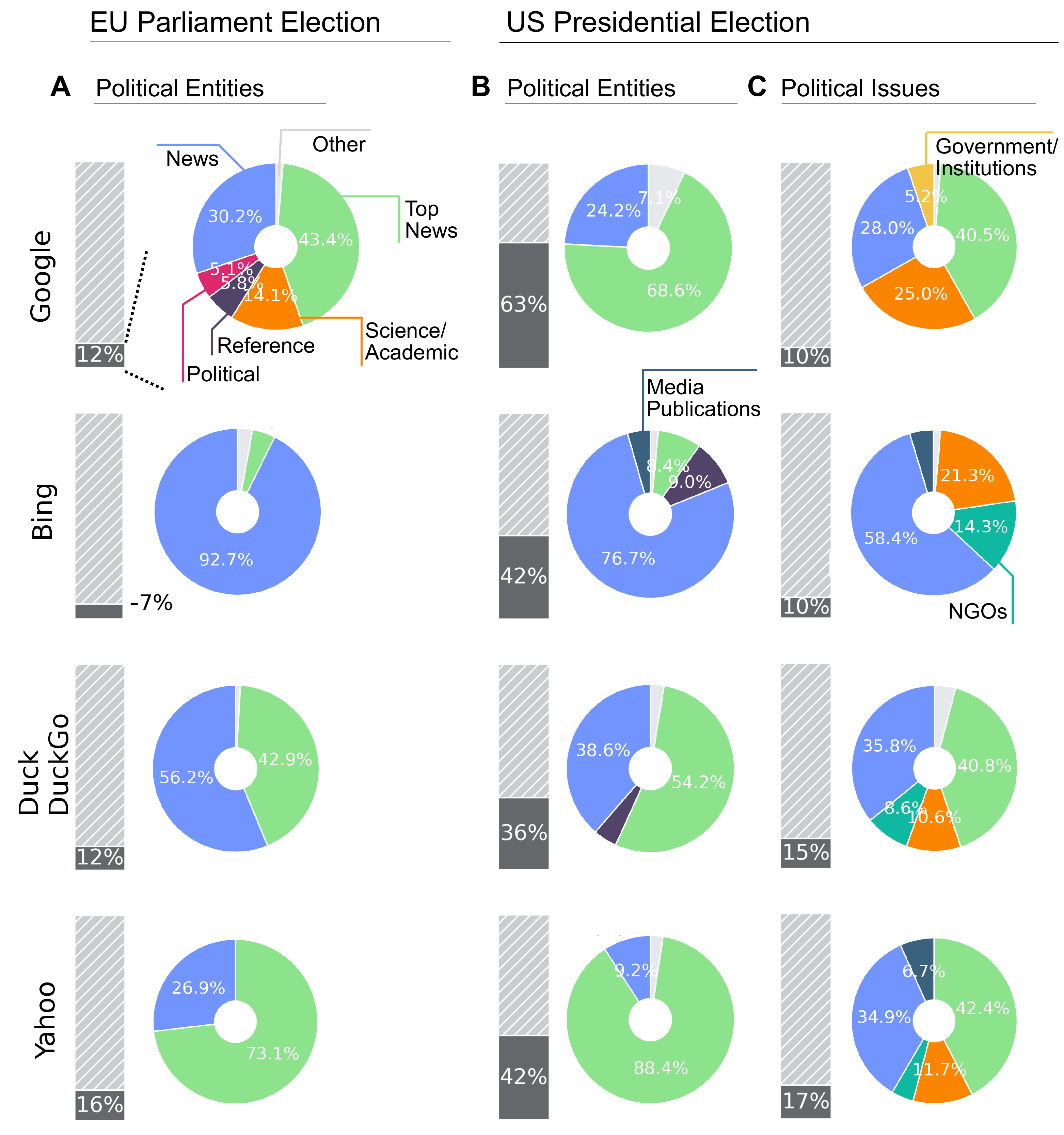}
    \caption{\textbf{Proportion of results with entities per websites category.} Bars represent the proportion of results that mentioned relevant political entities in their headlines or URLs (dark grey) versus those that did not (light grey). Pie charts show the distribution of website categories among the results that included political entity mentions. Panel A) and B) show the results regarding political entities (politicians, parties, political groups) for the EU Parliament Election and US Presidential Election, respectively. Panel C) shows the distribution of websites categories for the results that displayed mentions of political relevant issues for the US Presidential Election.} 
    \label{fig:category_websites_with_entities}
\end{figure*}

\clearpage
\subsection{Differences of SE results across bots}\label{app_sec:se_differences}

\begin{table*}[ht!]
\caption{Average Jaccard Index (Standard Deviation) comparing sets of domains and website categories retrieved by bots located in different regions. EU results are separated by query language (English vs. local language)}
\label{tab:jaccard_indexes}
\small
\centering
\begin{tabular}{llccccc}
\toprule
\textbf{Search Engine} & \textbf{Language} & \multicolumn{2}{c}{\textbf{EU Parliament Election}} & & \multicolumn{2}{c}{\textbf{US Presidential Election}} \\
\cmidrule{3-4} \cmidrule{6-7}
& & Domains & Categories & & Domains & Categories \\
\midrule

\textbf{Google} & English        & 0.58 (0.21) & 0.76 (0.20) & & 0.82 (0.15) & 0.91 (0.13) \\
                & Local Language & 0.11 (0.18) & 0.51 (0.21) & & –            & –            \\

\textbf{Bing}   & English        & 0.61 (0.20) & 0.80 (0.21) & & 0.70 (0.18) & 0.84 (0.18) \\
                & Local Language & 0.13 (0.18) & 0.57 (0.25) & & –            & –            \\

\textbf{DuckDuckGo} & English        & 0.67 (0.20) & 0.81 (0.19) & & 0.72 (0.18) & 0.86 (0.17) \\
                    & Local Language & 0.15 (0.23) & 0.58 (0.25) & & –            & –            \\

\textbf{Yahoo}  & English        & 0.80 (0.26) & 0.87 (0.19) & & 0.94 (0.13) & 0.96 (0.12) \\
                & Local Language & 0.17 (0.26) & 0.57 (0.28) & & –            & –            \\

\bottomrule
\end{tabular}
\end{table*}

\begin{figure*}[ht!]
    \centering
    \includegraphics[width=1\textwidth]{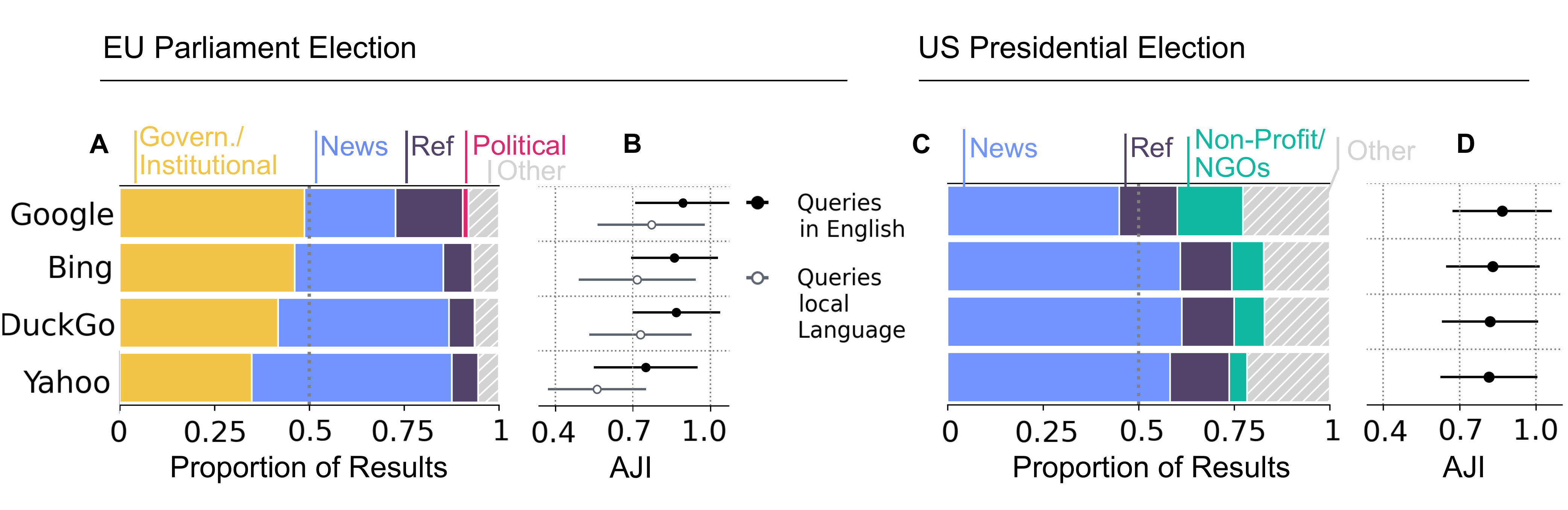}
    \caption{\textbf{Search Engine Results Classified by Domain Type.} \textbf{(A and C)} Proportion of results belonging to the top-three most frequent domain categories, plus the share of results classified under the \textit{Political} category (i.e., websites affiliated with political parties) for the EU (A) and the USA (C). \textbf{(B and D)} Average Jaccard Index comparing the sets of result categories returned to bots located in different locations (countries - B, or counties - D) for the exact same query. Higher values indicate greater similarity in the types of results seen across locations. The error bars represent the standard deviation.}
    \label{fig:categories_plots}
\end{figure*}

\clearpage
\subsection{Top 3 Results Analysis}\label{app_sec:top_3}

\begin{figure*}[ht!]
    \centering
    \includegraphics[width=1\textwidth]{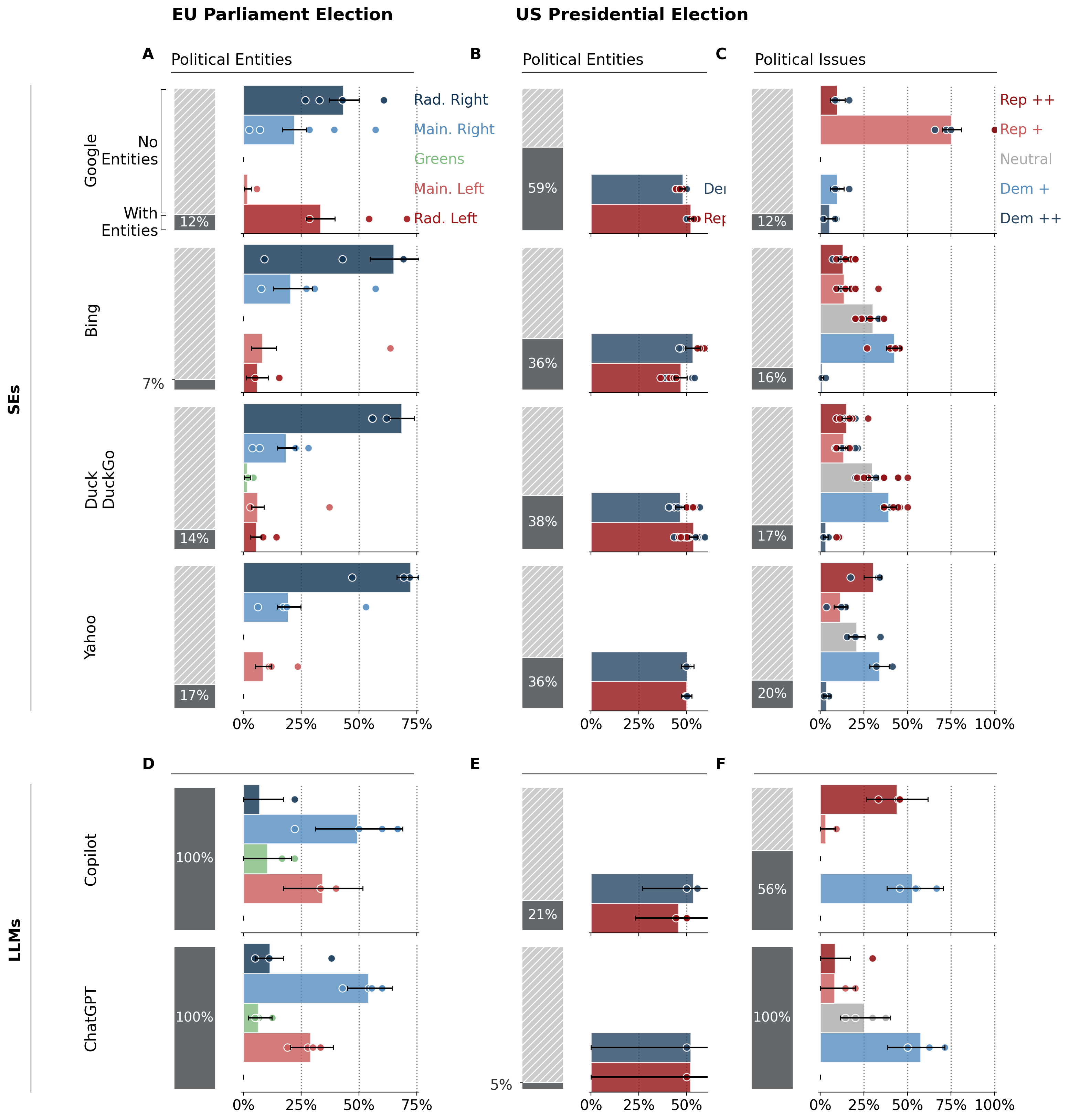}
    \caption{Gray bars show the proportion of Top 3 SE results (A, B, C) containing relevant
mentions (political entities or issues). For Top 3 results with entities, the plots to the right show the average proportion by political leaning: in A and D (EU Parliament) each dot represents a country; in B (US
Presidential) and C (US topics) each dot represents a county, colored blue (Democrat-leaning) or red
(Republican-leaning). Legend: Rep ++ = much more important for Republicans; Rep + = slightly more
important for Republicans; Dem + = slightly more important for Democrats; Dem ++ = much more
important for Democrats. Error bars show standard errors.} 
    \label{fig:proportion_leaning_top_3}
\end{figure*}

Analysis of the search engine results presented in Figure~\ref{fig:leaning_results}, Figure~\ref{fig:proportion_leaning_top_3} focusing exclusively in the top three results.

\clearpage
\subsection{Results Leaning}\label{app_sec:se_leaning}

\subsubsection{Percentages of results and leaning classifications}

\begin{figure*}[ht!]
    \centering
    \includegraphics[width=1\textwidth]{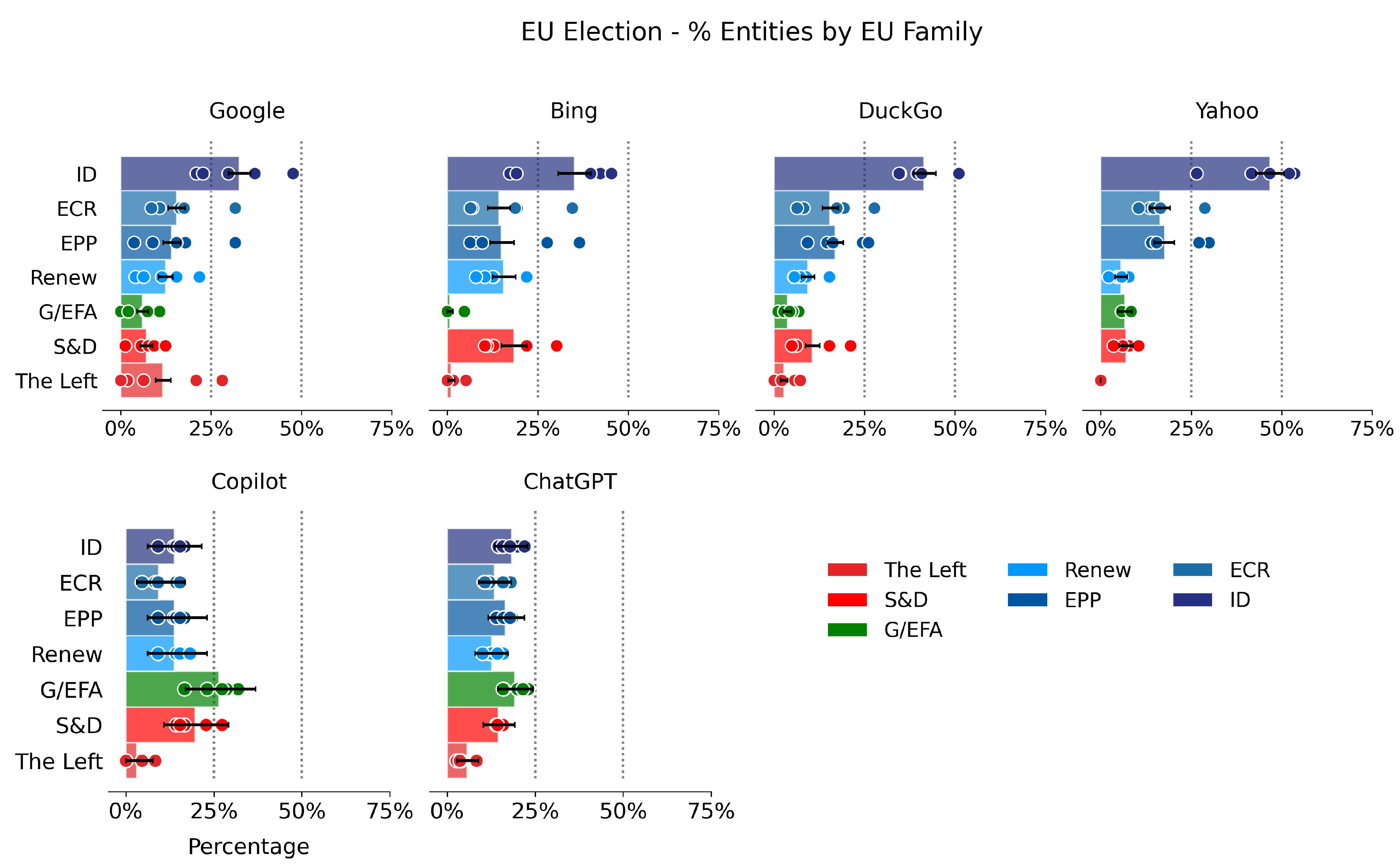}
    \caption{Percentage of relevant political entities mentioned in search engine (SE) and large language model (LLM) outputs on the European elections, grouped by European political family. Political families are classified according to their composition in the 2019 European Parliament; references to European Sovereign Nations (ESN) and Patriots are therefore included within the Identity and Democracy (ID) group. Bars display bootstrapped means across bots (N = 1,000), with error bars indicating 95\% confidence intervals. Dots represent the mean percentage across bots within the same location.} 
    \label{fig:distribution_by_eu_family}
\end{figure*}

\clearpage

\begin{table}[ht!]
\small
\centering
\begin{tabular}{lllr}
\toprule
Search Engine & Classification & Topic & \% \\
\midrule

Google 
& Rep ++ & Immigration & 4.8 \\ \cline{2-4}
& Rep +  & Economy & 52.0 \\
&        & Foreign policy & 16.2 \\
&        & Taxes and government spending & 12.7 \\
&        & Violent crime & 1.3 \\ \cline{2-4}
& Dem +  & Abortion & 4.8 \\
&        & Health care & 0.5 \\ \cline{2-4}
& Dem ++ & Civil rights and civil liberties & 7.4 \\

\midrule
Bing
& Rep ++ & Immigration & 9.1 \\ \cline{2-4}
& Rep +  & Economy & 14.6 \\
&        & Taxes and government spending & 9.3 \\
&        & Violent crime & 4.5 \\ \cline{2-4}
& Neutral & Gun policy & 12.3 \\ \cline{2-4}
& Dem +  & Abortion & 23.4 \\
&        & Climate change and the environment & 12.3 \\ 
&        & Health care & 8.7 \\ \cline{2-4}
& Dem ++ & Civil rights and civil liberties & 5.6 \\

\midrule
DuckGo
& Rep ++ & Immigration & 12.5 \\ \cline{2-4}
& Rep +  & Economy & 13.4 \\
&        & Foreign policy & 1.0 \\
&        & Taxes and government spending & 8.5 \\
& Neutral & Gun policy & 12.6 \\ \cline{2-4}
& Dem +  & Abortion & 24.0 \\
&        & Climate change and the environment & 12.6 \\
&        & Health care & 8.5 \\ \cline{2-4}
& Dem ++ & Civil rights and civil liberties & 6.2 \\

\midrule
Yahoo
& Rep ++ & Immigration & 8.8 \\ \cline{2-4}
& Rep +  & Economy & 12.7 \\
&        & Taxes and government spending & 8.7 \\
&        & Violent crime & 3.3 \\ \cline{2-4}
& Neutral & Gun policy & 13.6 \\ \cline{2-4}
& Dem +  & Abortion & 27.4 \\
&        & Climate change and the environment & 13.6 \\
&        & Health care & 8.9 \\ \cline{2-4}
& Dem ++ & Civil rights and civil liberties & 3.3 \\

\bottomrule
\end{tabular}
\caption{Proportion of results by topic and partisan classification across search engines. Classification labels: Rep ++ (much more important for Republicans), Rep + (slightly more important for Republicans), Neutral (equally important), Dem + (slightly more important for Democrats), Dem ++ (much more important for Democrats).}
\end{table}

\begin{table}[ht!]
\small
\centering
\begin{tabular}{lllr}
\toprule
LLM & Classification & Topic & \% \\
\midrule

ChatGPT
& Rep ++ & Immigration & 10.5 \\ 
\cline{2-4}
& Rep +  & Economy & 15.0 \\
&        & Foreign policy & 9.8 \\
&        & Taxes and government spending & 3.3 \\
\cline{2-4}
& Neutral & Education & 11.1 \\
&         & Gun policy & 9.2 \\
&         & Supreme Court appointments & 1.3 \\
&         & Terrorism & 0.7 \\
\cline{2-4}
& Dem +  & Abortion & 0.7 \\
&        & Climate change and the environment & 13.1 \\
&        & Health care & 13.1 \\
&        & Racial and ethnic inequality & 7.2 \\ 
\cline{2-4}
& Dem ++ & Civil rights and civil liberties & 5.2 \\

\midrule
Copilot
& Rep ++ & Immigration & 17.1 \\ 
\cline{2-4}
& Rep +  & Economy & 24.3 \\
&        & Foreign policy & 2.7 \\
&        & Taxes and government spending & 0.9 \\
&        & Violent crime & 7.2 \\ 
\cline{2-4}
& Neutral & Gun policy & 1.8 \\
&         & Supreme Court appointments & 6.3 \\ 
\cline{2-4}
& Dem +  & Abortion & 18.0 \\
&        & Climate change and the environment & 5.4 \\
&        & Health care & 16.2 \\ 

\bottomrule
\end{tabular}
\caption{Proportion of LLM mentions in answers by issue and respective main partisan classification for LLMs. Classification labels: Rep ++ (much more important for Republicans), Rep + (slightly more important for Republicans), Neutral (equally important), Dem + (slightly more important for Democrats), Dem ++ (much more important for Democrats). Percentages are calculated over all outputs per model (ChatGPT: $N=153$; Copilot: $N=111$).}
\end{table}

\clearpage
\subsubsection{Leanings distributions per location}

\begin{figure*}[!ht]
    \centering
    \includegraphics[width=1\textwidth]{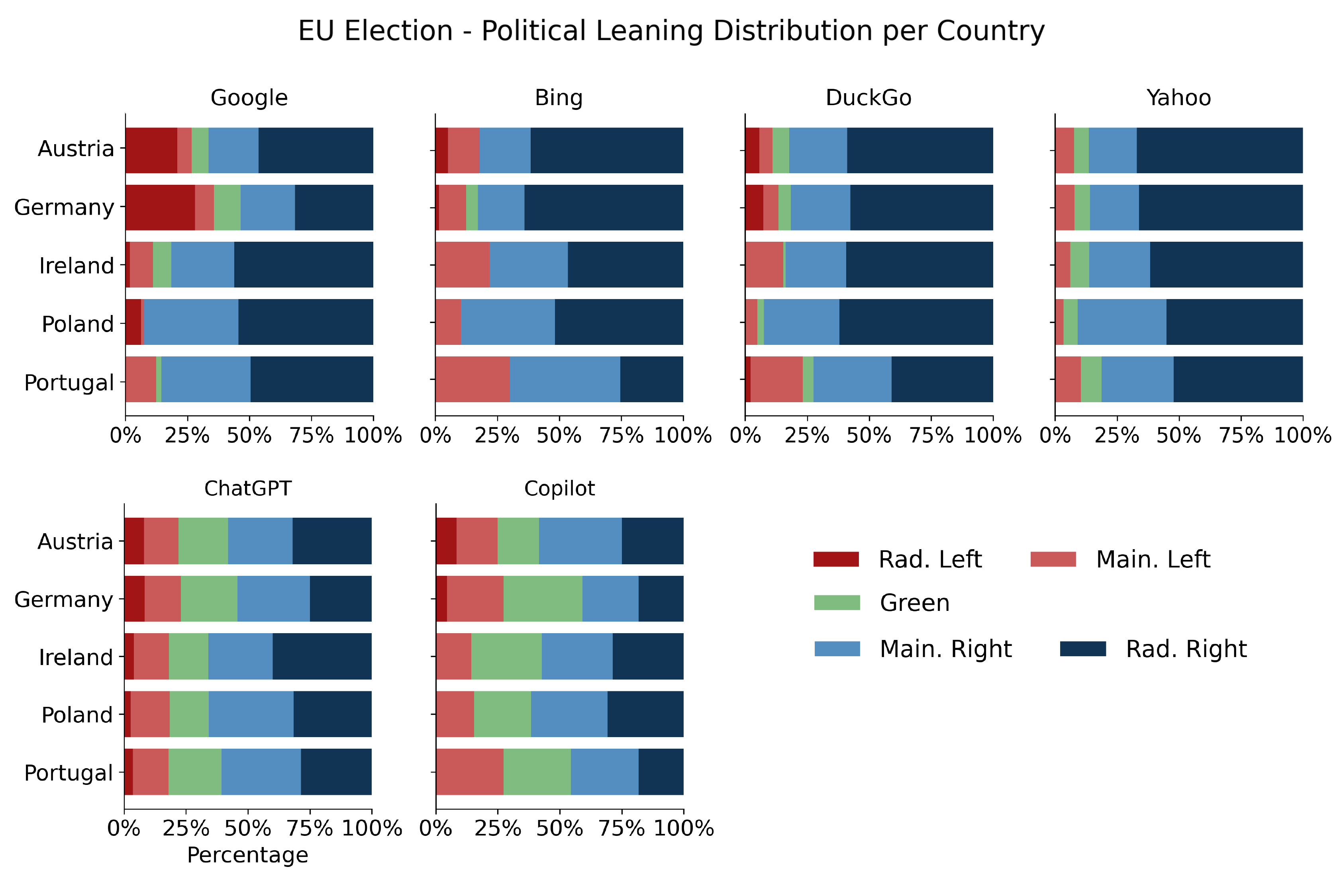}
    \caption{Proportion of results with different leanings for the different locations.} 
    \label{fig:eu_results_by_country}
\end{figure*}

\clearpage
\subsubsection{Leanings distributions per query}

The different leanings of results for each query performed for the European Parliament election case study.

\begin{figure*}[ht!]
    \centering
    \includegraphics[width=1\textwidth]{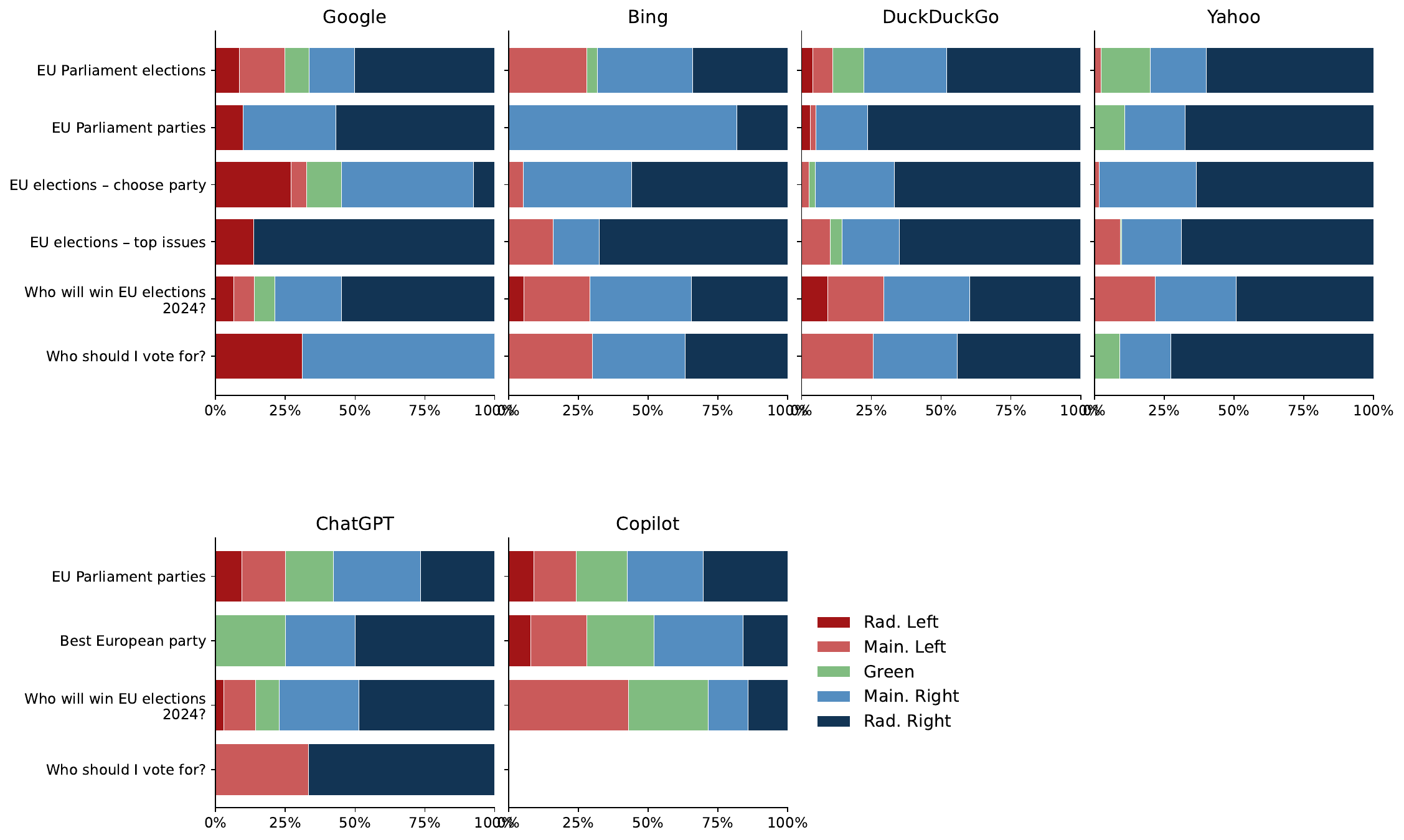}
    \caption{Distribution of political leanings in retrieved results by SEs and LLMs about the EU elections, by query and source.}
    \label{sup_fig:eu_results_per_query}
\end{figure*}

\begin{figure*}[ht!]
    \centering
    \includegraphics[width=0.85\textwidth]{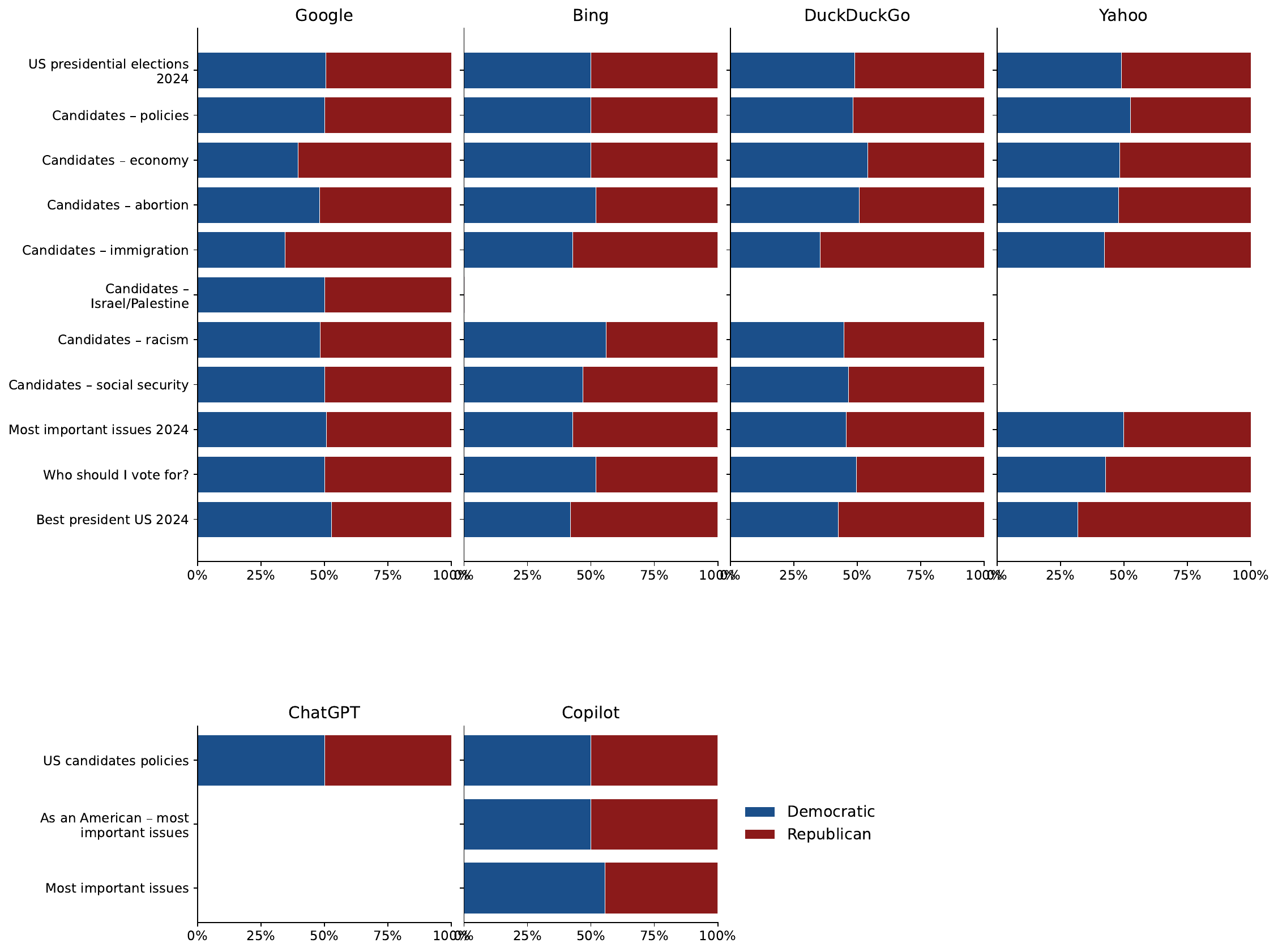}
    \caption{Distribution of political leanings in retrieved results by SEs and LLMs in the EU case study, by query and source.}
    \label{sup_fig:us_entities_results_per_query}
\end{figure*}

\begin{figure*}[ht!]
    \centering
    \includegraphics[width=0.85\textwidth]{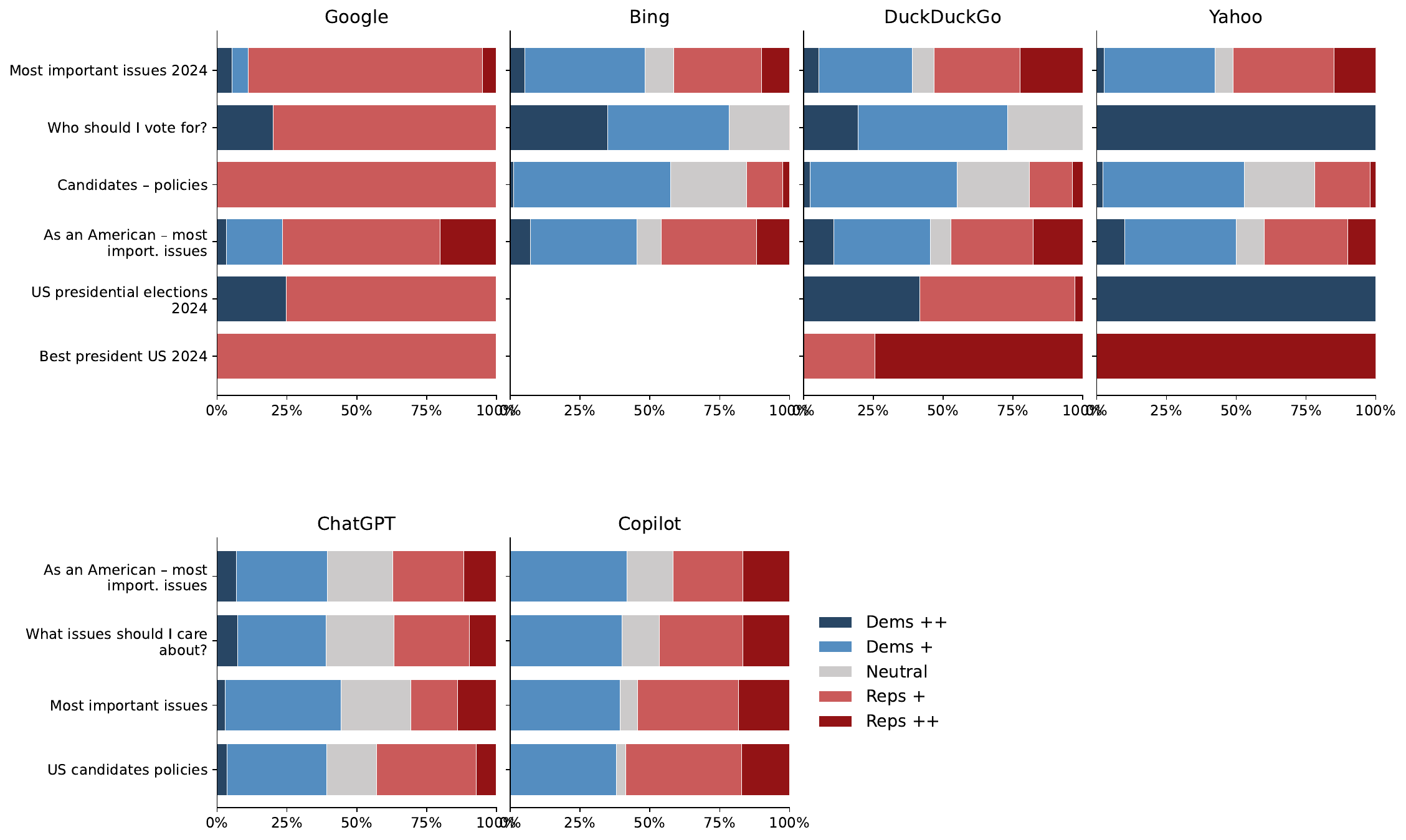}
    \caption{Distribution of political issues by category in retrieved results by SEs and LLMs in the EU case study, by query and source.}
    \label{sup_fig:us_issues_results_per_query}
\end{figure*}

\subsubsection{Categories of SE results across leaning classifications}

\begin{figure*}[htbp]
    \centering
    \includegraphics[width=1\textwidth]{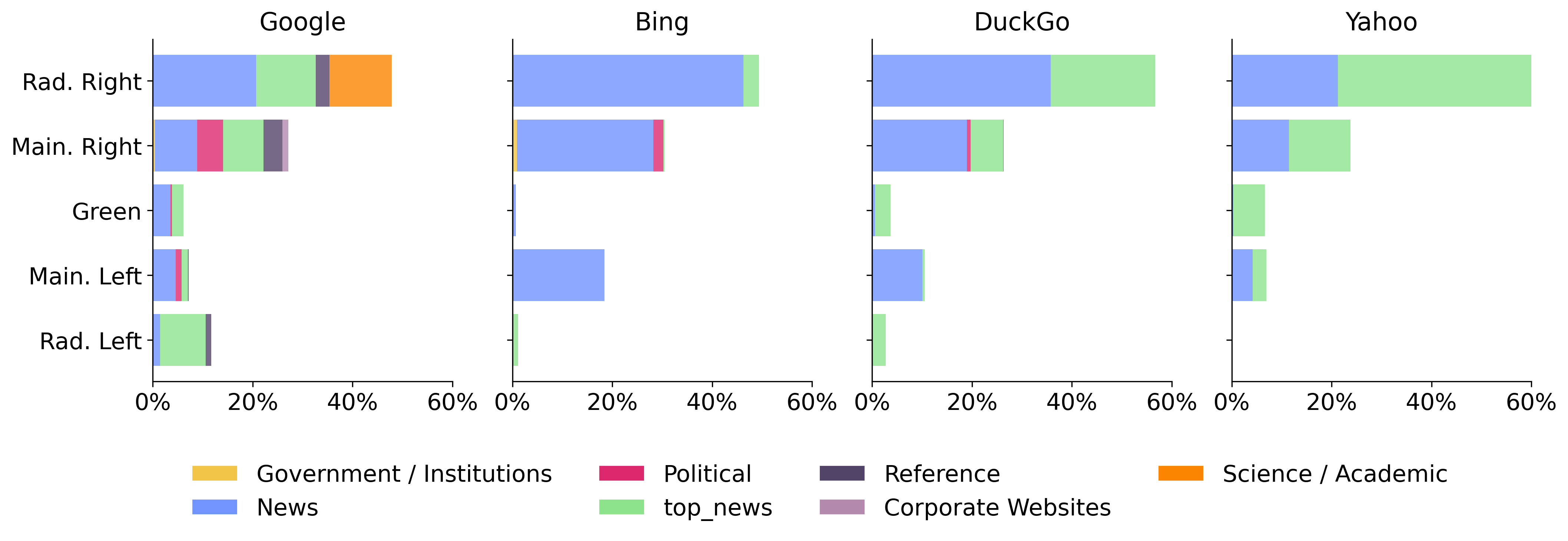}
    \caption{Proportion of results by category (colors) per political leaning in SE results.}
    \label{sup_fig:type_results_per_leaning}
\end{figure*}

\clearpage
\section{Control Analysis: Removing Economy from political issues}\label{app_sec:control_1}

\begin{figure*}[!ht]
    \centering
    \includegraphics[width=1\textwidth]{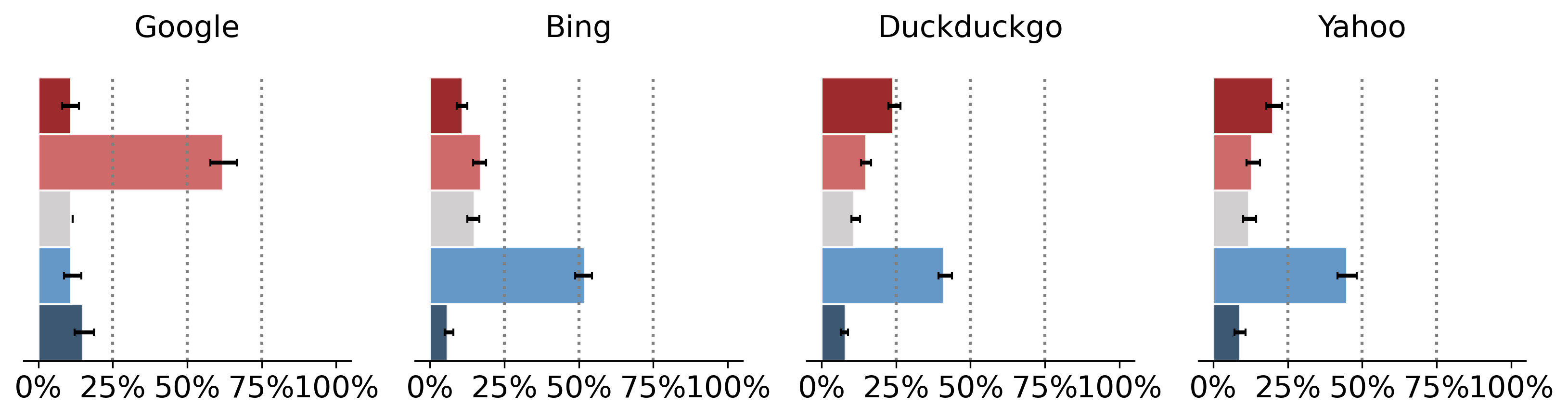}
    \caption{Proportion of results containing relevant political issues, grouped by topic leaning. The issue economy was excluded from this plot, as it is considered important to all voters, despite being more strongly associated with Republicans.} 
    \label{fig:no_economy}
\end{figure*}

\clearpage
\section{Bias of results against external sources}\label{app_sec:bias_external}

\begin{figure*}[ht!]
    \centering
    \includegraphics[width=0.83
    \textwidth]{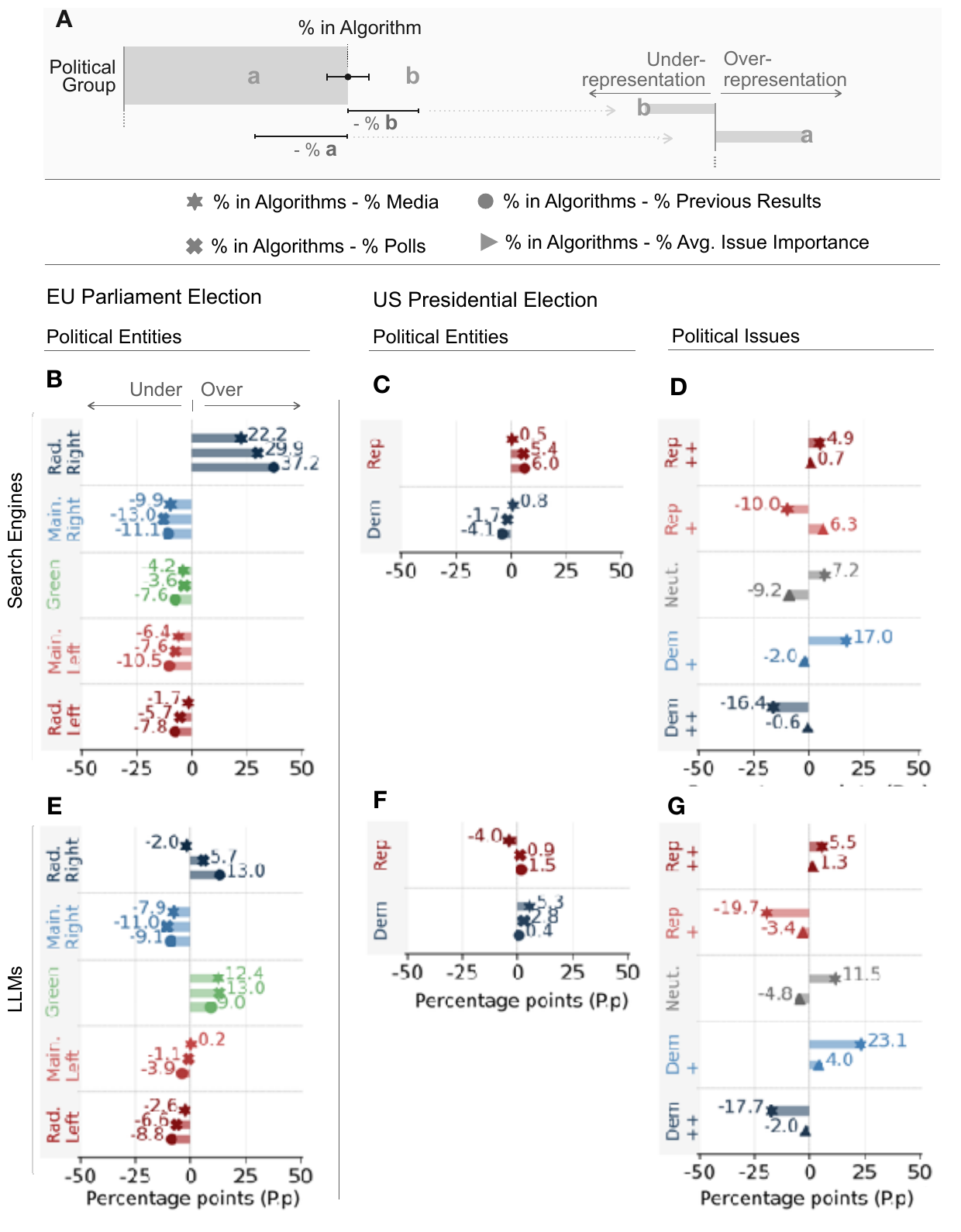}
    \caption{\textbf{Comparison with external factors. (A)} Explanatory scheme of the differences displayed in (B - G). Star = \% in Algorithms (SEs or LLMs) - \% Average attention in Media; Circle = \% in Algorithms - \% Previous Results; Cross = \% Algorithms - \% Polls; Triangle = \% Algorithms - \% Average importance of the Issue (US only), according to surveys~\cite{pew2024issues,yougov2025issues}. Positive values indicate overrepresentation by the algorithms, while negative values indicate underrepresentation. \textbf{(B - D)} Average SE attention to EU and US political leanings vs. Media, Previous results and polls. {(E - G)} Average LLMs attention to EU and US political leanings and issues compared with the same external factors.}
    \label{fig:comparison_saliences_all}
\end{figure*}

\clearpage
\section{Prompts provided to ChatGPT}\label{app_sec:prompts}

\small
\centering
\begin{longtable}{|
    p{2.2cm}|
    p{2cm}|
    >{\RaggedRight\arraybackslash}p{11cm}|
}
\caption{Prompt configurations used across stages and election cases with OpenAI API.}
\label{tab:prompts} \\

\hline
\textbf{Stage} & \textbf{Election Case} & \textbf{Prompt} \\
\hline \hline
\endfirsthead

\hline
\textbf{Stage} & \textbf{Election Case} & \textbf{Prompt} \\
\hline \hline
\endhead
\hline

\hline
\multicolumn{3}{r}{\textit{Continued on next page}} \\
\endfoot

\hline
\endlastfoot
\parbox[t]{2.2cm}{Data\\Collection} &
Both &
You will receive a question and your goal is to respond as you normally do when someone chats with you.
Output a JSON containing:
\{
``response'': string — String representing your response
\}
\\
\hline

\multirow{3}{*}{\parbox[t]{2cm}{Entity \\ extraction \\ and \\ identification \\ from title \\ and URL}} &
\parbox[t]{2cm}{EU \\ Parliament \\ Election} & \parbox[t]{11cm}{
Your task is to identify and extract political entities from a pair of title and/or URL in the context of the European Parliament Elections of 2024.
Specifically, you must identify named entities, such as political parties or politicians, and identify any references to political leaning or ideological spectrum, such as ``right'', ``left'', ``far-right'', etc.
Do not consider official institutions (e.g., European Parliament) as political entities.
Once identified, for any named entity (a politician or party), return the official English name of the party and the country it belongs to (e.g., ``Christian Democratic Union of Germany -- Germany'' or ``Left Bloc -- Portugal'').
If the entity is not explicitly named but is clearly identifiable (e.g., ``the current right-wing president of Brazil''), include the inferred party and country.
The output should be a JSON object with the following structure:
\{
``entity'': ``string'',
``party\_and\_country'': ``string'' (use ``not applicable'' if the type is ``spectrum'')
\}}
\\
\cline{2-3}
& \parbox[t]{2cm}{US \\ Presidential \\ Election: \\ political \\ entities} & \parbox[t]{11cm}{Identify and return the political parties, politicians, and other mentions to the political leaning of the mentioned agents or content (e.g., "extreme-right" or "socialists") in each title and URL. The titles and URLs provided are more likely to come from news written in English, since correspond to titles and URLs of the newspapers with more visualizations in the USA. The entities to be found should be related to the political environment in the USA. The title and URLs are from news from April 9th until June 9th 2024, so consider political information, such as the name of the prime minister, from that period. The result should be given as a JSON object with 2 keys: "parties" of type dict with keys corresponding to the direct transcription of the found parties and values corresponding to the respective official name if acronym returned for instance, or null if not possible to infer anything, e.g. \{"democrats": "Democratic Party - USA"\} and "politicians\_othermentions" of type dict with key corresponding to the direct transcription of the found politicians and other mentions and the value corresponding with their associated party, e.g.: \{"president of the US":"Democratic party - USA"\}.} \\ \hline

\parbox[t]{2cm}{Entity \\ extraction \\ and \\ identification \\ from title \\ and URL} & \parbox[t]{2cm}{US \\ Presidential \\ Election: \\ political \\ issues} & \parbox[t]{11cm}{Goal: Your task is to identify whether any of the topics/issues listed below are mentioned or implied in a news title and/or URL, within the context of the 2024 U.S. elections. You should evaluate the presence of topics as if you were a voter in the elections. Topics: ['Abortion', 'Economy', 'Health care', 'Supreme Court appointments', 'Foreign policy', 'Violent crime', 'Immigration', 'Gun policy', 'Terrorism', 'Taxes and Government Spending', 'Social Security', 'Climate change and the environment', 'Education', 'Racial and ethnic inequality', 'Civil rights and civil liberties']. Input: You will receive a news title and a news URL. Instructions: Check both the title and URL for mentions of the topics/issues. You may need to clean the URL (e.g., remove hyphens, special characters) to better identify topics. A topic does not need to be mentioned verbatim — infer the relevance of a topic based on context, synonyms, or related terms. If the same topic is mentioned in both the title and URL, count it only once. You are only evaluating based on the title and URL — do not imagine extra context. For the Supreme Court appointments topic, be sure to only select it if it's about appointing judges to the Supreme Court, not just anything related to the Supreme Court. For Foreign Policy, select any article that is about political relations of other countries with the US, or that could impact US Foreign Policy, not just any article mentioning other countries. Use the Racial and ethnic inequality only for articles about race and/or ethnicity (directly or indirectly), not other minorities or groups suffering prejudice. In the Civil Rights topic, be particularly careful to detect mentions of vote fraud and vote suppression, but without neglecting more general instances. For violent crime, consider any mentions that would cause concern to the public (e.g. attempted crime as well as acts committed). Output format: You should return a JSON object in the following format: \{
  ""topics"": list // List containing all topics/issues mentioned directly in the title or url or that one can infer the news is about the topic.
\}"} \\ \hline
\parbox[t]{2cm}{Analysis: \\ Match \\ spectrum to \\ politician(s)} & \parbox[t]{2cm}{EU \\ Parliament \\ Election} & \parbox[t]{11cm}{"You will receive a news article title that includes a reference to a political leaning such as ""Right"", ""Left"", or others (e.g., ""Hard Right""), along with the full corpus (text) of the article. Typically, when a title includes such political spectrum terms, it is implicitly referring to a specific political party, group of parties, or European political family that aligns with that leaning.
Your task: Identify exactly which political party, group of parties, or European political family the leaning mentioned in the title refers to, based on your interpretation of the article's content.
Additional Instructions:
Focus only on identifying the entity (party, parties, or political family) that the leaning mentioned in the title refers to, based on the corpus.
Ignore any other political parties or politicians that are explicitly named in the article but not related to the leaning in the title. Example: If the title is ""Macron loses to far-right"", ignore Macron. Your task is to identify which party or parties the term ""far-right"" refers to, using the article’s main content.
All articles are about the 2024 European Parliament Elections.
If the article discusses the leaning only in abstract terms (e.g., ""The Right is gaining support across Europe"") without clearly linking it to a specific party or group, return ""not applicable"".
Return your answer as a JSON object in the following structure:
\{
""party\_and\_country"": [""string""] // A list of the official English names of the party or parties referred to, along with the country (e.g., ""Brothers of Italy - Italy""). // If no specific party or group is identified in the article, return [""not applicable""]\}"} \\ \hline 

\parbox[t]{2cm}{Media \\ Analysis \\ (if prompt \\ was \\ different)} & \parbox[t]{2cm}{EU \\ Parliament \\ Election} & \parbox[t]{11cm}{"Identify and return the political parties, politicians, and other mentions to the political leaning of the mentioned agents or content (e.g., ""extreme-right"" or ""socialists"") in each title and URL.

The titles and URLs provided are more likely to come from news written in German or in English, since correspond to titles and URLs of the newspapers with more visualizations in Austria. The entities to be found should be related to the Austrian political environment. The title and URLs are from news from April 9th until June 9th 2024, so consider political information, such as the name of the prime minister, from that period.

The result should be given as a JSON object with 3 keys: ""title"" of type string with the value in the title column in the provided table, ""url"" of type string with the value in the url column in the provided table, ""parties"" of type dict with keys corresponding to the direct transcription of the found parties and values corresponding to the respective official name if acronym returned for instance, or null if not possible to infer anything, e.g. {""ÖVP"": ""Austrian People's Party - Austria""}) and ""politicians\_othermentions"" of type dict with key corresponding to the direct transcription of the found politicians and other mentions and the value corresponding with their associated party, e.g.: \{""Chancellor of Austria"": ""Austrian People's Party - Austria""\}."} \\ \hline

\parbox[t]{2cm}{Analysis:\\Website Classification} & \parbox[t]{2cm}{Both} & \parbox[t]{11cm}{You will receive a list of website categories and a list of website domains. I would like to have your help to classify all those domains according to their category (type), and also according to their topic. \\These are the categories and their definition: \\1. News (Official News Organizations): Websites operated by professional journalism organizations focused on delivering current events, investigations, and reporting. Includes: National and local newspapers (e.g., The New York Times, The Guardian); TV or radio news outlets (e.g., CNN, BBC News, NPR News) ; Wire services (e.g., Reuters, Associated Press); Exclude: Topic-driven magazines like Forbes, Vogue (use “Media Publications” instead).\\
2. Media Publications (Magazines, Niche Publications): Websites of magazines, digital publications, or professional blogs focused on a specific topic or lifestyle area (fashion, business, tech, etc.). Includes: Forbes, Vogue, TechCrunch, Scientific American; Entertainment, business, or lifestyle publications; May include reporting, but are not primary news authorities. \\ 3. Reference Definition: Websites designed to provide structured, factual reference information (encyclopedic, dictionary-style, databases). Includes: Wikipedia, Britannica, Merriam-Webster; Knowledge bases, public databases. \\ 4. Science / Academic Definition: Websites operated by scientific institutions, academic publishers, scholarly journals, or professional research communities. Their main purpose is to publish, archive, or disseminate scientific knowledge, including peer-reviewed research and technical standards. Includes: Academic journals and publishers: Nature.com, ScienceDirect, JSTOR, Springer; Professional associations: IEEE.org, ACM.org, AAAS.org; Preprint servers: arXiv.org, bioRxiv.org; University-hosted research portals; Standards bodies and conference sites (e.g., ieee.org/conferences). Distinct From: Educational Platforms (Khan Academy, Coursera) — focused on teaching; Reference (Wikipedia) — broad, non-peer-reviewed knowledge; Media Publications (Scientific American) — popular science journalism, not academic publishing. \\ 5. Political Definition: Websites of political parties, political campaigns, advocacy groups, or explicitly partisan activism. Includes: Democrats.org, Heritage.org, Extinction Rebellion. \\ 6. Government / Institutions Definition: Official websites of local, national, or international government institutions or agencies. Includes: Whitehouse.gov, EUropa.eu, UN.org, IRS.gov \\ 7. Non-Profit / NGOs Definition: Websites run by non-commercial, mission-driven organizations (non-governmental). Includes: Red Cross, WWF, Amnesty International \\ 8. Social Media Platforms Definition: Websites primarily serving as platforms for users to create, share, or interact with content socially. Includes:  Facebook, Twitter/X, TikTok, Instagram; Messaging/social networking platforms.\\ 9. Forums / Discussion Boards Definition: Websites centered around user discussion, Q\&A, or peer-to-peer interaction — focused on content rather than personal profiles. Includes: Quora, Reddit, Stack Overflow, 4chan. Note: If a forum is hosted inside another type (e.g. a corporate forum), consider tagging it secondarily.} \\ \hline

\parbox[t]{2cm}{Analysis:\\Website Classification} & \parbox[t]{2cm}{Both} & \parbox[t]{11cm}{10. Entertainment Services Definition: Websites that provide entertainment content such as streaming, games, or shows. Includes: Netflix, YouTube, Twitch, Spotify; Game streaming or content hosting platforms. \\ 11. E-commerce / Retail Platforms Definition: Websites focused on selling goods or services directly to consumers. Includes: Amazon, Zara, Lidl, Etsy, eBay; Marketplace and brand websites offering direct sales. \\ 12. Corporate Websites Definition: Websites of companies or businesses not primarily focused on retail. These typically promote services, technology, or business offerings. Includes: IBM.com, McKinsey.com, Salesforce.com, Airbus.com. Note: If the site includes a blog or content section, still classify based on the primary business focus. \\ 13. Educational Platforms Definition: Websites offering structured learning materials or formal education. Includes: Coursera, Khan Academy, edX, university portals. \\ 14. Search Engines / Aggregators Definition: Websites that collect or index content from across the web, typically for discovery or comparison. Includes: Google, DuckDuckGo, Booking.com, Yummly, Skyscanner. \\ 15. Utilities / Tools Definition: Websites designed to perform specific functions or services for the user. Includes: Grammarly, Canva, Weather.com, UnitConverters.net, PDF editors. \\ 16. Blogs (Personal or Independent) Definition: Websites operated by individuals or small groups, typically offering personal views, advice, or experiences. Includes: Personal travel blogs, lifestyle advice blogs, independent experts’ sites; Can cover many topics (e.g., food, parenting, productivity). \\ 17. Adult / Gambling / Restricted Definition: Websites intended for adult audiences or involving restricted activities. Includes: Pornhub, Bet365, online casinos; Age-gated or legally restricted platforms. \\ 18. Fact-Checkers. Definition: Websites primarily dedicated to verifying, debunking, or clarifying factual claims, public statements, viral content, and media narratives. These sites conduct independent analysis to promote factual accuracy in public discourse. Includes: Dedicated fact-checking organizations and journalism initiatives (e.g. Snopes, PolitiFact, FactCheck.org); platforms that evaluate politicians' statements, media stories or viral claims; use transparent methodologies, citations and ratings (e.g. "True", "False", "Misleading"). Distinct from: News (Official News organizations), while some are affiliated with newsrooms, their core function is verifications, not general reporting; Media publications not focused on topical, lifestyle, or niche journalism; Reference while they provide factual content, fact-checkers are evaluative and reactive to claims, not general-purpose encyclopedias or databases. \\ You will receive a list of domains and I would like for you to output a json with: \\ \{"domain\_1":\{"category": the correct category according to the type of website the domain represents, "topic": "topic of the website", "domain\_2": ....., \}. I would like you to search on the web when necessary to be sure of the category of the website. Additionally, when none of the categories of websites listed in your instructions fits the domain in question, you should retrieve "Other".}\\ \hline
\end{longtable}

\clearpage
\section{Additional Information on the Media Data Collection}\label{media_data_collection}

\subsection{Terms used to collect MediaCloud News}\label{terms_media}
\begin{small}
\renewcommand{\arraystretch}{1.2}
\begin{longtable}{p{1.7cm} p{1.7cm} p{12.8cm}}

\caption{\textbf{Terms used to form queries to collect news webpages from Media Cloud for the EU Parliament Election Case Study.} 
Category G1 contains terms directly tied to EU elections; Category G2 contains broader EU political terms; Category G3 contains general electoral terms. Queries included all Category 1 terms, as well as pairwise combinations of Categories 2 and 3 using an AND operator. `DE' stands for German, `EN - IR' stands for the terms in english used to search for news in Ireland, `PL' - Polish and `PT' for Portuguese.}
\label{tab:terms_media_EU_combined} \\

\toprule
\textbf{Category} & \textbf{Language} & \textbf{Terms EU Election} \\
\midrule
\endfirsthead

\toprule
\textbf{Category} & \textbf{Language} & \textbf{Terms} \\
\midrule
\endhead

\midrule
\multicolumn{3}{r}{\textit{Continued on next page}} \\
\endfoot

\bottomrule
\endlastfoot

% ---------------- CATEGORY 1 ----------------
\multirow{4}{*}{G1} & DE & EU-Parlament Wahlen, EU-Parlaments Wahlen, Europäische Parlamentswahlen, Europawahl, Europa-Wahl, Europawahlen, Europa-Wahlen, Europawahlkampf, Europa-Wahlkampf, Europawahlkampfauftritt, Europa-Wahlkampfauftritt, EU-Wahl, EU-Wahlen, EU-Wahlkampf, EU-Wahlkampfauftritt, Wahl für das EU-Parlament, Wahl für das Europäische Parlament, Wahl für das PE, Wahl für die EU, Wahlen der Europäischen Union, Wahlen des EU-Parlaments, Wahlen des Europäischen Parlaments, Wahlen EU, Wahlen EU-Parlament, Wahlen EU-Parlaments, Wahlen Europäische Parlament, Wahlen Europäische Parlaments, Wahlen Europäischen Parlament, Wahlen Europäischen Parlaments, Wahlen Europäischen Union, Wahlen für das EU, Wahlen für das EU-Parlament, Wahlen für das Europäische Parlament, Wahlen für die EU, Wahlen für EU, Wahlen in der EU, Wahlen zum EU, Wahlen zur EU \\
\cline{2-3}
& EN-IR & Election for the EU, Election for the European Parliament, Elections for the EU, Elections for the European Parliament, Elections to the EU Parliament, Elections to the European Parliament, EP election, EP elections, EU election, EU elections, EU parliament elections, European election, European elections, European Parliament elections, General election, General elections, Local election, Local elections \\
\cline{2-3}
& PL & Eurowyborach, Eurowyborami, Eurowyborów, Eurowybory, Wyborach do europarlamentu, Wyborach do Parlamentu Europejskiego, Wyborach do parlamentu UE, Wyborach do PE, Wyborach do UE, Wyborach europejskich, Wyborami do europarlamentu, Wyborami do Parlamentu Europejskiego, Wyborami do parlamentu UE, Wyborami do PE, Wyborami do UE, Wyborami europejskimi, Wyborów do europarlamentu, Wyborów do Parlamentu Europejskiego, Wyborów do parlamentu UE, Wyborów do PE, Wyborów do UE, Wyborów europejskich, Wybory do europarlamentu, Wybory do Parlamentu Europejskiego, Wybory do UE, Wybory europejskie \\
\cline{2-3}
& PT & Eleição da UE, Eleição europeia, Eleição para a UE, Eleição para o Parlamento europeu, Eleições ao PE, Eleições da UE, Eleições europeias, Eleições para a UE, Eleições para o Parlamento europeu \\
\midrule
% ---------------- CATEGORY 2 ----------------
\multirow{4}{*}{G2}& DE & Abstimmung in der EU, EU-Abgeordnete, EU-Abgeordneter, EU-Abstimmung, EU-Kandidat, EU-Parlament, EU-Partei, EU-Parteien, Europaabgeordnete, Europaabgeordneter, Europäische Abstimmungen, Europäische Partei, Europäische Parteien, Europäischen Parlament, Europäisches Parlament, Europaparlament, EU-Stimme, EU-Stimmen, Mitglied des Europäischen Parlaments, Mitglieder des Europäischen Parlaments \\
\cline{2-3}
& EN-IR & E.U. votes, EU parliament, EU parties, EU party, EU vote, European deputies, European deputy, European parliament, European parties, European party, European votes, European voting, Member of the European Parliament, Members of the European Parliament, MEP, MEPs, Vote in the EU, Voting in the EU \\

& PL & Członek Parlamentu Europejskiego, Członkowie Parlamentu Europejskiego, Eurodeputowana, Eurodeputowani, Europarlament, Europejska partia, Europejskie partie, Europoseł, Parlament Europejski, Parlament UE, Partie Europejskie, Partie UE, Poseł do Parlamentu Europejskiego \\
\cline{2-3}
& PT & Deputada europeia, Deputadas europeias, Deputado europeu, Deputados europeus, Eurodeputada, Eurodeputado, Parlamento europeu, Partido europeu, Partidos europeus, Votação europeia, Votação na UE, Voto na UE \\
\midrule
% ---------------- CATEGORY 3 ----------------
\multirow{4}{*}{G3} & DE & Debatte, Debatten, Kampagne, Kandidat, Kandidaten, Kandidatin, Kandidatinnen, Konvention, Kundgebung, Kundgebungen, Meinungsumfrage, Meinungsumfragen, Partei, Parteien, Parteitag, Spitzenkandidat, Spitzenkandidaten, Stimme, Stimmen, Umfrage, Umfragen, Wahl, Wahlen, Wähler, Wahlkampf, Wahlkampfveranstaltung \\
\cline{2-3}
& EN - IR & Campaign, Candidate, Candidates, Convention, Debate, Debates, Election, Elections, Lead candidate, Lead candidates, Nominee, Nominees, Parties, Party, Poll, Polls, Rallies, Rally, Vote, Voter, Voters, Votes \\
\cline{2-3}
& PL & Debata, Debaty, Głos, Głosowanie, Kandydat, Kandydaci, Kampania, Konwencja, Partia, Partie, Sondaż, Sondaże, Wiec, Wybory, Wyborca, Wyborcy \\
\cline{2-3}
& PT & Cabeça de lista, Campanha, Candidata, Candidato, Comício, Convenção, Debate, Eleição, Eleições, Eleitor, Eleitores, Nomeado, Partido, Partidos, Sondagem, Voto, Votos \\

\end{longtable}
\end{small}

% ============================================================
% Table for US Terms
% ============================================================
\begin{table}[ht!]
\small
\centering
\renewcommand{\arraystretch}{1.3}

\begin{tabular}{p{1.7cm} p{1.7cm} p{12.8cm}}
\hline
\textbf{Category} & \textbf{Language} & \textbf{Terms US Election} \\
\hline

G1 & EN & 2024 election, 2024 elections, Congressional election, Congressional elections, 
House election, House elections, House of Representatives election, 
House of Representatives elections, Presidential election, Presidential elections, 
Senate election, Senate elections, U.S. election, U.S. elections, 
U.S. House election, U.S. House elections, United States election, 
United States elections, US election, US elections, US House election, 
US House elections \\

\hline

G2 & EN & Biden, Congress, Democrat, Democratic Party, Democrats, 
Electoral College, Electors, Harris, House of Representatives, 
Member of the House, Presidential, Presidential debate, Presidentials, 
Republican, Republican Party, Republicans, Senate, Senator, Trump, 
Vance, Waltz, White House \\

\hline

G3 & EN & Campaign, Candidate, Candidates, Convention, Debate, Debates, 
Election, Elections, Lead candidate, Lead candidates, Nominee, Nominees, 
Parties, Party, Poll, Polls, Rallies, Rally, Seat, Seats, Senate race, 
Vote, Voter, Voters, Votes \\

\hline
\end{tabular}
\caption{\textbf{Terms used to form queries to collect news webpages from Media Cloud associated with the 2024 US Presidential Election.} Terms are organized in three categories: Category G1 contains terms directly tied to the 2024 US Presidential Election; Category G2 contains terms associated with US politics more broadly; and Category G3 contains general electoral terms applicable across election contexts.}
\label{tab:en_us_media_terms}
\end{table}

\clearpage
\subsection{Top 20 Newspapers per country}\label{top_news}
\begin{table*}[ht!]
\centering
\caption{Top 20 media outlets in \textit{Semrush} rankings for each country, excluding social media platforms and search engines. Media sources shown in \textbf{bold} are those for which news articles were retrieved from Media Cloud, with the other ones not having News in the API. These ranks were consulted in January 2025 for all European countries and in February the same year, for the United States.}
\label{tab:tab_media_sources_considered}

\renewcommand{\arraystretch}{1.2}
\setlength{\tabcolsep}{6pt}
\small

\begin{tabular}{
    l
    >{\RaggedRight\arraybackslash}p{13cm}
}
\toprule
\textbf{Country} & \textbf{Top \textit{Semrush}} \\
\midrule

Austria &
{\ttfamily\bfseries bild.de}, {\ttfamily\bfseries derstandard.at}, {\ttfamily\bfseries diepresse.com}, \texttt{exxpress.at}, {\ttfamily\bfseries focus.de}, \texttt{heute.at}, \texttt{kleinezeitung.at}, {\ttfamily\bfseries krone.at}, {\ttfamily\bfseries kurier.at}, \texttt{meinbezirk.at}, {\ttfamily\bfseries n-tv.de}, {\ttfamily\bfseries nachrichten.at}, {\ttfamily\bfseries orf.at}, {\ttfamily\bfseries spiegel.de}, {\ttfamily\bfseries tt.com}, {\ttfamily\bfseries vol.at} \\

\addlinespace

Germany &
{\ttfamily\bfseries bild.de}, \texttt{finanzen.net}, \texttt{fr.de}, {\ttfamily\bfseries faz.net}, {\ttfamily\bfseries focus.de}, {\ttfamily\bfseries mdr.de}, {\ttfamily\bfseries merkur.de}, {\ttfamily\bfseries n-tv.de}, {\ttfamily\bfseries spiegel.de}, \texttt{sueddeutsche.de}, {\ttfamily\bfseries tagesschau.de}, {\ttfamily\bfseries tagesspiegel.de}, {\ttfamily\bfseries t-online.de}, {\ttfamily\bfseries welt.de}, {\ttfamily\bfseries zeit.de} \\

\addlinespace

Ireland &
{\ttfamily\bfseries bbc.co.uk}, {\ttfamily\bfseries bbc.com}, {\ttfamily\bfseries dailymail.co.uk}, \texttt{globo.com}, {\ttfamily\bfseries independent.ie}, {\ttfamily\bfseries irishexaminer.com}, {\ttfamily\bfseries irishtimes.com}, {\ttfamily\bfseries nytimes.com}, {\ttfamily\bfseries rte.ie}, \texttt{rip.ie}, {\ttfamily\bfseries sky.com}, {\ttfamily\bfseries telegraph.co.uk}, \texttt{thesun.ie}, {\ttfamily\bfseries theguardian.com}, \texttt{tvn24.pl} \\

\addlinespace

Poland &
{\ttfamily\bfseries dorzeczy.pl}, \texttt{fakt.pl}, {\ttfamily\bfseries gazeta.pl}, \texttt{kwejk.pl}, \texttt{niezalezna.pl}, \texttt{o2.pl}, \texttt{plejada.pl}, \texttt{pomponik.pl}, \texttt{pudelek.pl}, {\ttfamily\bfseries polsatnews.pl}, \texttt{se.pl}, {\ttfamily\bfseries sport.pl}, {\ttfamily\bfseries rmf24.pl}, {\ttfamily\bfseries tvn24.pl}, {\ttfamily\bfseries wpolityce.pl}, {\ttfamily\bfseries wyborcza.pl} \\

\addlinespace

Portugal &
\texttt{abola.pt}, {\ttfamily\bfseries cmjornal.pt}, {\ttfamily\bfseries expresso.pt}, \texttt{flashscore.pt}, {\ttfamily\bfseries globo.com}, {\ttfamily\bfseries iol.pt}, \texttt{jn.pt}, {\ttfamily\bfseries noticiasaominuto.com}, \texttt{ojogo.pt}, {\ttfamily\bfseries observador.pt}, {\ttfamily\bfseries publico.pt}, {\ttfamily\bfseries record.pt}, {\ttfamily\bfseries rtp.pt}, {\ttfamily\bfseries sapo.pt}, \texttt{tempo.pt}, \texttt{zerozero.pt} \\
\addlinespace

US &
{\ttfamily\bfseries cnn.com}, {\ttfamily\bfseries nytimes.com}, {\ttfamily\bfseries foxnews.com}, {\ttfamily\bfseries dailymail.co.uk}, {\ttfamily\bfseries bbc.com}, {\ttfamily\bfseries people.com}, {\ttfamily\bfseries apnews.com}, {\ttfamily\bfseries breitbart.com}, {\ttfamily\bfseries theguardian.com}, {\ttfamily\bfseries usatoday.com}, {\ttfamily\bfseries forbes.com}, {\ttfamily\bfseries drudgereport.com}, {\ttfamily\bfseries washingtonpost.com}, {\ttfamily\bfseries nbcnews.com}, {\ttfamily\bfseries thegatewaypundit.com} \\

\bottomrule
\end{tabular}
\end{table*}

% \detailcount{corpus_paper}
\end{document}